\documentclass[12pt]{article}
\usepackage{amsmath}
\usepackage{graphicx}
\usepackage{enumerate}
\usepackage{natbib}
\usepackage{url} 
\usepackage{algorithm} 
\usepackage{algpseudocode} 
\usepackage{amsmath, amsthm, amssymb, appendix, bm, graphicx, hyperref, booktabs, verbatim}
\usepackage{color,xcolor}
\usepackage{multirow,array}
\usepackage{soul}

\makeatletter
\newcommand\customfontsize{\@setfontsize\customfontsize{10}{8}} 
\makeatother

\usepackage[textwidth=2.5cm,size=customfontsize]{todonotes}
\usepackage[normalem]{ulem}

\newcommand{\blind}{1}

\newcommand{\indep}{{\stackrel{\rm indep.}{\sim}}}
\newcommand{\ind}{{\stackrel{\rm ind}{\sim}}}

\newtheorem{theorem}{Theorem}

\newtheorem{lemma}[theorem]{Lemma}

\newtheorem{proposition}[theorem]{Proposition}
\newtheorem{remark}[theorem]{Remark}

\definecolor{deletedcolor}{RGB}{255,0,0} 
\definecolor{replacedcolor}{RGB}{0,0,255} 

\def\cmin{$C_n(\hat{\boldsymbol{\theta}})$}
\def\ctrue{$C_n(\boldsymbol{\theta}^\star)$}

\def\pg{PG~1116+215}

\newcommand{\dvd}[1]{\textcolor{red}{[DvD: {#1}]}}

\newcommand{\HEAGOF}{{\tt HEAGOF}}

\begin{document}

\def\spacingset#1{\renewcommand{\baselinestretch}%
{#1}\small\normalsize} \spacingset{1}
\def\C{$C$}


\if1\blind
{
\title{Making high-order asymptotics practical: correcting goodness-of-fit test for astronomical count data}
  \author{
    Xiaoli Li\hspace{.2cm}\\
    Department of Statistics, The University of Chicago\hspace{.2cm}\\
    Yang Chen\thanks{All correspondence should be sent to ychenang@umich.edu.}\hspace{.2cm}\\
    Department of Statistics, University of Michigan, Ann Arbor\hspace{.2cm}\\
    Xiao-Li Meng\hspace{.2cm}\\
    Department of Statistics, Harvard University\hspace{.2cm}\\
    David A. van Dyk\hspace{.2cm}\\
     Department of Mathematics, Imperial College London\hspace{.2cm}\\
    Massimiliano Bonamente\hspace{.2cm}\\
    Department of Physics and Astronomy, University of Alabama in Huntsville\hspace{.2cm}\\
      Vinay L.\ Kashyap\hspace{.2cm}\\
    Center for Astrophysics $|$ Harvard \& Smithsonian}
  \maketitle
} \fi

\if0\blind
{
  \bigskip
  \bigskip
  \bigskip
  \begin{center} \LARGE
{Making high-order asymptotics practical: correcting goodness-of-fit test for astronomical count data}
\end{center}
  \medskip
} \fi

\bigskip
\begin{abstract}




The \C\ statistic is a widely used likelihood-ratio statistic for model fitting and goodness-of-fit assessments with Poisson data in high-energy physics and astrophysics. Although it enjoys convenient asymptotic properties, the statistic is routinely applied in cases where its nominal null distribution relies on unwarranted assumptions. Because researchers do not typically carry out robustness checks, their scientific findings are left vulnerable to misleading significance calculations. With an emphasis on low-count scenarios, we present a comprehensive study of the theoretical properties of \C\ statistics and related goodness-of-fit algorithms. We focus on common ``plug-in'' algorithms where moments of \C\ are obtained by assuming the true parameter equals its estimate. To correct such methods, we provide a suite of new principled user-friendly algorithms and well-calibrated p-values that are ready for immediate deployment in the (astro)physics data-analysis pipeline. Using both theoretical and numerical results, we show (a) standard $\chi^2$-based goodness-of-fit assessments are invalid in low-count settings, (b) na\"ive methods (e.g., vanilla bootstrap) result in biased null distributions, and (c) the corrected Z-test based on conditioning and high-order asymptotics gives the best precision with low computational cost. We illustrate our methods via a suite of simulations and applied astrophysical analyses. An open-source Python package is provided in a GitHub repository.

\end{abstract}

\noindent%
\vfill

\spacingset{1.9} 

\section{Introduction}
\label{sec:intro}

Astrophysics and high-energy physics make extensive use of count data, often in the form of the number of particles or photons detected as a function of time, energy, and other independent variables \citep[e.g.][]{eadie1971}. For 50 years advances in X-ray and $\gamma$-ray astronomy have been accelerated by state-of-the-art space-based  observatories, from early missions such as Uhuru~\citep[e.g.][]{giacconi1971} and HEAO-1 \citep[e.g.][]{rothschild1979} to the current generation, including the \textit{XMM-Newton} and \textit{Chandra} X-ray Telescopes \citep{aschenbach2002orbit,weisskopf2000} and the \textit{Fermi} Gamma-ray Space Telescope \citep{michelson2010fermi}. These missions have dramatically increased the volume of available data and generated a need for principled regression methods for extensive but sparse Poisson-distributed data. Given that the most distant and faint sources are often of particular scientific interest, the paucity of photons detected from such sources, and the inherent
difficulty of obtaining long and high count-rate data, it is fundamental that the Poisson nature of such counts is properly accounted for. The pioneering work of early X-ray astronomers laid the foundations for the use of likelihood-ratio methods for parameter estimation and goodness-of-fit assessment \citep[e.g.][]{cash1976,cash1979,baker1984}. As a result, the most used goodness-of-fit statistics based on the Poisson distribution is the \C\ statistic, which is based on the deviance of the Poisson log-likelihood.

In recent years, researchers in astronomy and astrophysics have investigated the benefits of using  \C\ statistics instead of the $\chi^2$-statistic for both model fitting and hypothesis testing, particularly goodness-of-fit tests. 
In a typical setting, the dataset is composed of a large number of low-count observations, e.g., the photon counts in each of a large number of high-resolution energy bins. In this paper, we focus on the analysis of the distribution of photon energy, known as spectral analysis, 
but the same considerations apply to images and light curves where the independent variables are location and time. More precisely, let $N_i$ be the photon count with expected value $s_i (\boldsymbol{\theta})$ 
in energy bin $i$, where $\boldsymbol{\theta}\in\boldsymbol{\Theta}\subset \mathbb{R}^d$, energy bin $i$ spans the energy range $[E_i, E_{i+1})$, and $i = 1, \ldots, n$. Given the nature of the astrophysical sources and the photon collection mechanisms, 
it is generally assumed~\citep[e.g.,][]{cash1979} that the $N_i$ follow independent Poisson distributions,
\begin{equation}\label{eqn:Poiss_ind_intro}
N_i \mid \boldsymbol{\theta} \stackrel{\rm indep.}{\sim} {\rm Poisson}(s_i (\boldsymbol{\theta}))
\quad \hbox{for} \quad i = 1,\ldots, n,
\end{equation}
where $s_i(\boldsymbol{\theta})$ is the spectral model and is typically regarded as a \textit{known} function of unknown parameters $\boldsymbol{\theta}$. We leave the full description of spectral models and instrumental effects to Section~\ref{sec:spectrum} and 
the data analyses in Section~\ref{sec:real_data}. 
Here we simply emphasize that the evaluation of $s_i(\boldsymbol{\theta})$ 
is often expensive, involving the summation over hundreds of terms, due to convolution with instrument photon distribution matrices. 

We focus on two aspect of the statistical analysis of (\ref{eqn:Poiss_ind_intro}): estimating  $\boldsymbol{\theta}$ with $\hat{\boldsymbol{\theta}}$ 
and assessing the goodness-of-fit of the model.  In the astronomy community, the $\chi^2$-statistic,
\begin{equation}
    \label{eq:chisq}
    \chi^2_n (\hat{\boldsymbol{\theta}}) := \sum_{i=1}^n \frac{(N_i - s_i(\hat{\boldsymbol{\theta}}))^2}{{\sigma_i}^2},
\end{equation}
is typically adopted for both model fitting (by finding the minimizer of $\chi^2_n (\boldsymbol{\theta})$) and goodness-of-fit assessment, where $\sigma_i^2 = s_i(\hat{\boldsymbol{\theta}})$ or $N_i$, which corresponds to the Neyman $\chi^2$ statistic \citep[e.g.][]{neyman1949} and the Pearson statistic \citep{pearson1900} respectively. In astronomy, people also use $\sigma_i^2 = \sqrt{N_i+1}+0.75$. 

The $\chi^2$-statistic is popularly employed by observational astronomers for fitting Poisson-distributed data despite the fact that it is known to produce seriously biased estimates in both the high-count and the low-count regimes \citep{humphrey2009chi2}.  To reduce this bias, \citeauthor{humphrey2009chi2} went on to advocate the use of the \C\ statistic, given by 
\begin{equation}
C_n (\boldsymbol{\theta}) = 2 \sum_{i=1}^n \left[s_i({\boldsymbol{\theta}}) - N_i \log s_i({\boldsymbol{\theta}}) - N_i + N_i \log N_i\right]  
\label{eqn:cstat}
\end{equation}
The maximum likelihood estimate $\hat{\boldsymbol{\theta}}$ under (\ref{eqn:Poiss_ind_intro}) is the minimizer of $C_n (\boldsymbol{\theta})$. Historically, the use of the \C\ statistic can be found in and is often attributed to \citet{cash1979}, although its early use can traced at least as far back as \citet{bishop1975}, where it is called the $G^2$-statistic. As pointed out by \citet{cash1979}, the terms in (\ref{eqn:cstat}) not involving $\boldsymbol{\theta}$ may be omitted for the purpose of parameter estimation; this simplified version of the statistic is sometimes called the Cash-statistic.  


{
\citet{kaastra2017use} proposes numerical approximations of the mean and variance of the plug-in \C\ statistic, $C_n(\hat{\boldsymbol{\theta}})$, under the assumption that the true parameter value $\boldsymbol{\theta}^\star$ is equal to $\hat{\boldsymbol{\theta}}$. Then, assuming the standardized statistic is Gaussian, \citet{kaastra2017use} shows how $C_n(\hat{\boldsymbol{\theta}})$ can be used in goodness-of-fit tests. Unfortunately, as we shall discuss, this and similar strategies break down with the moderately sized data sets that are common in practice.} \citet{bonamente2019distribution} studies the asymptotic distribution of $C_n(\boldsymbol{\theta}^\star)$ as the Poisson rate goes to infinity and also provides asymptotic approximations for $C_n(\boldsymbol{\theta}^\star)$, $C_n(\hat{\boldsymbol{\theta}})$, and 
$C_n(\boldsymbol{\theta}^\star) - C_n(\hat{\boldsymbol{\theta}})$ in the special case where the $s_i(\boldsymbol{\theta})$ are equal.
Furthermore, \citet{bonamente2019distribution} discusses implications (but not solutions) {for the theoretical limitations of plug-in \C\ statistics}
in the low to moderate-count regime. 

The rigorous application of the aforementioned methods requires high counts in each bin. 
However, in high-energy astrophysical practice, asymptotic results are often applied to low-count data \citep{perkins2000introduction} with potentially misleading results. {The primary contributions of this paper are new theoretical results and algorithms that are appropriate to the low-count regime.}


Another thread of relevant literature deploys Bayesian approaches and re-binning schemes to model low-count data in high-energy astrophysics. \citet{wood2002} proposes grouping bins to better justify the asymptotic $\chi^2$ distribution of the likelihood ratio test statistic in low-count settings. \citet{van2001analysis}, for example, model photon arrivals as a Poisson process and adopts Bayesian posterior simulation for parameter inference.  Model checking and Bayesian goodness-of-fit is assessed via posterior predictive model checks \citep{protassov2002}. Similarly, \citet{primini2014determining} and \citet{dona:etal:24} adopt a Bayesian approach under a Poisson model for high-energy astrophysical photon-count images. The Bayesian nature of these approaches means that they do not rely on large-count asymptotic results and can be applied to low-count data in a principled manner. This does not mean that estimates and error bars derived with these methods enjoy well-calibrated frequentist properties, but rather that they are founded on Bayesian principles rather than frequentist calibration.  
\citet{pollack2017bayesian} uses Bayesian Blocks \citep{scargle2013} to improve the binning of histograms in high-energy physics. In practice, however, most researchers simply bin the data with the aim of maintaining a minimum accumulated counts in a set of consecutive bin
and adopt the $\chi^2$ or \C\ statistic for goodness-of-fit assessment. This can be problematic from a statistical perspective, as we detail below.

Given the popularity of the \C\ statistic in high-energy astrophysics and the fact that the default software for processing and analyzing count data is based on the \C\ statistic \citep[e.g., XSPEC and SPEX,][]{arnaud1996,kaastra1996}, there is an urgent need for a thorough theoretical study of the \C\ statistic when adopted for model fitting and goodness-of-fit assessment. Novel methods for goodness-of-fit assessment based on the \C\ statistic, but which enjoy well-calibrated statistical properties, high precision, and low computational cost, are highly desirable and if available, will have immediate impact on the day-to-day practice of research astrophysicists. A natural outcome is updated default model fitting and goodness-of-fit assessment software that can be made available to all research astronomers through software environments such as 
{XSPEC} \citep{arnaud1996astronomical} and/or \texttt{sherpa} \citep{freeman2001sherpa}.  To meet this need, this paper builds the theoretical foundations and gives numerical demonstrations of an open source software package \texttt{High-Energy Astronomy Goodness-of-Fit (HEAGOF)}. We developed \HEAGOF\ for the broader  astronomical community and other researchers that need goodness-of-fit assessments for heterogeneous low-count Poisson data.
\HEAGOF\ enables researchers to directly obtain well-calibrated principled goodness-of-fit assessment in a straightforward manner without needing a full understanding of the technical details required
either to determine an appropriate approximation to the null distribution or to
define and fit a fully Bayesian model.


In astronomy, $C_n({\boldsymbol{\theta}})$ is widely used for both model fitting and goodness-of-fit testing, typically with explicit or implicit large-sample  Gaussian approximations. However, no thorough theoretical or numerical analysis has been performed to examine the validity of these approximations. We close this gap with a new theoretical framework and practical solutions in the settings where they are commonly used, e.g., with low-count high-resolution data. Based on the derived theoretical properties and extensive numerical studies, we give clear guidelines for the efficient and well-calibrated use of \C\ statistics in practice. In Section~\ref{sec:meth}, we present theoretical properties of the \C\ statistic and four algorithms to determine the statistical significance and/or $p$-value of goodness-of-fit tests based on the \C\ statistic.  Our theoretical results and algorithms are illustrated, validated, and compared via simulation studies in Section~\ref{sec:sim} and real data applications in Section~\ref{sec:real_data}. 

\section{Spectral Analysis in Astronomy}
\label{sec:spectrum}

High-energy astrophysical detectors record a photon count in each of a large number of digitized energy channels. The observed channel in which an individual photon is recorded is governed by a multinomial distribution that depends strongly on the photon's energy, or equivalently, on its wavelength. Wavelength is inversely proportional to energy and is the more appropriate independent variable in some settings like data collected with gratings in the light path. Thus, the channel counts can be viewed as a noisy manifestation of the true {spectrum}. 
The detector's ``Redistribution Matrix File'' (or RMF) tabulates the multinomial probabilities as a function of energy {or wavelength}, specifically ${\rm R}({X},i)$, where $i$ indexes the detector channels {where $X$ represents either scaled photon energy $E$ or wavelength $w$.}

The sensitivity of detectors vary with photon energies and channel counts are subject to background contamination, i.e., among the recorded counts are counts from other sources.  Sensitivity is quantified by the effective area $A({X})$ in units of cm$^2$ under the notion that sensitivity increases proportionally with the detector area; see \citet{vand:kang:04} for a statistical review of high-energy spectral analysis. Generally, ${\rm R}({X},i)$ and $A({X})$ are assumed known from instrumental calibration \citep[but see][]{chen:19:jasa, Xu_2014}

The expected Poisson counts under (\ref{eqn:Poiss_ind_intro}) can be formulated as
\begin{align}
s_i (\boldsymbol{\theta}) &= T\left(\int_{\underline{{X}}}^{\overline{{X}}}  {\rm R}({X},i) A({X}) g({X}, \boldsymbol{\theta}) \ {\rm d}{X} + B_i\right), 
\label{eq:si1}
\end{align}
where $T$ is the exposure time (seconds), $\boldsymbol{\theta}$ collects the parameters of an astrophysical spectral model $g({X}, \boldsymbol{\theta})$ 
 {(typically with units photons per cm$^{2}$ per second),} 
$(\underline{{X}}, \overline{{X}})$ is the range of energies {or wavelengths} considered, and $B_i$ is the expected background count in channel $i$. In practice, the integral in (\ref{eq:si1}) is replaced by the sum,
\begin{align}
    \hat{s}_i(\boldsymbol{\theta}) &= \sum_{j=1}^{J} R(\tilde{{X}}_j, i) A(\tilde{{X}}_j) g(\tilde{{X}}_j, \boldsymbol{\theta}) [{X}_{j+1}-{X}_j] + B_i.
    \label{eq:si2}
\end{align}
where the $[{X}_j, {X}_{j+1})$ are energy/wavelength bins with midpoints $\tilde{{X}}_j = ({X}_j + {X}_{j+1})/{2}$.

\theoremstyle{definition}
\newtheorem{examplex}{Example}
\begin{examplex}[Power Law] A 
power-law model can be written as
\begin{align}
    g(X,\boldsymbol{\theta}) = K 
    \cdot {\left({\frac{X}{X_0}}\right)^{\displaystyle -\Gamma}},
    \label{eq:po}
\end{align}
where 
$\boldsymbol{\theta} = (K,\Gamma)$, 
{with $K>0$ being a normalization constant, $\Gamma$ the characteristic index of the power-law spectrum, and $X_0$ is a reference value with the same units as $X$.  For energy spectra in the X-ray regime, $X=E$ and the reference point often used is $E_0=1$~keV.  In this scaling, $\Gamma$ tends to be in the range $\sim1.4-3.5$ for many astrophysical sources.}
Model (\ref{eq:po}) is used in a wide variety of standard astronomical data analyses including spectral and light curve (time series) analyses \citep[e.g., for segments of gamma-ray bursts,][]{nousek2006}. For spectra, (\ref{eq:po}) is commonly used for non-thermal emission, such as those originating from synchrotron or inverse-Compton radiation \citep[e.g.,][]{laor1994}, for other smooth spectra without sharp features such as emission or absorption lines, and when analyzing a narrow range of energies. \citet{wong:etal:14} provides statistical discussion of power-laws and broken power-laws as generalized Pareto distributions. 
\end{examplex}

\begin{examplex}[Power Law + Spectral Line] Spectral lines are narrow features in spectra that represent either excess emission or absorption of photons relative to the baseline spectrum (e.g., the power-law). Such features can be used to infer the composition, density, {velocity} and/or temperature of a source \citep{yu:etal:24}. When electrons in a hot plasma transition from a higher energy level to a lower level of an atom, for example, a photon with energy equal to the difference in level energies may be emitted. This results in an excess 
of photons with energies in a specific narrow range and appears as a positive line relative to the baseline spectrum \citep[e.g.,][]{protassov2002,vand:kang:04}. Since the energy levels and their differences are idiosyncratic to specific ion species the central energies of these lines indicate the presence of a particular ion species \citep[via quantum mechanics calculations, e.g.,][]{verner1996}. The relative strength of the lines measures the relative frequency of electron transitions between particular levels which can be used to estimate the composition, temperature, and density of the emitting plasma. Photons passing through a gas can be absorbed by atoms in the gas at differential rates depending on the composition of the gas and energy of the photons. This leads to narrow negative aberrations from the baseline spectrum in the observed data. Such negative features are known as absorption and lines occur in a variety of astrophysical contexts \citep[e.g.][]{werk2013}. 
Although the theoretical line profile can be quite complex, 
both emission and absorption lines can be modeled as delta functions, step-functions, narrow Gaussian, or $t$ distributions that are added to or subtracted from the baseline spectral model, see Section~\ref{sec:sim}. 
\end{examplex}

In this article, we focus on the goodness-of-fit test of the null model 
\begin{equation}
    \label{eqn:model}
    H_0:s_i(\boldsymbol{\theta})
    =\sum^{J+1}_{j=1}c_{ij}\tilde g_j(\boldsymbol{\theta}), 
\end{equation}
against the fully saturated alternative in (\ref{eqn:alt}), where the $c_{ij} \geq0$ are constants and the $\tilde g_j(\cdot)$ are non-negative smooth {functions of $\boldsymbol{\theta}$}; both the $c_{ij}$ are $\tilde g_j(\cdot)$ depend on the binning scheme. Setting $\tilde g_j(\boldsymbol{\theta}) = g(\tilde{{X}}_j, \boldsymbol{\theta})$ and $c_{ij}=R(\tilde{{X}}_j, i) A(\tilde{{X}}_j) [{X}_{j+1}-{X}_j]$ for $j=1,\ldots, J$, $g_{J+1}(\boldsymbol{\theta})=1$ and $c_{i,J+1}=B_i$ we see that Model~(\ref{eq:si2}) is a special case of (\ref{eqn:model}). These models satisfy the relevant Regularity Conditions (B) described in Section~\ref{sec:asymptotic} and we assume that the $s_i(\boldsymbol{\theta}^\star)$ are uniformly bounded from below and above (see Lemma~\ref{lemma:divide_into_infinite_bins}), and that the number of estimated parameters, $d$, is fixed.  

Evaluation of the likelihood function with high-resolution data can be computationally demanding (with $J$ ranging from $\sim$100 to $\sim$100,000). Thus, the computational efficiency of fitting and evaluating the model with the \C\ statistic is a concern in practice. Researchers have attempted to tackle this problem by introducing data-driven (block) binning algorithms; this reduces the number of independent observations, ideally without significant loss of estimation efficiency \citet[e.g.,][]{kaastra2016optimal,pollack2017bayesian}.

\section{Model and Methodology}
\label{sec:meth}


{Following the notation in Section~\ref{sec:intro}}, we assume {the observed counts in each bin, $N_1, \ldots, N_n$, can be modeled as} independent Poisson variables as in (\ref{eqn:Poiss_ind_intro}). 
The binning of the detector, 
is fixed and determined in advance of data collection.  Typically the observations are obtained at a high resolution, i.e., there are numerous narrow bins, resulting in the counts in many of the bins being small. In other words, we are faced with high-resolution low-count data. Narrow bins are used in the hope of identifying high-resolution spatial, spectral, or temporal features -- such features
would be indistinguishable if fewer larger bins were used \citep[e.g., for the detection of narrow absorption lines,][]{nicastro2018,spence2023}.

\subsection{\C\ statistics as Likelihood Ratio Statistics}

The \C\ statistic can be defined in terms of a likelihood ratio statistics for the null hypothesis  
\begin{equation}\label{eqn:null}
   H_0: N_i \mid \boldsymbol{\theta} \stackrel{\rm indep.}{\sim} {\rm Poisson}(s_i (\boldsymbol{\theta})) 
   \ \hbox{ with }  \ \boldsymbol{\theta}\in \mathbb{R}^d
\end{equation}
against the alternative
\begin{equation}\label{eqn:alt}
   H_1: N_i  \stackrel{\rm indep.}{\sim} {\rm Poisson}(s_i)
   \ \hbox{ with }  \ (s_1,\ldots, s_n) \in \mathbb{R}^n_+,
\end{equation}
for $i=1\ldots, n$
where $d$ is the dimension of $\boldsymbol{\theta}$ under the null.
Let $\hat{\boldsymbol{\theta}}$ denote the maximum likelihood estimate for $\boldsymbol{\theta}$ under the null model. 
It is easy to show that the log-likelihood ratio statistic 
is the \textit{\C\ statistic}, 
obtained by plugging the MLE, $\hat{\boldsymbol{\theta}}$, into the \textit{\C\ function} defined in (\ref{eqn:cstat}). 

Although $C_n (\boldsymbol{\theta})$ is often referred to in the astrophysics literature as the \C\ statistic, we emphasize that $C_n (\boldsymbol{\theta})$ is not a statistic, but a random variable with unknown parameter $\boldsymbol{\theta}$. Hence, we refer to it as the \textit{\C\ function}, where the random variables are $N_i$, for $i=1,\dots,n$.  
{(For consistency with the literature,  $C_n(\boldsymbol{\theta})$ suppresses the dependence of both the \textit{\C\ function} 
and the \textit{\C\ statistic} on $N_1,\ldots, N_n$.)}
Confusing the \textit{\C\ statistic} 
with the \textit{\C\ function}, 
e.g., by assuming $C_n(\hat{\boldsymbol{\theta}})$ equals $C_n(\boldsymbol{\theta}^\star)$
where $\boldsymbol{\theta}^\star$ is the true parameter value, has led to an erroneous stream of exact inference results \citep[e.g.,][]{kaastra2017}.

Wilk's theorem \citep{wilks1943, wilks1938}, which states that under certain regularity conditions, the likelihood ratio statistic converges to a chi-square distribution, is closely associated with the routine application of \C\ statistics in astrophysics.  
{While this is a useful approach when the expected counts are moderate to large in \emph{every} bin (see Section~\ref{sec:sim}), it can be highly misleading when applied to a data set with a few counts spread over many bins. Following~\cite{mccullagh1986} and ~\cite{Paul2000JRSSB}, we instead consider results that are asymptotic in the number of bins. In this framework, }
the dimension of the parameter under the saturated model is $n$, which is not fixed, as required by Wilk's theorem. Thus, Wilk's theorem 
does not apply and the distribution of $C_n(\hat{\boldsymbol{\theta}})$ does not converge to a $\chi^2_{n-d}$ distribution as $n\rightarrow\infty$.  



Consider a dataset generated under (\ref{eqn:Poiss_ind_intro}) with $\boldsymbol{\theta}^{\star}$ and
assume that $\sum_{i=1}^n s_i(\boldsymbol{\theta}^{\star})$ is large. This may correspond to a set of  $s_i(\boldsymbol{\theta}^{\star})$ that includes large values and/or a set with a large $n$. 
The infinite divisibility property of the Poisson distribution allows an equivalent representation of model \eqref{eqn:Poiss_ind_intro}, i.e., a representation with the same likelihood, where each $s_i(\boldsymbol{\theta}^{\star})$ is bounded above but $n$ is large (or in the limit, goes to infinity); see Lemma~\ref{lemma:divide_into_infinite_bins} 
in Appendix~\ref{appendix:proof_divide_into_infinite_bins}. Thus, without loss of generality, for the rest of the paper, we may assume that the $s_i(\boldsymbol{\theta}^\star)$ are bounded and that $n$ is large.

\subsection{Theoretical Properties of \C\ statistics}
\label{sec:asymptotic}

To derive the asymptotic distributions of $C_n(\boldsymbol{\theta}^\star)$ and $C_n(\hat{\boldsymbol{\theta}})$, 
we define
\begin{equation*}
    C^{(i)}(\boldsymbol{\theta}) = 2[s_i(\boldsymbol{\theta})-N_i\log s_i(\boldsymbol{\theta})-N_i+N_i\log N_i],\quad i = 1,\ldots, n
\end{equation*}
so that $C_n(\boldsymbol{\theta})=\sum_{i=1}^n C^{(i)}(\boldsymbol{\theta})$.
Let the log-likelihood 
under $H_0$ be
\begin{align}
\label{eq:loglike}
    \ell (\boldsymbol{\theta}) =\sum_{i=1}^n \ell_i(\boldsymbol{\theta}) =\sum_{i=1}^n \log f_i(N_i\mid \boldsymbol{\theta}),
\end{align}
where $f_i(\cdot\mid \boldsymbol{\theta})$ is the probability mass function of the Poisson distribution with rate $s_i(\boldsymbol{\theta})$. We write the derivative of the log-likelihood as the $d\times 1$ vector
\begin{equation}
\label{eq:derivative_loglike}
\boldsymbol{D}\ell (\boldsymbol{\theta}) = \sum_{i=1}^n \boldsymbol{D}\ell_i(\boldsymbol{\theta})=\sum_{i=1}^n\frac{\partial \ell_i (\boldsymbol{\theta})}{\partial\boldsymbol{\theta}},
\end{equation}
and the expected information matrix as the $d\times d$ matrix
\begin{equation}
    \label{eq:information}
    I(\boldsymbol{\theta}) =\lim\limits_{n\to\infty}I_n(\boldsymbol{\theta})=\lim\limits_{n\to\infty}\frac1n{\rm Cov}_{\boldsymbol{\theta}}(\boldsymbol{D}\ell(\boldsymbol{\theta}))=\lim\limits_{n\to\infty}\frac1n\sum^n_{i=1}{\rm Cov}_{\boldsymbol{\theta}}(\boldsymbol{D}\ell_i(\boldsymbol{\theta})).
\end{equation}

\noindent
We assume standard regularity conditions \citep[e.g.,][]{mccullagh1986,muller2003Statistics}.

\noindent
{\bf Regularity Conditions (B)}\label{conditionB}
\begin{description}
    \item (B1) $\sqrt{n}(\hat{\boldsymbol{\theta}}-\boldsymbol{\theta}^\star)\to N(\boldsymbol{0},I^{-1}(\boldsymbol{\theta}^\star)),\quad \mathrm{as}\quad n\to\infty.$
    \item (B2)  
    There exists $\epsilon>0$ such that $s_i (\boldsymbol{\theta})\ge\epsilon$ for each $i$ and any $\boldsymbol{\theta}$.
\end{description}

We denote unconditional expectations as $\mathbb{E}_{\boldsymbol{\theta}}[\cdot]$ and conditional expectations as $\mathbb{E}_{\boldsymbol{\theta}}[\cdot\vert\hat{\boldsymbol{\theta}}]$, where the subscript is the parameter value that defines the underlying measure.  For example, the expectation of the \C\ function and the \C\ statistic evaluated at $\hat{\boldsymbol{\theta}}$ are
\begin{align}
  \mathbb{E}_{\boldsymbol{\theta}^\prime} [ \, C_n({\boldsymbol{\theta}}) \, ]
  &=\int   C_n({\boldsymbol{\theta}})
   dP_{\boldsymbol{\theta}^\prime} (N_1,\cdots,N_n) \quad \hbox{and}
   \label{eq:marg.mean} \\
  \mathbb{E}_{\boldsymbol{\theta}^\prime} [ \, C_n(\hat{\boldsymbol{\theta}}) \, ]
  &=\int   C_n(\hat{\boldsymbol{\theta}})
   dP_{\boldsymbol{\theta}^\prime} (N_1,\cdots,N_n 
   ), 
   \label{eq:marg.mean.at.mle}
\end{align} 
respectively.
Similarly, we define the conditional expectation of the \C\ statistic as
\begin{equation}
  \mathbb{E}_{\boldsymbol{\theta}^\prime} 
  [\, C_n(\hat{\boldsymbol{\theta}})
        \mid \hat{\boldsymbol{\theta}}\,]
  =\int   C_n(\hat{\boldsymbol{\theta}})
dP_{\boldsymbol{\theta}^\prime} (N_1,\cdots,N_n \mid \hat{\boldsymbol{\theta}})  .
   \label{eq:cond.mean}
\end{equation}
The variances 
${\rm Var}_{\boldsymbol{\theta}^\prime} [ \, C_n({\boldsymbol{\theta}}) \, ]$,
${\rm Var}_{\boldsymbol{\theta}^\prime} [ \, C_n(\hat{\boldsymbol{\theta}}) \, ]$,
and 
${\rm Var}_{\boldsymbol{\theta}^\prime} 
  [\, C_n(\hat{\boldsymbol{\theta}})
        \vert \hat{\boldsymbol{\theta}}\,]$
are defined using the same measures as in (\ref{eq:marg.mean}), (\ref{eq:marg.mean.at.mle}) and (\ref{eq:cond.mean}), respectively. 
%
Because $\boldsymbol{\theta}^\star$ is unknown, we can only compute moments that depend solely on an estimate of $\boldsymbol{\theta}$, such as $\hat{\boldsymbol{\theta}}$. For example, we can compute $\mathbb{E}_{\hat{\boldsymbol{\theta}}} [ \, C_n({\hat{\boldsymbol{\theta}}}) \, ]$, 
and
${\rm Var}_{\hat{\boldsymbol{\theta}}}[C_n(\hat{\boldsymbol{\theta}})]$,
but not
$\mathbb{E}_{{\boldsymbol{\theta}}^\star}[C_n(\hat{\boldsymbol{\theta}})]$, 
or
${\rm Var}_{{\boldsymbol{\theta}^\star}}[C_n(\hat{\boldsymbol{\theta}})]$.

Our asymptotic theory is divided into non-computable results that require $\boldsymbol{\theta}^\star$ (Theorem~\ref{theorem:CAN of C_true}) and computable results  that require only $\hat{\boldsymbol{\theta}}$ (Theorem~\ref{thm:cmin}).
When $\boldsymbol{\theta}^\star$ is known and $N_i$ are independently Poisson distributed, $\{C^{(i)}(\boldsymbol{\theta}^\star), i=1,\ldots, n\}$ are independently distributed random variables. 
Consequently, we can prove the following (see Appendix~\ref{proof:CAN of C_true}).
\begin{theorem}[Non-Computable Asymptotic Normality of \C\ statistics]
\label{theorem:CAN of C_true}
    Under $H_0$ and Regularity Conditions (B), the \C\ function evaluated at $\boldsymbol{\theta}^\star$ is asymptotic normal:
\begin{equation}\label{eq:marginal_normal_ctrue}
    {\rm(i)} \quad\frac{C_n(\boldsymbol{\theta}^\star)-\mathbb{E}_{\boldsymbol{\theta}^*}[C_n(\boldsymbol{\theta}^\star)]}{\sqrt{{\rm Var}_{\boldsymbol{\theta}^*}[C_n(\boldsymbol{\theta}^\star)]}}\stackrel{D}{\to} N(0,1) 
    \ \ \hbox{as} \ \ n\to\infty,
\end{equation}
and the \C\ statistic is asymptotic normal:
\begin{equation}\label{eq:marginal_normal_cmin}
{\rm(ii)} \quad\frac{C_n(\hat{\boldsymbol{\theta}})-\mathbb{E}_{\boldsymbol{\theta}^*}[C_n(\hat{\boldsymbol{\theta}})]}{\sqrt{{\rm Var}_{\boldsymbol{\theta}^*}[C_n(\hat{\boldsymbol{\theta}})]}}\stackrel{D}{\to} N(0,1)
\ \ \hbox{as} \ \ n\to\infty.
\end{equation}
\end{theorem}

In practice $\boldsymbol{\theta}^\star$ is unknown and we cannot directly use Equation~\eqref{eq:marginal_normal_ctrue} or \eqref{eq:marginal_normal_cmin} to derive a goodness-of-fit test.
In Theorem~\ref{thm:cmin}, we analyze the asymptotic distribution of $C_n(\hat{\boldsymbol{\theta}})$ and its limiting moments, whose expressions are functions of $\hat{\boldsymbol{\theta}}$. A proof appears in Appendix~\ref{proof:cmin}. 

\begin{theorem}[Computable Asymptotic Normality of \C\ statistics]
\label{thm:cmin}
Under $H_0$ and Regularity Condition (B), as $n\to\infty$, we have

\begin{equation}\label{eq:T_normal}
   {\rm(i)} \quad T_1=\frac{C_n(\hat{\boldsymbol{\theta}})-\mathbb{E}_{\hat{\boldsymbol{\theta}}} [ \, C_n(\hat{{\boldsymbol{\theta}}}) \, ]}{\sqrt{{\rm Var}_{\hat{\boldsymbol{\theta}}}(C_n(\hat{\boldsymbol{\theta}}))-Q(\hat{\boldsymbol{\theta}})}}\stackrel{D}{\to} N(0,1),
\end{equation}
and
conditional on $\hat{\boldsymbol{\theta}}$,
\begin{equation}\label{eq:conditional_normal_cmin}
    {\rm(ii)} \quad T_2=\frac{C_n(\hat{\boldsymbol{\theta}})-\mathbb{E}_{{\boldsymbol{\theta}}^*}[C_n(\hat{\boldsymbol{\theta}})\mid\hat{\boldsymbol{\theta}}]}{\sqrt{{\rm Var}_{{\boldsymbol{\theta}}^*}[C_n(\hat{\boldsymbol{\theta}})\mid\hat{\boldsymbol{\theta}}]}}\stackrel{D}{\to} N(0,1),
\end{equation}
where 
$Q({\boldsymbol{\theta}})
=\boldsymbol{c}^\top({\boldsymbol{\theta}}){I}_n^{-1}({\boldsymbol{\theta}})\boldsymbol{c}({\boldsymbol{\theta}})$, 
and
$\boldsymbol{c}({\boldsymbol{\theta}})
={\rm Cov}_{{\boldsymbol{\theta}}}\left\{
C_n({\boldsymbol{\theta}}),
\boldsymbol{D}\ell(\boldsymbol{\theta})
\right\}$. 
\end{theorem}



As noted in \citet{mccullagh1986}, if we only know $\hat{\boldsymbol{\theta}}$, it is the conditional distribution given $\hat{\boldsymbol{\theta}}$ that is relevant for assessing goodness of fit, not the marginal distribution. Thus, once we derive expressions for the conditional expectation and variance of $C_n(\hat{\boldsymbol{\theta}})$, we can use
Equation~\eqref{eq:conditional_normal_cmin} to formulate an appropriate goodness-of-fit hypothesis test and to quantify its significance via a $p$-value. Proposition~\ref{rmk:1_order_same} states that the expressions of these conditional moments coincide to first order with the unconditional moments $\mu(\hat{\boldsymbol{\theta}})=\mathbb{E}_{\hat{\boldsymbol{\theta}}} [ \, C_n(\hat{{\boldsymbol{\theta}}}) \, ]$  
and $\sigma^2(\hat{\boldsymbol{\theta}})={\rm Var}_{\hat{\boldsymbol{\theta}}}(C_n(\hat{\boldsymbol{\theta}}))-Q(\hat{\boldsymbol{\theta}})$ of Equation \eqref{eq:T_normal}. This result is also reported in \cite{osius1992} for multinomial models. A detailed proof is given in Appendix~\ref{proof:con_uncon_same}.
\begin{proposition}
    \label{rmk:1_order_same}
    To the first order, the unconditional moments in Equation~\eqref{eq:T_normal} are the same as the conditional moments in Equation~\eqref{eq:conditional_normal_cmin}, i.e.,
\begin{equation}\label{eq:con_error}
\begin{aligned}
    \mathbb{E}_{{\boldsymbol{\theta}}^*}[C_n(\hat{\boldsymbol{\theta}})\mid\hat{\boldsymbol{\theta}}]&=\mu(\hat{\boldsymbol{\theta}})+O_p(1),\\
    {\rm Var}_{{\boldsymbol{\theta}}^*}(C_n(\hat{\boldsymbol{\theta}})\mid\hat{\boldsymbol{\theta}})&=\sigma^2(\hat{\boldsymbol{\theta}})+O_p(n^{1/2}).
\end{aligned}    
\end{equation}
\end{proposition}

The quadratic form $Q$ in \eqref{eq:T_normal}, which is $O_p(n)$, involves the $d$-dimensional covariance vector of the sum $C_n(\boldsymbol{\theta})$ and the score vector $\boldsymbol{D}\ell(\boldsymbol{\theta})$; and $Q\to0$ provided the expected counts per bin $s_i(\boldsymbol{\theta})\to\infty$ uniformly.
Thus, if we simply use the unconditional results in Theorem~\ref{theorem:CAN of C_true}, e.g., by replacing $\boldsymbol{\theta}^\star$ with $\hat{\boldsymbol{\theta}}$ in Theorem~\ref{theorem:CAN of C_true}(ii),
asymptotic normality does not hold. {Unfortunately, this means that na\"ive plug-in methods, including the parametric bootstrap, which compute the mean and variance in \eqref{eq:marginal_normal_cmin} assuming $\boldsymbol{\theta}^\star =\hat{\boldsymbol{\theta}}$ are not valid for goodness-of-fit testing;}
see Appendix~\ref{proof:bootstrap_invalid}.


\begin{proposition}[Invalidity of Na\"ive Plug-in Methods]
\label{prop:bootstrap_invalid}
Under $H_0$ and Regularity Conditions (B), methods based on Theorem~\ref{theorem:CAN of C_true}, but with the mean and variance in \eqref{eq:marginal_normal_cmin}
computed assuming $\boldsymbol{\theta}^\star =\hat{\boldsymbol{\theta}}$ exhibit
a bias of order $O_p(1)$, which decreases to $0$ as $s_i\to\infty$ uniformly.
\end{proposition}


In practice, simulation studies show that this bias can be ignored when $s_i>1$ uniformly, but it is significant when most $s_i\le0.5$. Thus, the bootstrap method is not recommended for relatively sparse data; see Sections~\ref{sec:algorithm} and~\ref{sec:numerical} for details.

\subsection{High-order Approximations for Moments of \C\ statistics}
In this section, for simplicity, moments with respect to distributions given $\boldsymbol{\theta}^*$ are written
without a subscript, and, moments with respect to distributions given $\hat{\boldsymbol{\theta}}$ are written with a hat. For example, $\mathbb{E}[C_n(\boldsymbol{\hat\theta})]=\mathbb{E}_{\boldsymbol{\theta}^*}[C_n(\hat{\boldsymbol{\theta}})]$ and $\hat{\mathbb{E}}[C_n(\hat{\boldsymbol{\theta}})]=\mathbb{E}_{\hat{\boldsymbol{\theta}}}[C_n(\hat{\boldsymbol{\theta}})]=\mathbb{E}_{\boldsymbol{\theta}}[C_n(\hat{\boldsymbol{\theta}})]\mid_{\boldsymbol{\theta}={\hat{\boldsymbol{\theta}}}}$.

To improve the first-order normal-theory approximations to the distribution of \C\ statistic, it is necessary to compute the 
conditional mean and variance in \eqref{eq:con_error} up to and including terms that are $O_p(1)$ and
$O_p(n^{1/2})$, respectively, and to provide explicit formulae for computation. \citet{Paul2000JRSSB} derive the conditional distribution of $C_n(\hat{\boldsymbol{\theta}})$ under a generalized linear model via a high-order expansion. Because Model~(\ref{eqn:model}) does not fall under this framework, we extend the work of \citet{Paul2000JRSSB} without being restricted to the generalized linear model by using formulae for computing conditional cumulants provided by \cite{farrington1996JASA} and \cite{mccullagh1986}.
\begin{theorem}[Unconditional Mean and Variance of \cmin]
\label{thm:uncon_highorder}
Assume $N_i$ are independent Poisson variables with mean 
   $s_i(\boldsymbol{\theta})$ are functions with second-order continuous derivatives. Define the cumulants $\kappa_1^{(i)}=\mathbb{E}(C^{(i)})$, $\kappa_2^{(i)}=\mathbb{E}(C^{(i)}-\kappa_1^{(i)})^2$, where all expectations are with respect to $N_i\indep{\rm Poisson}(s_i)$, and we suppress the dependence on $\boldsymbol{\theta}$. Under Condition (B1),
    \begin{equation}
\begin{aligned}\label{eq:moments_uncon_Cmin}
    &\mathbb{E}[C(\hat{\boldsymbol{\theta}})]=\mathbb{E}[C({\boldsymbol{\theta}}^*)]-d+O(n^{-1})=\kappa_1^{(\cdot)}-d+O(n^{-1}),\\
    &{\rm Var}(C(\hat{\boldsymbol{\theta}}))={\rm Var}(C({\boldsymbol{\theta}}^*))+O(1)=\kappa_2^{(\cdot)}+O(1),
\end{aligned}
\end{equation}
where $\kappa_1^{(\cdot)}=\sum^n_{i=1}\kappa_1^{(i)} $ and $\kappa_2^{(\cdot)}=\sum^n_{i=1}\kappa_2^{(i)}$.
\end{theorem}

\begin{theorem}[Conditional Mean and Variance of \cmin]
    Let $X=(\nabla s_1^\top,\cdots,\nabla s_n^\top)^\top$, $V=\mathrm{diag}(s_i)$, 
    $Q=(Q_{ij})=X(X^\top V^{-1} X)^{-1}X^\top $ and define the cumulants $\kappa_3^{(i)}=\mathbb{E}(C^{(i)}-\kappa_1^{(i)})^3$, $\kappa_{11}^{(i)}=\mathbb{E}\{(C^{(i)}-\kappa_1^{(i)})(N_i-s_i)\}$, $\kappa_{12}^{(i)}=\mathbb{E}\{(C^{(i)}-\kappa_1^{(i)})(N_i-s_i)^2\}$, $\kappa_{21}^{(i)}=\mathbb{E}\{(C^{(i)}-\kappa_1^{(i)})^2(N_i-s_i)\}$ and $\kappa_{03}^{(i)}=\mathbb{E}(N_i-s_i)^3$. (For simplicity, we suppress the dependence of all these expressions on $\boldsymbol{\theta}$.) Under Condition (B1),
\begin{align}
\label{eq:thm5.mean}
    \mathbb{E}(C_n(\hat{\boldsymbol{\theta}})\mid \hat{\boldsymbol{\theta}})&=\hat{\kappa}_1^{(\cdot)}-\frac12\bm1^\top \hat{X}^\top \hat\Sigma \hat{X} (\hat{X}^\top\hat V^{-1}\hat{X})^{-1}\bm1+O(n^{-1/2}),\\
\label{eq:thm5.var}
    {\rm Var}(C_n(\hat{\boldsymbol{\theta}})\mid\hat{\boldsymbol{\theta}}) &=\hat{\kappa}_2^{(\cdot)}-\hat\kappa_{11}^\top \hat{X}(\hat{X}^\top\hat{V}^{-1}\hat{X})^{-1}\hat{X}^\top\hat\kappa_{11}+O(1),
\end{align}
    where $\Sigma=\mathrm{diag}\{\kappa_{12}^{(i)}/V_i^2-(\sum_j\kappa_{11}^{(j)}Q_{ji}/V_j)\kappa_{03}^{(i)}/V_i^3\}$, $\kappa_{11}=(\kappa_{11}^{(1)}/V_1,\cdots,\kappa_{11}^{(n)}/V_n)^\top$,
    and $\hat s_i, \hat X, \hat V, \hat Q, \hat \Sigma$ and the various $\hat\kappa$ are defined as above, but evaluated at $\hat{\boldsymbol{\theta}}$ rather than $\boldsymbol{\theta}$. For example, $\hat\kappa_1^{(i)}=\kappa_1^{(i)}(\hat{\boldsymbol{\theta}})= \mathbb{E}_{\boldsymbol{\hat\theta}}(C^{(i)})$.

    \label{thm:highorder} 
\end{theorem}
As a special case, when $s_i$ satisfies a log-linear regression model $\boldsymbol{\eta}=\tilde{X}\boldsymbol{\theta}$, where $\eta_i=\log s_i$ and $X=V\tilde{X}$, we have the following proposition. 
\begin{proposition}[Conditional Mean and Variance for log-linear Models]
    Under $H_0$ and Condition (B1), if the $s_i$ satisfy a log-linear regression $\boldsymbol{\eta}=X\boldsymbol{\theta}$ where $\eta_i=\log s_i$,  then 
    $$\mathbb{E}(C_n(\hat{\boldsymbol{\theta}})\mid \hat{\boldsymbol{\theta}})=\hat{\kappa}_1^{(\cdot)}-\frac12\bm1^\top X^\top \hat\Sigma X (X^\top\hat VX)^{-1}\bm1+O(n^{-1/2}),$$
    $${\rm Var}(C_n(\hat{\boldsymbol{\theta}})\mid\hat{\boldsymbol{\theta}})=\hat{\kappa}_2^{(\cdot)}-\hat\kappa_{11}^\top X(X^\top\hat{V}X)^{-1}X^\top\hat\kappa_{11}+O(1),$$
    where $Q=X(X^\top VX)^{-1}X^\top$, $\kappa_{11}=(\kappa_{11}^{(1)},\cdots,\kappa_{11}^{(n)})^\top$, $\Sigma=\mathrm{diag}\{\kappa_{12}^{(i)}-(\sum_j\kappa_{11}^{(j)}Q_{ji})\kappa_{03}^{(i)}\}$, and we use the same notational conventions as in Theorem~\ref{thm:highorder}.\label{prop:highorder}
\end{proposition}
The expressions for the conditional moments of $C_n(\hat{\boldsymbol{\theta}})$ obtained above are consistent with those given in \cite{Paul2000JRSSB} and \cite{mccullagh1986}.
Because the counts are independent Poisson variables, the cumulants can easily be obtained via direct summation. 

\subsection{Goodness-of-fit Algorithms based on  \C\ statistics}
\label{sec:algorithm}

{
Given the theoretical properties of $C_n(\boldsymbol{\theta}^\star)$ and $C_n(\hat{\boldsymbol{\theta}})$, we derive four sets of goodness-of-fit algorithms based on \C\ statistics: 

\paragraph{1) LR-$\chi^2$ test.} \texttt{Algorithm~1} uses the likelihood ratio test with a $\chi^2$ approximation based on
Wilks' theorem. Although 
\texttt{Algorithm~1} {is inappropriate for data consisting of a moderate number of counts spread across many bins, it}
is used in numerous scientific platforms and packages, including the current standard default astrophysical data processing package \texttt{sherpa} developed by~\cite{freeman2001sherpa}. 

\paragraph{2) Na\"ive Z-test.} We consider  theree variants of \texttt{Algorithm~2}, all are based on Theorem~\ref{theorem:CAN of C_true} but with the mean and variance in \eqref{eq:marginal_normal_cmin} computed assuming $\boldsymbol{\theta}^\star =\hat{\boldsymbol{\theta}}$. Different versions compute the moments differently. For instance, a popular common choice is bootstrap estimation (Bootstrap-normal, \texttt{Algorithm 2a}). ~\cite{kaastra2017use} uses polynomials to empirically approximate the moments (Kaastra's method, \texttt{Algorithm 2b}); \cite{bonamente2019distribution} subsequently provided a similar approximation with more complex functions. Finally, the equations in \eqref{eq:moments_uncon_Cmin} derived in this paper provide a new approximation for the moments (High-order, \texttt{Algorithm 2c}). We compare the performances of \texttt{Algorithm 2a, 2b, 2c} in Section~\ref{sec:compare_null_alg2} and conclude that the performances are similar. Thus, we refer to \texttt{Algorithm 2c} as the Na\"ive Z-test unless otherwise specified. 

\paragraph{3) Corrected Z-test.} \texttt{Algorithm~3} includes Z-tests based on the asymptotic normality of $C_n(\hat{\boldsymbol{\theta}})$, with corrected moments calculated from the expansions given in Theorems~\ref{thm:cmin} and~\ref{thm:highorder}. Either the first-order approximations in~\eqref{eq:con_error}, named \texttt{Algorithm 3a}, or the high-order approximations in~\eqref{eq:thm5.mean} and~\eqref{eq:thm5.var}, named \texttt{Algorithm 3b}, can be used. We use \texttt{Algorithm~3b} for the Corrected Z-test in our numerical studies, as its high-order expansions provide superior accuracy at the expense of an increase in computational cost relative to \texttt{Algorithm 3a}. 

\paragraph{4) Parametric Bootstrap.} The commonly used parametric bootstrap with empirical distribution does not assume normality. Instead, \texttt{Algorithm~4} uses the empirical distribution of a generated bootstrap sample to approximate the true distribution. Alternatively, the double bootstrap, which is much more time consuming and can be impractical, is discussed in Appendix~\ref{sec:algorithms_description} with computational time comparison in~\ref{sec:Label_time_appendix}.






Table~\ref{tab:algorithms_list} describes all four algorithms, including, \texttt{Algorithms 1, 2c, 3b, 4} that are used in numerical studies in the next section. Full details of all variants of the four types of algorithms appear in Appendix~\ref{sec:algorithms_description}.
}

\begin{table}[p!]
\spacingset{1.1}
        \caption{Goodness-of-Fit Algorithms compared in numerical studies.}
    \label{tab:algorithms_list}
        \centering
    \begin{tabular}{p{0.2cm}p{0.2cm}p{14.575cm}}
    \hline \hline
\multicolumn{3}{l}{{\bf Algorithm~1:} Likelihood ratio test with $\chi^2$ approximation (LR-$\chi^2$ test)} \\
&\multicolumn{2}{l}{\emph{Input:} $C_n(\hat{\boldsymbol{\theta}})$, number of bins $n$, and number of parameters $d$}\\
& 1.&Compute the $p$-value$^a$,
  $$
  \setlength{\abovedisplayskip}{0pt}
  \hat{p}=\sup_{p\in[0,1]}\{ C_n(\hat{\boldsymbol{\theta}})\le\chi^2_{n-d}(1-p)\}.$$\\
%
%
\multicolumn{3}{l}{{\bf Algorithm~2:}  $Z$-test with plug-in first-order moments (Na\"ive $Z$ test)} \\
&\multicolumn{2}{l}{\emph{Input:} $C_n(\hat{\boldsymbol{\theta}})$, $\hat{\boldsymbol{\theta}}$ and number of bins $n$.}\\
& 1. & Formulate  
$\mathbb{E}_{\boldsymbol{\theta}^\star}(C_n(\hat{\boldsymbol{\theta}}))$ and 
$\hbox{Var}_{\boldsymbol{\theta}^\star}(C_n(\hat{\boldsymbol{\theta}}))$
to approximate the unconditional mean and variance {of $C_n({\boldsymbol{\theta}})$}, 
e.g., via 
\texttt{Algorithm~2a}: the bootstrap,  \texttt{Algorithm~2b}: \cite{kaastra2017}'s empirical polynomial approximation, or \texttt{Algorithm~2c}: given by Eq. \eqref{eq:moments_uncon_Cmin}. 
\\
& 2.& Compute the estimated mean and variance by plugging in the estimator $\hat{\boldsymbol{\theta}}$, i.e., 
$$
\hat\mu= \mathbb{E}_{{\boldsymbol{\theta}^\star}}(C_n(\hat{\boldsymbol{\theta}}))\,\big|_{{\boldsymbol{\theta}^\star} = \hat{\boldsymbol{\theta}}},
\quad \hbox{and} \quad 
\hat{\sigma}^2= 
\hbox{Var}_{{\boldsymbol{\theta}^\star}}(C_n(\hat{\boldsymbol{\theta}})) \,\big|_{{\boldsymbol{\theta}^\star} = \hat{\boldsymbol{\theta}}}.$$\\
%
& 3. &Compute the $p$-value$^a$,
  $$
  \setlength{\abovedisplayskip}{0pt}
  \hat{p}=\sup_{p\in[0,1]}\left\{\frac{C_n(\hat{\boldsymbol{\theta}})-\hat{\mu}}{\hat{\sigma}}\le Z \left(1-p\right)\right\}.$$\\
%
\multicolumn{3}{l}{{\bf Algorithm~3:} $Z$-test with corrected moments (Corrected $Z$ test; Theorems~\ref{thm:cmin}, \ref{thm:highorder})}\\
&\multicolumn{2}{l}{\emph{Input:} $C_n(\hat{\boldsymbol{\theta}})$, $\hat{\boldsymbol{\theta}}$, and number of bins $n$.}\\
& 1. & Compute the asymptotic (a) mean and variance given in (\ref{eq:con_error}), i.e., \texttt{Algorithm~3a}; or (b) conditional mean and variance given in (\ref{eq:thm5.mean}) and (\ref{eq:thm5.var}), i.e., \texttt{Algorithm~3b} (omitting the $\cal O$ terms).\\
& 2. & Compute the $p$-value$^a$ with Z-test, for example, in \texttt{Algorithm 3b}, 
$$
\setlength{\abovedisplayskip}{0pt}
\hat{p}=\sup_{p\in[0,1]}\left\{\frac{C_n(\hat{\boldsymbol{\theta}})-\mathbb{E}[C_n(\hat{\boldsymbol{\theta}})\vert \hat{\boldsymbol{\theta}}]}{\sqrt{\mathrm{Var}(C_n(\hat{\boldsymbol{\theta}})\vert \hat{\boldsymbol{\theta}})}}\le Z\left(1-p\right)\right\}.$$\\
%
\multicolumn{3}{l}{{\bf Algorithm~4:} Parametric-Bootstrap} \\
&\multicolumn{2}{l}{\emph{Input:} $C_n(\hat{\boldsymbol{\theta}})$, $\hat{\boldsymbol{\theta}}$, number of bins $n$, and number of bootstrap replicates $B$.}\\
& 1. & For $m=1,\dots, B$ and $i=1,\ldots, n$, sample
$N_i^{[m]} \; \ind \; \rm{Poisson}(s_i(\hat{\boldsymbol{\theta}}))$. 
\newline
For each bootstrap replicate, compute $\hat{\boldsymbol{\theta}}^{(m)}$, 
$s_i (\hat{\boldsymbol{\theta}}^{[m]})$, and 
$C_n^{[m]}(\hat{\boldsymbol{\theta}}^{[m]})$.\\

& 2. & Compute the $p$-value$^a$,
  $$
  \setlength{\abovedisplayskip}{0pt}
  \hat{p}={1\over B}\sum^{B}_{m=1}1_{\{C_n^{[m]}(\hat{\boldsymbol{\theta}}^{[m]})\ge C_n(\hat{\boldsymbol{\theta}})\}}.$$\\
       \hline
    \end{tabular}
\footnotesize{
$^a$ $\chi^2_{n-d}(\cdot)$ and
$Z(\cdot)$ are the $\chi^2_{n-d}$ and
a standard normal cumulative distribution functions, respectively.
}
\end{table}

\if false
\dvd{Can we guide the reader through this a bit more? For example: Algorithm~1 is based on a flawed application of Wilk's theorem and thus is a straw-man baseline. Algorithm~2 is based on a Guassian approximation for the null distribution of the \C\ stat. Algorithm~3 is based on the higher-order asymptotics in Thm X and Y. Algorithm~4 uses variants of the bootstrap with no distributional assumptions. Xiaoli: Yes. Thank you!}

{
Given the theoretical properties of $C_n(\boldsymbol{\theta}^\star)$ and $C_n(\hat{\boldsymbol{\theta}})$, we derive four different sets of algorithms for the goodness-of-fit assessment using the \C\ statistics: (1) a likelihood ratio test with $\chi^2$ statistics based on the Wilk's theorem, detailed in Algorithm~\ref{algo:chisq}; (2) a Z-test based on the asymptotic normality of $C_n(\hat{\boldsymbol{\theta}})$, with mean and variance calculated from a parametric bootstrap, detailed in Algorithm~\ref{algo:bootstrap_normal} and Polynomial expansions as given in \citet{kaastra2017,bonamente2019}, detailed in Algorithm~\ref{algo:Kaastra&Max}; (3) a Z-test based on asymptotic normality of $C_n(\hat{\boldsymbol{\theta}})$, with mean and variance calculated from high-order expansions as given in Theorem~\ref{thm:cmin} and~\ref{thm:highorder}, detailed in Algorithm~\ref{algo:Ztest_conditional} and \ref{algo:Ztest_marginal}; and (4) a parametric bootstrap using $C_n(\hat{\boldsymbol{\theta}})$ as the test statistic, detailed in Algorithm~\ref{algo:bootstrap}, a bias-corrected bootstrap as detailed in Algorithm~\ref{algo:boot_b.c.}, and a double bootstrap as detailed in Algorithm~\ref{algo:P.DoubleBoot}. Table~\ref{tab:algorithms_list} gives the list of the algorithms considered with names and a short description of the methods. Appendix~\ref{sec:algorithms_description} gives full descriptions of the algorithms.
} 
\begin{table}[t!]
    \centering
    \begin{tabular}{|c|c|}
    \hline
        Algorithm Number (Name) & Method \\
        \hline
       Algorithm~1 (LR-$\chi^2$)  & Likelihood ratio with $\chi^2$ statistics\\
       \hline
       Algorithm~2a (K-B-Expansion) & Z-test with Polynomial approximation\\
       Algorithm~2b (Bootstrap-Gaussian) & Z-test with bootstrap mean \& variance\\
       \hline
       Algorithm~3a (High-Order-Marginal) & Z-test based on (i) in Theorem~\ref{thm:cmin} and \ref{thm:highorder}\\
       Algorithm~3b (High-Order-Conditional) & Z-test based on (ii) in Theorem~\ref{thm:cmin}
       and \ref{thm:highorder}\\
       \hline
       Algorithm~4a (Parametric-Bootstrap) & Parametric bootstrap with estimated $p$-value\\
       Algorithm~4b (B-\C\ Bootstrap) & Parametric bootstrap with bias correction\\
       Algorithm~4c (Double-Bootstrap) & Double bootstrap with adjusted $p$-value\\
       \hline
    \end{tabular}
    \caption{List of algorithms considered in numerical studies.}
\end{table}
\fi

\section{Simulation Studies}
\label{sec:sim}
We investigate the Type~I error and the power of the goodness-of-fit algorithms in Table~\ref{tab:algorithms_list} using a set of simulations designed to study 
various settings (large count, small count, and mixed count) and scenarios (the null hypothesis holds or does not hold). 



     
We use a power-law model with and without an added emission or absorption line as the generative model for data simulation. This class of models is widely used in astrophysical spectral analysis; see Section~\ref{sec:spectrum}. 
We generated counts
in $n$ channels 
under Model (\ref{eqn:Poiss_ind_intro}) with 
    \begin{equation}
    \begin{aligned}
        \label{eq:si_spec}
        &s_i(\boldsymbol{\theta})=KE_i^{-\Gamma}, \quad  \text{for }1\le i\le m_1, \ m_2<i\le n, \ \text{ and}\\
        &s_i(\boldsymbol{\theta})=\Psi, \qquad\quad   \text{for } \ m_1< i\le m_2,
    \end{aligned}
    \end{equation}
where $K>0$, $\Psi>0$, and $\Gamma\in\mathbb{R}$ are fitted parameters, $\{E_i = 1 +i/n; \ i=1,\ldots, n\}$ define energy bins over the range $(1,2)$, and $1\le m_1\le m_2\le n$ define the location and width of the line and are assumed known. We denote the line width by $b=m_2-m_1$ and refer to $\Psi$ as the strength of the emission or absorption line. When $b=0$ (\ref{eq:si_spec}) reduces to a power-law without an emission/absorption line. 
(The $\Gamma$ in (\ref{eq:si_spec}) differs subtly from that in 
(\ref{eq:po}) because of the fixed range.  For an astronomical spectrum defined over a range $X\in[\underline{X},\overline{X}]$, the spectral index, $\Gamma_\mathrm{astro}$ is related to the $\Gamma$ in Eq.~\ref{eq:si_spec} via $\Gamma_\mathrm{astro}={0.693} \cdot \Gamma /\,{\log({\overline{X}}/{\underline{X}})}$.)

In our simulation design, we set $\Gamma= 3$, and treat $n\in\{10, 25, 50, 100, 200, 300, 400\}$, and $K\in\{0.1, 0.25, 0.5, 1, 1.6, 2.5, 5,10\}$ as factors. The entire set of simulations is repeated with (a) no line, i.e., $b=0$, (b) an emission line with $\Psi = 2K$, and (c) an absorption line with $\Psi = K/10$. In both (b) and (c) the line is located with $m_1=n/10$ and $b=n/10$ so that the line extends over the energy range $(1.1, 1.2]$
. We set $\Psi$ 
so that the expected emission (absorption) line counts are about 3 times (about one-seventh) that of the bin immediately to the left of the line.  
We denote the total expected count by $s_+=\sum_{i=1}^n s_i$ and the total observed counts by $N_+=\sum_{i=1}^n N_i$. If $K$ or $n$ is fixed, $s_+$ depends linearly on the other quantity. If $\Gamma=3$, for example, $\bar s=\frac{s_+}{n}\approx\frac{3}{8}K$.
Here we describe results from a few illustrative simulation settings; see Appendix~\ref{results_appendix} of the supplement for
more details. 

\subsection{Simulation designs and results}
\label{sec:numerical}
We present an extensive numerical study of the performance of the methods in Table~\ref{tab:algorithms_list} in different scenarios. 

\paragraph{Comparison of oracle distribution and true null distribution.}\label{sec:compare_null_alg2}

\begin{figure}[t!]
    \centering
    \includegraphics[width=.97\textwidth]{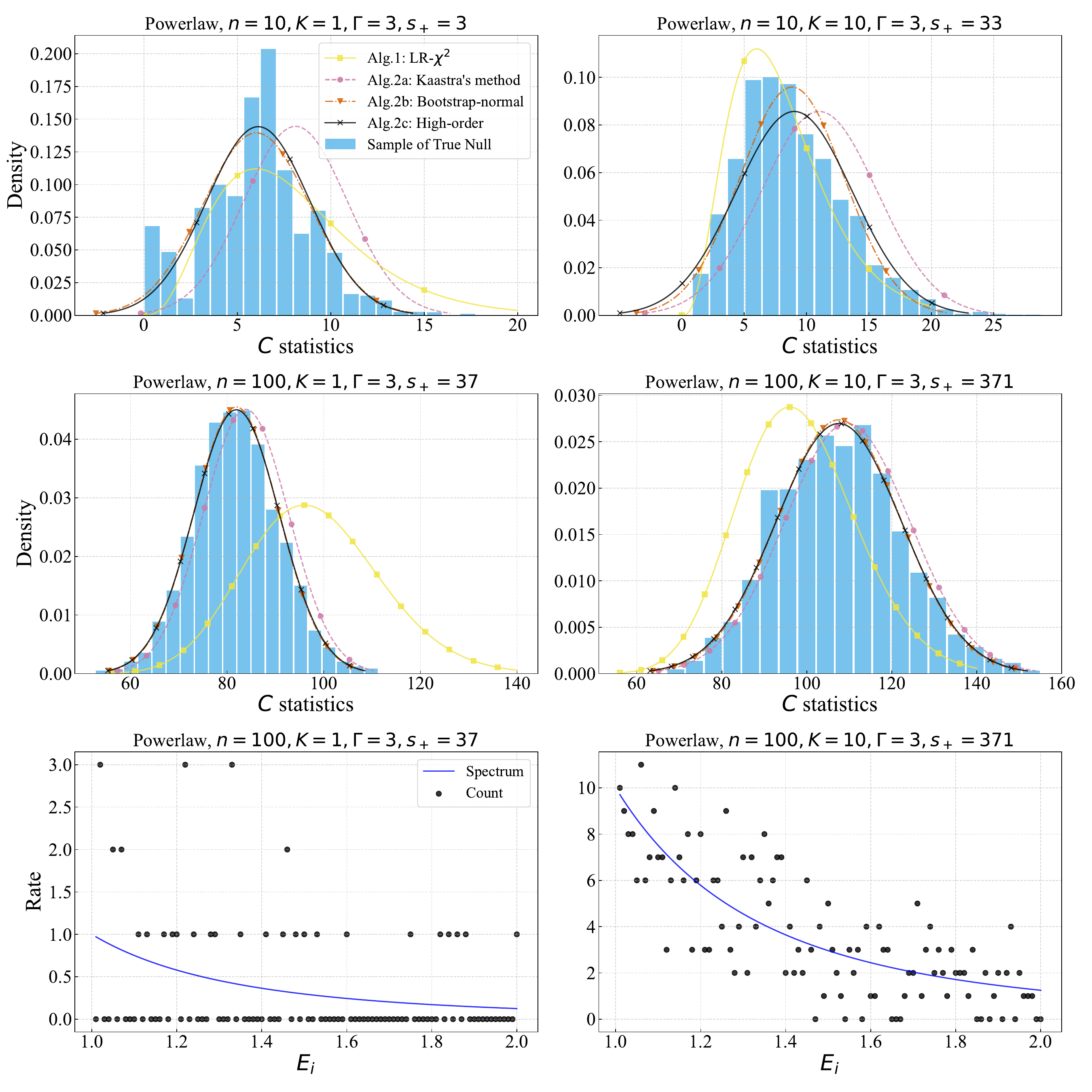}
    \spacingset{1.1}
    \caption{
    Comparison of a histogram of true null distribution of $C_n(\hat{\boldsymbol{\theta}})$ with the approximations of \texttt{Algorithm~1} (LR-$\chi^2$) and the three variants of \texttt{Algorithm~2}. Date are simulated from a Powerlaw model with $\Gamma=3$; the remaining parameters are listed in each panel; and the bootstrap size is $B=1000$. The bottom row plots the spectra $s_i(\boldsymbol{\theta})$ and the observed counts under the two simulation settings.}
    \label{fig:compare_K&M}
\end{figure}

We compare a Monte Carlo sample of the true null distribution of the \C\ statistic
with nominal null distributions obtained using four approximations, (i) \texttt{Algorithm~1}: the LR$-\chi^2$ approximation, (ii) \texttt{Algorithm~2a}: moments computed using Kaastra's \citeyear{kaastra2017} empirical polynomial approximation, (iii) \texttt{Algorithm~2b}: bootstrap mean and variance, and (iv) \texttt{Algorithm~2c}: moments computed with high-order expansions~\eqref{eq:moments_uncon_Cmin}.
Here we report results for the four simulation settings determined by crossing $n\in\{10, 100\}$  with $K\in\{1,10\}$ (with no emission/absorption line). 
Figure~\ref{fig:compare_K&M} compares the four approximations with histograms of the Monte Carlo sample of the true null distribution 
. Regardless of $K$, as $n$ increases, the distribution of the \C\ statistic behaves increasingly like a normal distribution; and the nominal distribution obtained with LR-$\chi^2$ is biased, relative to the true (null) distribution. 
The alignment between the results given by the unconditional high-order expansions~\eqref{eq:moments_uncon_Cmin} and the histogram, consequently, shows the accuracy of \eqref{eq:moments_uncon_Cmin} in practice, which improves as $n$ increases. Thus, we {do not 
include \texttt{Algorithms~2a-2b}}
in the remaining simulations.

\begin{figure}[t!]
    \centering
    \includegraphics[width=1\linewidth]{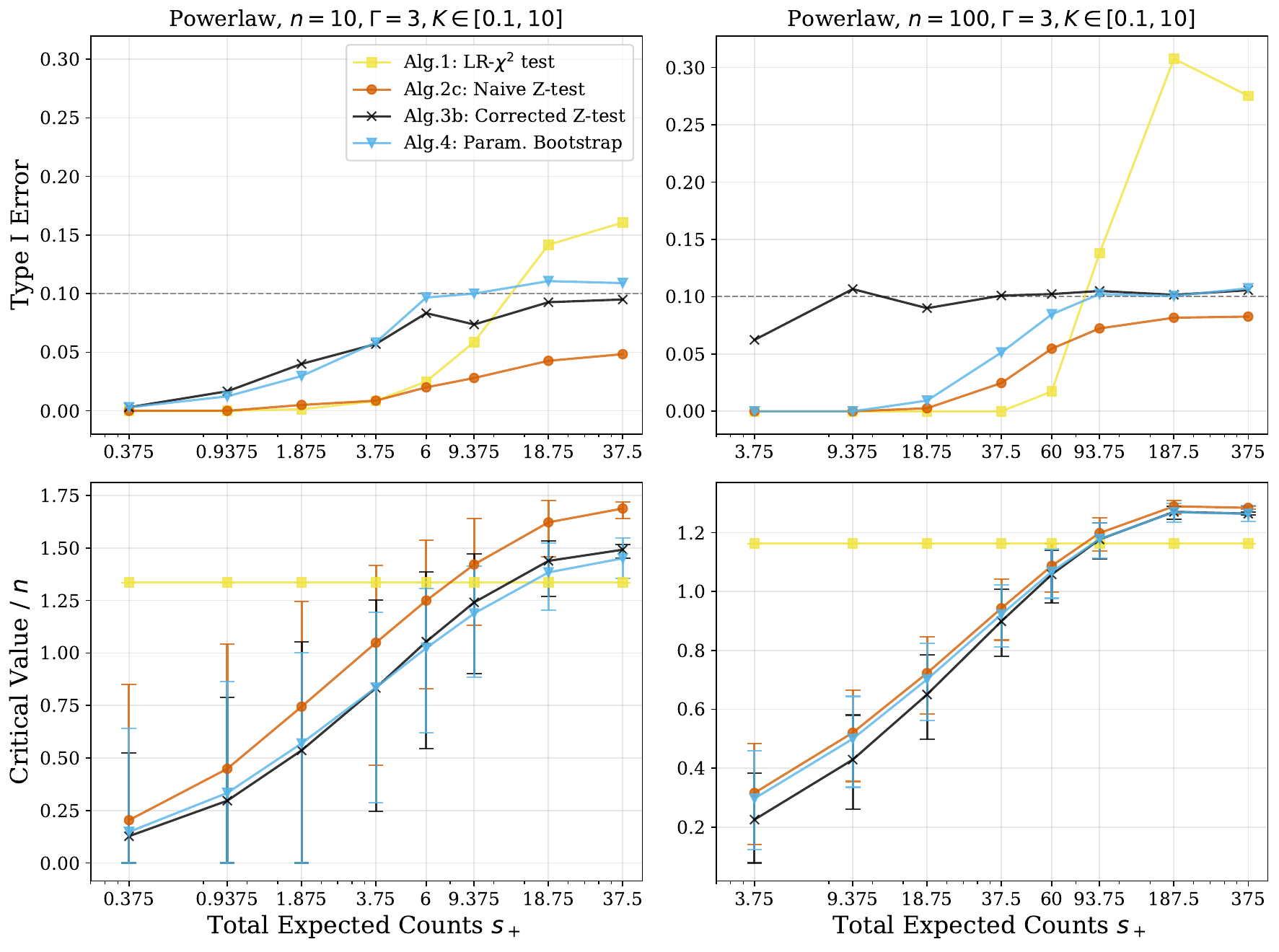}
    \spacingset{1.1}
    \caption{
    Performances of the four algorithms when $K\in\{0.1, 0.25, 0.5, 1, 1.6, 2.5, 5,10\}$ varies. The true models and null models are a Powerlaw with $\Gamma=3~(s_+\approx\frac{3}{8}nK)$. We simulated 3000 replicate data sets and used $B=300$ bootstrap replicates.  The dashed line in the first row is the nominal Type~I error rate. Tests with Type~I error rates close to the nominal rate and small critical values are preferred. 
    The simulation shows the overall strong performance of our recommended Corrected $Z$-test. 
    }
    \label{fig:cover_width_mu}
\end{figure}

\begin{figure}[t!]
    \centering
    \includegraphics[width=1\linewidth]{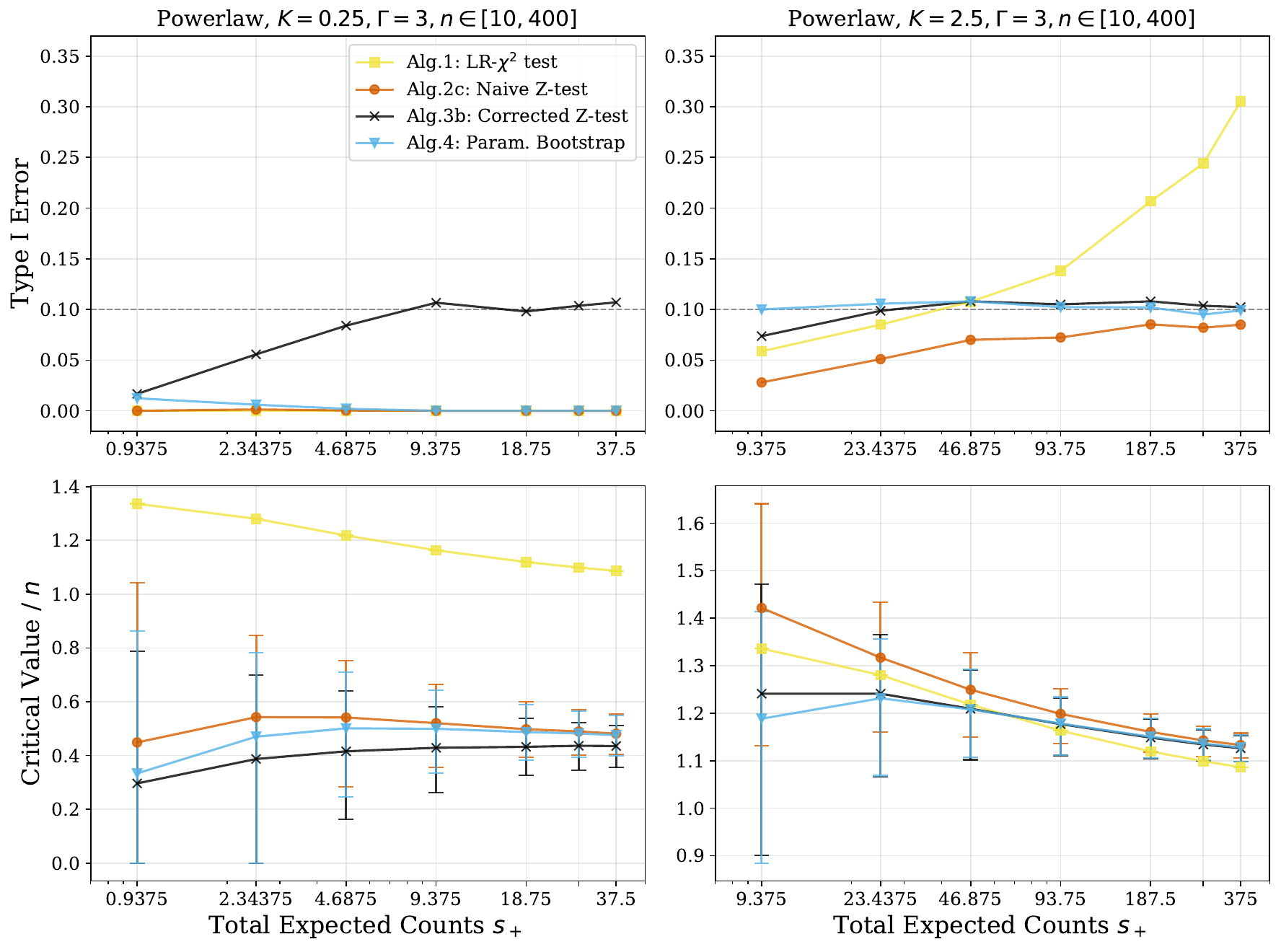}
    \spacingset{1.1}
    \caption{
    Performances of the four algorithms when $n\in\{10, 25, 50, 100, 200, 300, 400\}$ varies.  The simulations set up is as in Figure~\ref{fig:cover_width_mu}, but results are plotted as a function of $n$. The overall strong performance of our recommended Corrected $Z$-test is again evident. }
    \label{fig:cover_width_n}
\end{figure}

\paragraph{Comparison of Type~I Errors.}
Next, we analyze the Type~I errors given by the algorithms in Table~\ref{tab:algorithms_list} under different significance levels $\alpha$ and numbers of bins $n$. The Type~I error rate of a well-calibrated method is equal to the significance level and the distribution of its $p$-value under H$_0$ is uniform. In other words,
\begin{align}
p_{1-\alpha}^{(m)} (\boldsymbol{\theta}^\star) \equiv
P_{\boldsymbol{\theta}^\star} 
\left( C_n(\hat{\boldsymbol{\theta}}) \geq q^{(m)}_\alpha (\hat{\boldsymbol{\theta}}) \right)
& =\alpha\quad  \ \hbox{ and}
\label{eq:valid_coverage} \\
P_{\boldsymbol{\theta}^*}\left(p^{(m)}\le\alpha\right) &=\alpha, \quad
\text{for all } \ \alpha\in(0,1),
\label{eq:valid_pvalue} 
\end{align}
where  $p_{1-\alpha}^{(m)} (\boldsymbol{\theta}^\star)$ is the probability of a Type~I error, $q^{(m)}_\alpha(\hat{\boldsymbol{\theta}})$ is the $\alpha$-level critical value of the nominal null distribution, and 
$p^{(m)}$ is the $p$-value, 
all under \texttt{Algorithm~$m$}. 
Figures~\ref{fig:cover_width_mu} and~\ref{fig:cover_width_n} plot $p^{(m)}_{0.9}(\boldsymbol{\theta}^\star)$  and $q_{0.1}^{(m)}(\hat{\boldsymbol{\theta}})$ for the algorithms in Table~\ref{tab:algorithms_list} and show that when $s_+$ are moderately large, the Corrected $Z$-test and the Parametric Bootstrap work well. 
In contrast, regardless of $n$, the $p$-value given by the LR-$\chi^2$ test is far from uniformly distributed unless $K$ is extremely large;
and when $K$ is very small, 
the Na\"ive $Z$-test and the Parametric Bootstrap exhibit extremely low Type~I errors 
due to the plug-in effect.


\begin{figure}[t!]
    \centering
    \includegraphics[width=.93\linewidth]{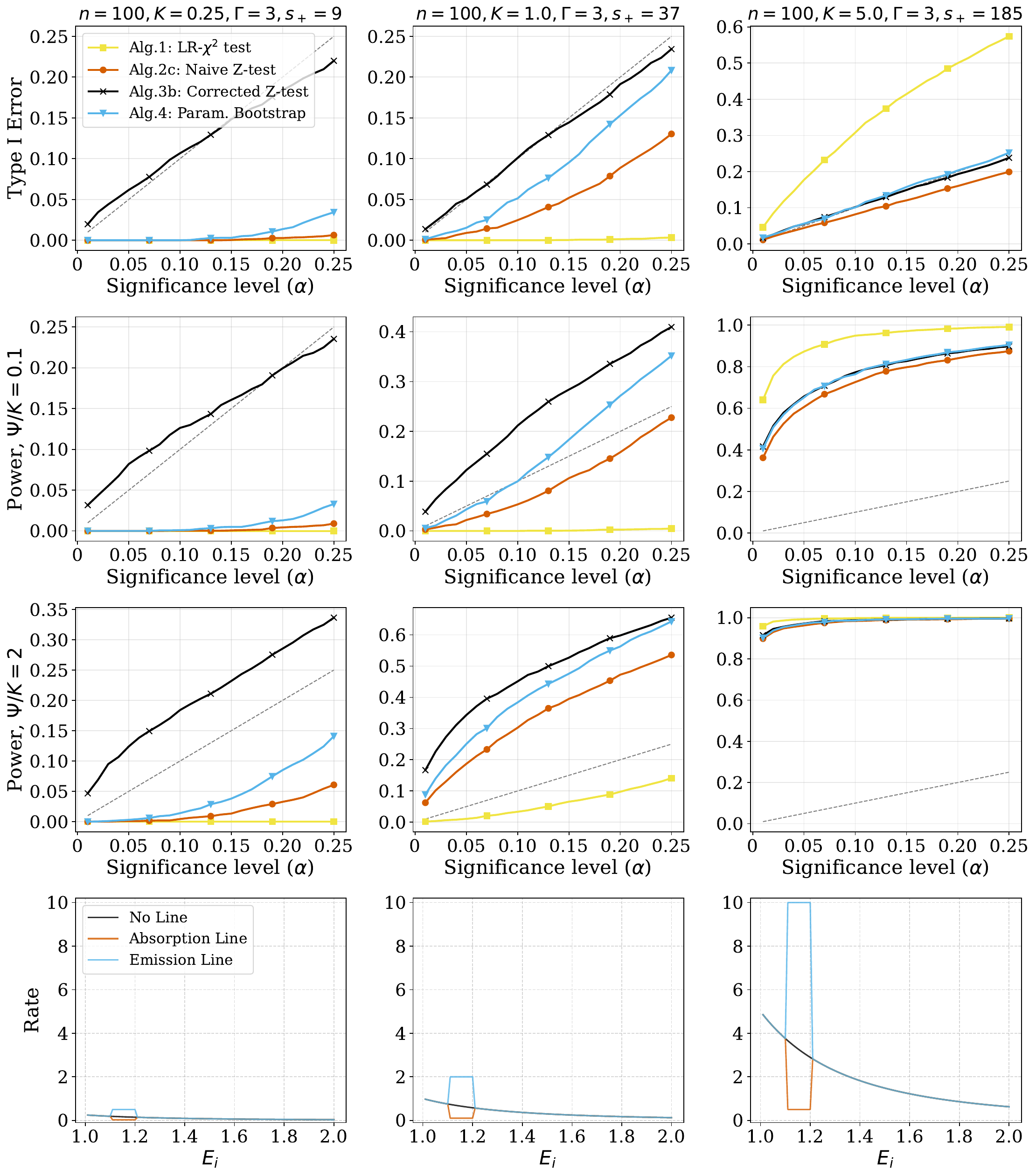}
    \spacingset{1.1}
    \caption{
    Comparison on the Type~I error rate and power of the four algorithms. We simulate 3000 replicate data sets under a Powerlaw (row 1), Powerlaw + Absorption Line  (row 2), and  Powerlaw + Emission Line (row 3).  In all cases $\Gamma=3$, $n=100$, we used $B=300$ bootstrap replicates, and $K$ varies among the columns. Each panels plots the probability of rejecting the null Powerlaw model (no emission/absorption line). This corresponds to Type~1 error in row 1 and power in rows 2 and 3. Row 4 plots the spectra used in each simulation setting, i.e., the Powerlaw model with/without a line. 
Ideally, the power should be as large as possible while maintaining the Type~I Error below $\alpha$. Overall, our recommended Corrected $Z$-test is best calibrated in terms of Type~I error rate and power.
}
    \label{fig4:spec_powervsalpha}
\end{figure}

\paragraph{Comparison of Power.}  Methods with greater power are preferred, so long as they successfully control the Type~I error rate. A comparison of the power of the four Algorithms  appears in  
Figure~\ref{fig4:spec_powervsalpha}. When both $n$ and $K$ are moderately large, the Corrected $Z$-test and the Parametric Bootstrap
enjoy similar power, which is the highest among the valid methods. However, when $K$ is very small, the Parametric Bootsrap 
exhibits extremely low power compared to the Corrected $Z$-test, as discussed in Proposition~\ref{prop:bootstrap_invalid}.

\subsection{Summary of Numerical Results}
\label{subsec:summary_sim}

When the Poisson rates $\{s_i\}$ are small, there is a meaningful difference between the distribution of $C_n(\hat{\boldsymbol{\theta}})$ and nominal $\chi^2$ distribution of the LR-statistic. The commonly used $\chi^2$ limit results in excessive false positive or false negative cases, especially when the data are relatively sparse.
These simulation studies also substantiate our theoretical results.
When both $n$ and $K$ are moderately large, the Corrected $Z$-test 
and the Parametric Bootstrap perform best. However, when $K$ is small, the Parametric Bootstrap exhibits low power regardless of $n$
due to the plug-in effect; see Figures~\ref{fig:cover_width_n} and~\ref{fig4:spec_powervsalpha}. This echoes Proposition~\ref{prop:bootstrap_invalid}.

In conclusion, the Corrected $Z$-test works well and gives a $p$-value approximately satisfying \eqref{eq:valid_coverage} when the data are extensive but sparse ($n\ge10$ and $s_+\ge10$ but $ s_i<1$), while the LR-$\chi^2$ test, the Na\"ive  $Z$-test and the Parametric Bootstrap test fail in this case. On the other hand, when the data is dense and the number of counts is large in all bins, such as $s_i\ge10$ uniformly, the commonly used LR-$\chi^2$ test is efficient with only a small bias. If the number of bins and the total expected photon count are both very small ($n<10$ and $s_+<10$), 
no valid test appears to enjoy high power.
In terms of computational costs, the Corrected Z-test is far more efficient than bootstrap methods, including robust variants such as the double bootstrap; see Section~\ref{sec:Label_time_appendix} in the supplemental material.

\def\chandra{\it Chandra\rm}
\def\letg{LETG}
\def\hrc{High-Energy Resolution Camera}

\section{Applications to Astrophysical Data}
\label{sec:real_data}

We apply the algorithms in Table~\ref{tab:algorithms_list} to X-ray spectra collected from a Quasar (\pg). Each observation consists of photon counts $N_i$ with $n=159$ energy channels indexed by $i$ that are appropriately modeled as independent Poisson variables as in \eqref{eqn:Poiss_ind_intro}.


\subsection{X-ray spectra of the Quasar \pg}
\label{sec:pg}

\begin{figure}[t!]
    \centering
    \includegraphics[width=0.49\textwidth]{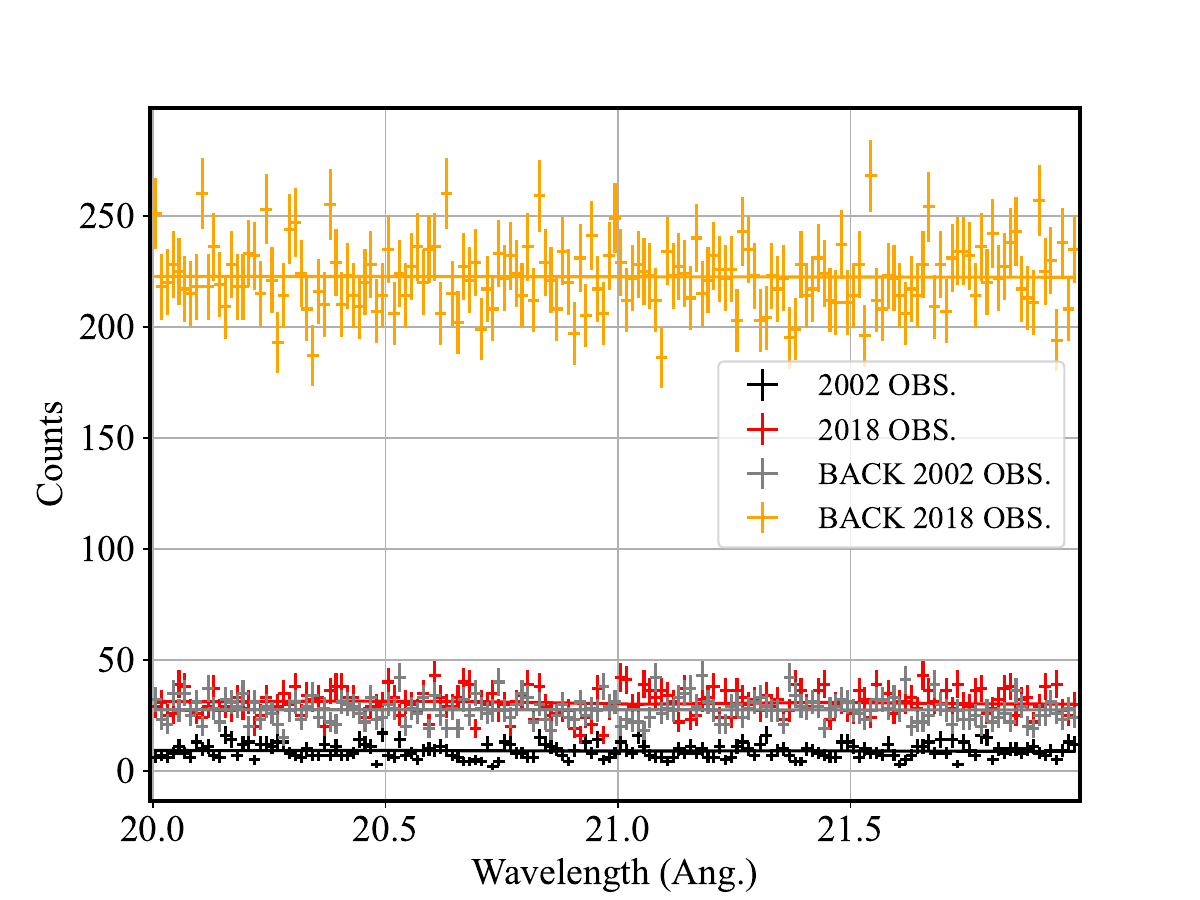}
    \includegraphics[width=0.49\textwidth]{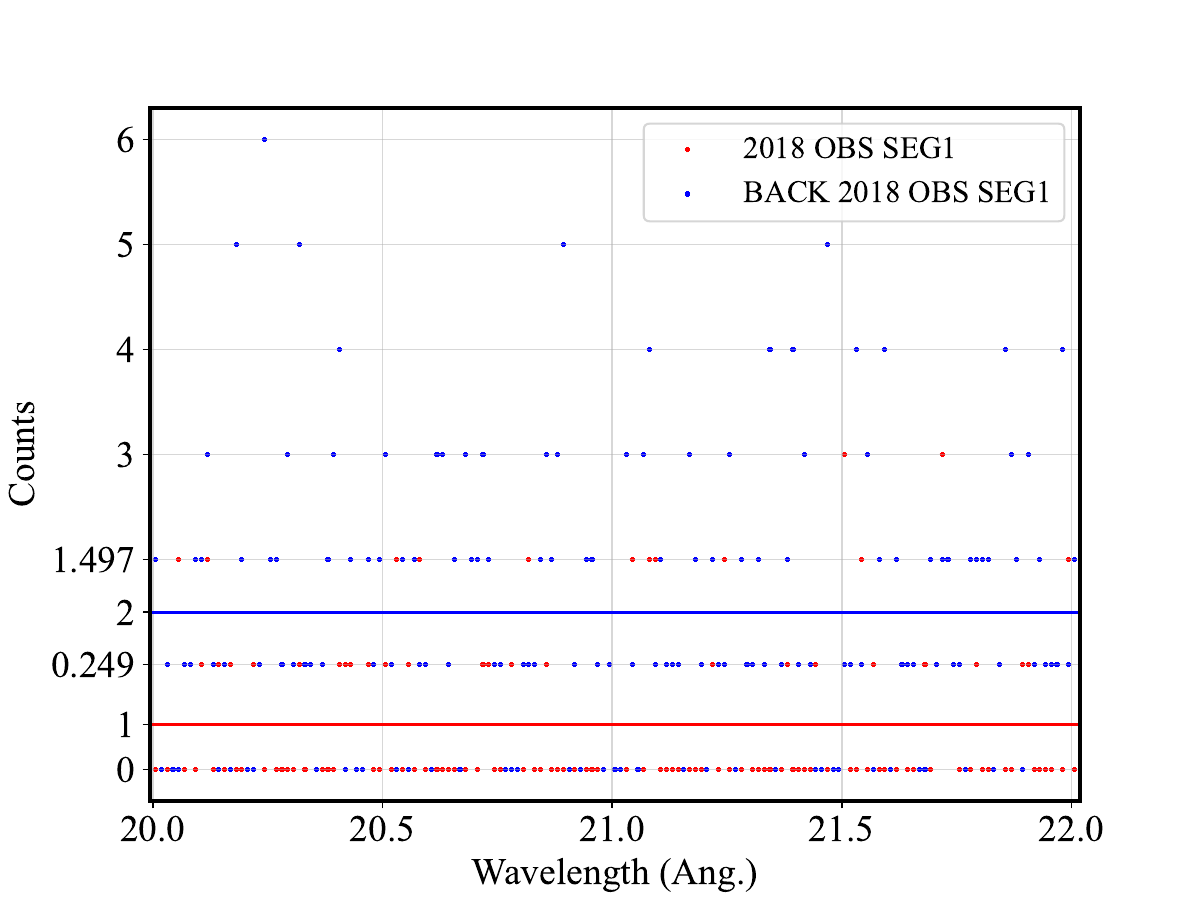}
    \spacingset{1.1}\caption{
    Left: X-ray count spectra of two observations of Quasar \pg, collected in 2002 and 2018, respectively. Each observation consists of a count spectrum collected in a small source region (consisting of a mixture of source and background photons, in black and red) and a count spectrum collected in a larger background region (consisting only of background counts, in grey and orange). The spectra are plotted with the default \chandra\ bin size (12.5~m \AA). Right: The count spectrum of Quasar \pg\, segment 1 (the first 1/20 of the exposure time of the full dataset; see text). The solid lines represent the averages of the source spectrum (red) and the background spectrum (blue).}
    
    \label{fig:fromMax}
\end{figure}

Quasars are very distant and luminous active galactic nuclei powered by supermassive black holes that are consuming their surrounding gas and dust. \pg\ has a redshift $z=0.177$, corresponding
to a luminosity distance of about 
2.812 billion light years, according to the currently favored cosmological model \citep[using the cosmology calculator of][]{wright2006}.
The left panel of Figure~\ref{fig:fromMax} displays two pairs of X-ray spectra of \pg, each obtained with the \hrc\ aboard the orbiting \chandra\ X-ray Observatory  in combination with the \letg\ grating spectrometer. (The processing of these data is described in \citet{bonamente2016}.)


The four spectra in the left panel of Figure~\ref{fig:fromMax} correspond to two exposures of a smaller source region and a larger background region. The source spectra are mixtures of source and background photons; the background exposures consist only of background photons. There is an 88~ks exposure taken in 2002 (labeled ``2002''; the source region is colored black and the background region gray), and
an a 267~ks exposure taken in 2018 (labeled ``2018''; the source region is colored red and the background region orange). The 2018 observation features substantially higher background (relative to the source intensity) than the 2002 observation; see
\citet{bonamente2019b} and Figure~\ref{fig:fromMax}. 


\subsection{Astrophysical Data Analysis}

In high-energy spectral analysis, the source spectra must be convolved with detector characteristics such as the redistribution matrix and effective area, so that the measured counts can be compared with the predicted model counts, see \eqref{eq:si1} and \eqref{eq:si2}. However, for the purpose of studying the properties of the \C\ statistic, we assume that the data are obtained with an ideal detector so that each photon was detected at its true
energy. This assumption can be justified because (a) the photon flux of the \pg\ data is rather constant within the band of interest, (b) the redistribution of photons is only energy-dependent in this band, and (c) the effective area of the detector varies by $\leq 10\%$ across the 20-22~\AA\ band. 
We test whether the constant model, i.e., $s_i(\theta) = \theta$, fits well for each of the four spectra by comparing the goodness-of-fit $p$-values obtained with the four algorithms in Table~\ref{tab:algorithms_list}.

\begin{table}[t!]
\spacingset{1.1}
\caption{$p$-values computed by applying Algorithms~1 - 4 to the data in Figure~\ref{fig:fromMax}.  
    \label{tab:Spectrum}}
\begin{tabular}{lccccc}
    \hline\hline
     &\phantom{Algorithm}& Algorithm &Algorithm & Algorithm & Algorithm \\
    Spectrum & $C_n(\hat{\boldsymbol{\theta}})$ 
    & 1 & 2c & 3b & 4$^a$\\
    \hline
    Source (2002)& 190.72&$0.039^\star$&0.061&0.054&0.055\\
    Source (2018) &167.67&0.284&0.332&0.312&0.305\\
    Background (2002) &171.39&0.221&0.263&0.245&0.240\\
    Background (2018) &153.46&0.587&0.624&0.603&0.587\\
    \hline
\end{tabular} \\
\footnotesize{
$^\star$ $p$-values $< 0.05$ are marked with a star.
\\
$^a$ Computed with $B=1000$ bootstrap replicates.}
    \end{table}


The resulting $p$-values are given in Table~\ref{tab:Spectrum}. \texttt{Algorithm}~2 to 4 produce relatively similar values and are different from those obtained with \texttt{Algorithm~1}, the commonly used LR-$\chi^2$ test. In particular, the LR-$\chi^2$ test with 
level $\alpha=0.05$ rejects $H_0$ for the 2002 source spectrum while other tests do not. This is because the counts in the 2002 source spectrum are small $(<10)$; thus the LR-$\chi^2$ test is prone to false rejections as discussed in Section~\ref{subsec:summary_sim}.
To study a case with 
low expected count in each channel, we split the 2018 exposure into 20 (approximately equal) time segments. The right panel of Figure~\ref{fig:fromMax} plots the counts in the first segment. 
For each segment, we performed goodness-of-fit tests of the constant spectral model.  
The LR-$\chi^2$ test (with level $\alpha=0.1$) does not reject $H_0$ for any of the 20 segments of the 2018 source spectrum, but rejects $H_0$ for 11 segments of the 2018 background spectrum. The other algorithms consistently reject 2 of the 20 segments for each spectrum, as expected. These results agree with those in Table~\ref{tab:Spectrum}, 
illustrating the poor statistical properties of the commonly used $\chi^2$ approximation of null distribution of \C. 

\section{Conclusions}
\label{sec:conc}

In this paper, we derive theoretical properties of the \C\ statistics and propose and compare four different computational algorithms for goodness-of-fit assessments using \C\ statistics. We establish the asymptotic normality of 
 \ctrue\ and \cmin\ (Theorems~\ref{theorem:CAN of C_true} and \ref{thm:cmin})
and provide new formulas for the conditional mean and variance of \cmin\ (Theorem~\ref{thm:highorder}) that extend the results available for generalized linear models
\citep[e.g.][]{Paul2000JRSSB, farrington1996JASA, mccullagh1986} to the significantly broader class
of parametric models that are in common use in astronomy and other fields.

Our numerical study shows that in large-count settings, all algorithms perform similarly; but in small-count settings, the corrected Z-test based on high-order asymptotics give superior performances compared to other methods including the bootstrap, especially when the data are extensive but relatively sparse. We implement all of our algorithms and make them available for astronomers via an open-source Python Module on Github. 

\section*{Acknowledgement}

This work was conducted under the auspices of the CHASC International Astrostatistics Center. It was supported by NSF grants DMS-21-13615, DMS-21-13397, and DMS-21-13605; by the UK Engineering and Physical Sciences Research Council [EP/W015080/1]. The authors thank CHASC members Aneta Siemiginowska, Herman Marshall, and Thomas Lee for many helpful discussions. YC acknowledges further support by the NSF grant PHY-20-27555 and NASA Federal Award No. 80NSSC23M0192 and No. 80NSSC23M0191. DvD was also supported in part by a Marie-Skodowska-Curie RISE Grant (H2020-MSCA-RISE-2019-873089) provided by the European Commission.   
VK further acknowledges support from NASA contract to the Chandra X-ray Center NAS8-03060. MB acknowledges support
from the NASA ADAP grant 80NSSC20K0885. 

\bibliographystyle{chicago-mine}

\bibliography{references}

\newpage
\begin{center}
{\large\bf SUPPLEMENTARY MATERIAL}
\end{center}

\appendix
\section{Details of Algorithms}
\label{sec:algorithms_description}

Here, we provide detailed descriptions of the algorithms
{summarized in Table~\ref{tab:algorithms_list} and deployed in the} numerical studies of Section~\ref{sec:sim}.  \texttt{Algorithms}~2a, 2b, and 2c correspond to the three approximations of the unconditional mean and variance of $C_n(\boldsymbol{\theta})$ described in Step~1 of \texttt{Algorithm~2} of Table~\ref{tab:algorithms_list}. Specifically, \texttt{Algorithm~2a} uses the empirical polynomial approximation of \citep{kaastra2017}; \texttt{Algorithm~2b} uses the parametric bootstrap; and \texttt{Algorithm~2c} {uses the formulae in \eqref{eq:moments_uncon_Cmin}.} \texttt{Algorithm~3} in Table~\ref{tab:algorithms_list} includes \texttt{Algorithm~3a} and \texttt{Algorithm~3b}, which are based on first-order and high-order expansions respectively. \texttt{Algorithm~4} corresponds to the Parametric Bootstrap in Table~\ref{tab:algorithms_list} ; while \texttt{Algorithm~5} uses the much more computationally demanding Double Bootstrap. 


\begin{algorithm}[htbp]
	\caption{Likelihood ratio test with $\chi^2$-approximation} 
 \label{algo:chisq}
	\begin{algorithmic}[1]
		\Require $C_n(\hat{\boldsymbol{\theta}})$, the number of bins $n$ and {  the number of free parameters $d$}.
  \State Determine the $p$-value by
  $$\hat{p}=\sup_{p\in[0,1]}\{ C_n(\hat{\boldsymbol{\theta}})\le\chi^2_{n-d}(1-p)\},$$
  {  where $\chi^2_{n-d}(\cdot)$ is the cumulative distribution function of a $\chi^2$ distribution with $n-d$ degrees of freedom.}
	\end{algorithmic} 
\end{algorithm}

\renewcommand{\thealgorithm}{2a}
\begin{algorithm}[htbp]
	\caption{Na\"ive Z-test with Polynomial approximation} 
 \label{algo:Kaastra&Max}
	\begin{algorithmic}[1]
	\Require $C_n(\hat{\boldsymbol{\theta}})$, $\hat{\boldsymbol{\theta}}$ and the number of bins $n$.
  \State Calculate $s_i(\hat{\boldsymbol{\theta}})$ for all $i$. Then calculate $\mathbb{E}(C_n(\hat{\boldsymbol{\theta}}))$ and $\mathrm{Var}(C_n(\hat{\boldsymbol{\theta}}))$ with the polynomial expansion formulas in \citet{kaastra2017} and \citep{bonamente2019}.
  \State Determine the $p$-value by
  $$\hat{p}=\sup_{p\in[0,1]}\left\{\frac{C_n(\hat{\boldsymbol{\theta}})-\mathbb{E}[C_n(\hat{\boldsymbol{\theta}})]}{\sqrt{\mathrm{Var}(C_n(\hat{\boldsymbol{\theta}}))}}\le Z \left(1-p\right)\right\},$$
  where $Z(\cdot)$ is the cumulative distribution function of a standard normal distribution.
	\end{algorithmic} 
\end{algorithm}

\renewcommand{\thealgorithm}{2b}
\begin{algorithm}
	\caption{Na\"ive Z-test with bootstrap mean \& variance} 
 \label{algo:bootstrap_normal}
	\begin{algorithmic}[1]
		\Require $C_n(\hat{\boldsymbol{\theta}})$, $\hat{\boldsymbol{\theta}}$, the number of bins $n$ and the bootstrap repetitions $B$.
    \State For $m\in\{1,2,\cdots,B\}$, based on the Poisson model with $\boldsymbol{\theta} =\hat{\boldsymbol{\theta}}$, generate $n$ Poisson bootstrap samples denoted by $N_i^{[m]}$, $i=1,\cdots,n$; for each sample $\{N_1^{[m]},\ldots, N_n^{[m]}\}$, obtain the MLE, $\hat{\boldsymbol{\theta}}^{[m]}$, and calculate $s_i^{[m]}(\hat{\boldsymbol{\theta}})$ and 
  $C_n^{[m]}(\hat{\boldsymbol{\theta}}^{[m]})$.
    
    \State Determine the bootstrap mean and variance with 
    $$\mathbb{E}_b(C_n(\hat{\boldsymbol{\theta}}))=\frac{\sum^B_{m=1}C_n^{[m]}(\hat{\boldsymbol{\theta}}^{[m]})}{B}, \quad \mathrm{Var}_b(C_n(\hat{\boldsymbol{\theta}}))=\frac{\sum^B_{m=1}(C_n^{[m]}(\hat{\boldsymbol{\theta}}^{[m]})-\mathbb{E}_b(C_n(\hat{\boldsymbol{\theta}})))^2}{B-1}.$$
  
  \State Determine the $p$-value by
  $$\hat{p}=\sup_{p\in[0,1]}\left\{\frac{C_n(\hat{\boldsymbol{\theta}})-\mathbb{E}_b(C_n(\hat{\boldsymbol{\theta}}))}{\sqrt{\mathrm{Var}_b(C_n(\hat{\boldsymbol{\theta}}))}}\le Z \left(1-p\right)\right\},$$
  where $Z(\cdot)$ is the cumulative distribution function of a standard normal distribution.
	\end{algorithmic} 
\end{algorithm}

\renewcommand{\thealgorithm}{2c}
\begin{algorithm}
	\caption{Na\"ive Z-test -- high order} 
 \label{algo:Ztest_marginal}
	\begin{algorithmic}[1]
		\Require $C_n(\hat{\boldsymbol{\theta}})$, $\hat{\boldsymbol{\theta}}$ and the number of bins $n$.
\State Calculate the cumulants and matrices $\hat{\kappa}_{1}^{(i)}$ and $\hat{\kappa}_{2}^{(i)}$ via direct summation over the Poisson data $N_i$, $i=1,...,n$ based on $s_i(\hat{\boldsymbol{\theta}})$. Note that $\kappa_1^{(i)}=\mathbb{E}(C^{(i)})$, $\kappa_2^{(i)}=\mathbb{E}(C^{(i)}-\kappa_1^{(i)})^2$.
    \State Determine ${\mu}(\hat{\boldsymbol{\theta}})$ and ${\sigma}^2(\hat{\boldsymbol{\theta}})$
    $${\mu}(\hat{\boldsymbol{\theta}})=\hat{\kappa}_1^{(\cdot)},$$
    $${\sigma}^2(\hat{\boldsymbol{\theta}})=\hat{\kappa}_2^{(\cdot)}.$$
  
  \State Determine the $p$-value by
  $$\hat{p}=\sup_{p\in[0,1]}\left\{\frac{C_n(\hat{\boldsymbol{\theta}})-{\mu}(\hat{\boldsymbol{\theta}})}{{\sigma}^2(\hat{\boldsymbol{\theta}})}\le Z\left(1-p\right)\right\},$$
  where $Z(\cdot)$ is the cumulative distribution function of a standard normal distribution.
	\end{algorithmic} 
\end{algorithm}

\renewcommand{\thealgorithm}{3a}
\begin{algorithm}
	\caption{Corrected Z-test -- first order} 
 \label{algo:Ztest_marginal_appendix}
	\begin{algorithmic}[1]
		\Require $C_n(\hat{\boldsymbol{\theta}})$, $\hat{\boldsymbol{\theta}}$ and the number of bins $n$.
\State Calculate the cumulants and matrices $\hat{\kappa}_{1}^{(i)}$, $\hat{\kappa}_{2}^{(i)}$, $\hat{\kappa}_{11}^{(i)}$, $\hat{X}$ and $\hat{V}$ via direct summation over the Poisson data $N_i$, $i=1,...,n$ based on $s_i(\hat{\boldsymbol{\theta}})$. Note that $X=(\nabla s_1^\top,\cdots,\nabla s_n^\top)^\top$, $V=\mathrm{diag}(s_i)$,  $\kappa_{11}^{(i)}=\mathbb{E}\{(C^{(i)}-\kappa_1^{(i)})(N_i-s_i)\}$.  The estimated quantities of these quantities are labeled with hats by plugging-in $\hat{\boldsymbol{\theta}}$. For example, in practice we calculate $\kappa_{1}^{(i)}$ by using summation $\sum_{k=0}^{K^*} C^{(i)}(N_i=k;s_i)f(N_i=k;s_i),$ where $f(\cdot;s_i)$ is the mass of Poisson distribution with rate $s_i$ and $K^*=\min\left\{K\ge s_i: \frac{C^{(i)}(N_i=K+1;s_i)f(N_i=K+1;s_i)}{\sum_{t=0}^K C^{(i)}(N_i=t;s_i)f(N_i=t;s_i)} \le\tau \right\}$ for a chosen threshold $\tau=10^{-30}$. 
To avoid repeated Poisson summation when we apply this algorithm to different datasets, we use a grid (from 0 to 100 for the Poisson rate, with a gap of $10^{-3}$) to get a convenient look-up table for the cumulants with relative accuracy $10^{-5}$, which is
similar to Table 1 in~\cite{mccullagh1986} but with a higher accuracy. This table is available from our GitHub page and via our Python module.
    \State Determine ${\mu}(\hat{\boldsymbol{\theta}})$ and ${\sigma}^2(\hat{\boldsymbol{\theta}})$
    $${\mu}(\hat{\boldsymbol{\theta}})=\hat{\kappa}_1^{(\cdot)},$$
    $${\sigma}^2(\hat{\boldsymbol{\theta}})=\left\{\hat{\kappa}_2^{(\cdot)}-\hat\kappa_{11}^\top \hat{X}(\hat{X}^\top\hat{V}^{-1}\hat{X})^{-1}\hat{X}^\top\hat\kappa_{11}\right\}.$$
  
  \State Determine the $p$-value by
  $$\hat{p}=\sup_{p\in[0,1]}\left\{\frac{C_n(\hat{\boldsymbol{\theta}})-{\mu}(\hat{\boldsymbol{\theta}})}{{\sigma}^2(\hat{\boldsymbol{\theta}})}\le Z\left(1-p\right)\right\},$$
  where $Z(\cdot)$ is the cumulative distribution function of a standard normal distribution.
	\end{algorithmic} 
\end{algorithm}

\renewcommand{\thealgorithm}{3b}
\begin{algorithm}
	\caption{Corrected Z-test -- high order} 
 \label{algo:Ztest_conditional}
	\begin{algorithmic}[1]
		\Require $C_n(\hat{\boldsymbol{\theta}})$, $\hat{\boldsymbol{\theta}}$ and the number of bins $n$.
\State Calculate the cumulants and matrices $\hat{\kappa}_{1}^{(i)}$, $\hat{\kappa}_{11}^{(i)}$, $\hat{\kappa}_{12}^{(i)}$, $\hat{\kappa}_{03}^{(i)}$, $\hat{X}$, $\hat{Q}$, and $\hat{\Sigma}$ via direct summation over the Poisson data $N_i$, $i=1,...,n$ based on $s_i(\hat{\boldsymbol{\theta}})$. Note that $X=(\nabla s_1^\top,\cdots,\nabla s_n^\top)^\top$, $V=\mathrm{diag}(s_i)$, $Q=(Q_{ij})=X(X^\top V^{-1} X)^{-1}X^\top $, $\kappa_3^{(i)}=\mathbb{E}(C^{(i)}-\kappa_1^{(i)})^3$, $\kappa_{11}^{(i)}=\mathbb{E}\{(C^{(i)}-\kappa_1^{(i)})(N_i-s_i)\}$, $\kappa_{12}^{(i)}=\mathbb{E}\{(C^{(i)}-\kappa_1^{(i)})(N_i-s_i)^2\}$, $\kappa_{21}^{(i)}=\mathbb{E}\{(C^{(i)}-\kappa_1^{(i)})^2(N_i-s_i)\}$ and $\kappa_{03}^{(i)}=\mathbb{E}(N_i-s_i)^3$, $\Sigma=\mathrm{diag}\{\kappa_{12}^{(i)}/V_i^2-(\sum_j\kappa_{11}^{(j)}Q_{ji}/V_j)\kappa_{03}^{(i)}/V_i^3\}$, $\kappa_{11}=(\kappa_{11}^{(1)}/V_1,\cdots,\kappa_{11}^{(n)}/V_n)^\top$. The estimated quantities of these quantities are labeled with hats by plugging-in $\hat{\boldsymbol{\theta}}$. The estimates are exactly the same as in Algorithm 3a. 
    \State Determine the theoretical asymptotic conditional mean and variance
    $$\mathbb{E}(C_n(\hat{\boldsymbol{\theta}})\vert \hat{\boldsymbol{\theta}})=\hat{\kappa}_1^{(\cdot)}-\frac12\bm1^\top \hat{X}^\top \hat\Sigma \hat{X} (\hat{X}^\top\hat V^{-1}\hat{X})^{-1}\bm1,$$
    $${\rm Var}(C_n(\hat{\boldsymbol{\theta}})\vert\hat{\boldsymbol{\theta}})=\left\{\hat{\kappa}_2^{(\cdot)}-\hat\kappa_{11}^\top \hat{X}(\hat{X}^\top\hat{V}^{-1}\hat{X})^{-1}\hat{X}^\top\hat\kappa_{11}\right\}.$$
  
  \State Determine the $p$-value by
  $$\hat{p}=\sup_{p\in[0,1]}\left\{\frac{C_n(\hat{\boldsymbol{\theta}})-\mathbb{E}[C_n(\hat{\boldsymbol{\theta}})\vert \hat{\boldsymbol{\theta}}]}{\sqrt{\mathrm{Var}(C_n(\hat{\boldsymbol{\theta}})\vert \hat{\boldsymbol{\theta}})}}\le Z\left(1-p\right)\right\},$$
  where $Z(\cdot)$ is the cumulative distribution function of a standard normal distribution.
	\end{algorithmic} 
\end{algorithm}


\renewcommand{\thealgorithm}{4}
\begin{algorithm}
	\caption{Parametric bootstrap} 
 \label{algo:bootstrap}
	\begin{algorithmic}[1]
		\Require $C_n(\hat{\boldsymbol{\theta}})$, $\hat{\boldsymbol{\theta}}$, the number of bins $n$ and the bootstrap repetitions $B$.

  \State For $m\in\{1,2,\cdots,B\}$, based on the Poisson model with $\boldsymbol{\theta} =\hat{\boldsymbol{\theta}}$, generate $n$ Poisson bootstrap samples denoted by $N_i^{[m]}$, $i=1,\cdots,n$; for each sample $\{N_1^{[m]},\ldots, N_n^{[m]}\}$, obtain the MLE, $\hat{\boldsymbol{\theta}}^{[m]}$, and calculate $s_i(\hat{\boldsymbol{\theta}}^{[m]})$ and 
  $C_n^{[m]}(\hat{\boldsymbol{\theta}}^{[m]})$.
    
   \State Determine the $p$-value by
  $$\hat{p}=\sum^{B_1}_{m=1}\mathbf{1}_{\{C_n^{[m]}(\hat{\boldsymbol{\theta}}^{[m]})\ge C_n(\hat{\boldsymbol{\theta}})\}}/B$$

	\end{algorithmic} 
\end{algorithm}


\renewcommand{\thealgorithm}{5}
\begin{algorithm}
	\caption{Double bootstrap} 
 \label{algo:P.DoubleBoot}
	\begin{algorithmic}[1]
		\Require $C_n(\hat{\boldsymbol{\theta}})$, $\hat{\boldsymbol{\theta}}$, the number of bins $n$ and the bootstrap repetitions $B_1$ and $B_2$.
\State Implement Algorithm~\ref{algo:bootstrap} with $C_n(\hat{\boldsymbol{\theta}})$, $\hat{\boldsymbol{\theta}}$ and $B_1$ to obtain the first-level $p$-value $\hat{p}$. Denote the estimation and corresponding Cash statistic in each bootstrap as $\hat{\boldsymbol{\theta}}^{[m]}$ and $C_n^{[m]}(\hat{\boldsymbol{\theta}}^{[m]}),m=1,\cdots,B_1$, respectively.

\State For each $\hat{\boldsymbol{\theta}}^{[m]}$, implement Algorithm~\ref{algo:bootstrap} with $C_n^{[m]}(\hat{\boldsymbol{\theta}}^{[m]})$, $\hat{\boldsymbol{\theta}}^{[m]}$ and $B_2$; denote the $p$-values as $\hat{p}^{[j]},j=1,\cdots,B_1$.
\State Determine the adjusted $p$-value by
$$\hat{p}_{adj}=\sum^{B_1}_{j=1}\mathbf{1}_{\{\hat{p}^{[j]}\le \hat{p} \}}/B_1.$$
	\end{algorithmic} 
\end{algorithm}

\newpage
\section{Theory and Proofs}\label{appendix:theory_and_proof}
For simplicity, moments are evaluated at $\boldsymbol{\theta}^*$ if the subscripts are not explicitly stated; and the moments with hat are evaluated at $\hat{\boldsymbol{\theta}}$. For example, $\mathbb{E}[C_n(\boldsymbol{\hat\theta})]=\mathbb{E}_{\boldsymbol{\theta}^*}[C_n(\hat{\boldsymbol{\theta}})]$ and $\hat{\mathbb{E}}[C_n(\hat{\boldsymbol{\theta}})]=\mathbb{E}_{\hat{\boldsymbol{\theta}}}[C_n(\hat{\boldsymbol{\theta}})]$.\label{set:default_subscript}

\subsection{Preliminary: Rewriting the Likelihood Function}
\label{appendix:proof_divide_into_infinite_bins}

\begin{lemma}[A large-count problem reduced to a large number of bins problem]
\label{lemma:divide_into_infinite_bins}
Assume that each $s_{i}(\boldsymbol{\theta})$ is continuous in an open neighbourhood of $\boldsymbol{\theta}^\star$, denoted by $B(\boldsymbol{\theta}^\star) = \{\boldsymbol{\theta}: |\boldsymbol{\theta} - \boldsymbol{\theta}^\star| < \delta\}$, $0<\delta <\infty$. If $\sum_{i=1}^n s_i(\boldsymbol{\theta}^{\star})\rightarrow \infty$, then there exists $\{m_1,\ldots, m_n\}$ such that (1) $\sum_{i=1}^n m_i \rightarrow\infty$, 
(2) $0 < {s_i(\boldsymbol{\theta}^{\star})}/{m_i} < 1$, (3) the likelihood given by model~\eqref{eqn:Poiss_ind_intro} is equivalent to the likelihood of the following model for $\boldsymbol{\theta}$ in $B(\boldsymbol{\theta}^\star)$,
\begin{align}
    \tilde{N}_{ij} \stackrel{\rm indep.}{\sim} {\rm Poisson} \left(\frac{s_i(\boldsymbol{\theta})}{m_i}\right),\quad \sum_{j=1}^{m_i} \tilde{N}_{ij} = N_i. 
\end{align}
Note that in this equivalent model, the new number of bins is $\sum_{i=1}^n m_i\rightarrow\infty$ and the new Poisson rate $s_i(\boldsymbol{\theta})/m_i$ is bounded from above when $\boldsymbol{\theta} = \boldsymbol{\theta}^\star$. 
\end{lemma}
\begin{proof}
Let $\tilde{m}_i = \lfloor s_i(\boldsymbol{\theta}^{\star}) \rfloor +1$, the largest integer that is smaller than $s_i(\boldsymbol{\theta}^\star)$, $i=1,\ldots, n$. Then $ s_i(\boldsymbol{\theta}^{\star}) / \tilde{m}_i < 1$ and $\sum_{i=1}^n \tilde{m}_i \geq \sum_{i=1}^n s_i(\boldsymbol{\theta}^\star) \rightarrow\infty$ as $\sum_{i=1}^n s_i(\boldsymbol{\theta}^\star)\rightarrow\infty$. 
Due to the continuity of $s_i(\boldsymbol{\theta})$, we can choose constants $m_i > \tilde{m}_i$ such that in the compact closure of $B(\boldsymbol{\theta}^\star)$, which contains $B(\boldsymbol{\theta}^\star)$, we have $s_i(\boldsymbol{\theta}) / m_i < 1$ for all $i = 1,\ldots, n$. The proof concludes by noticing that the likelihood of the new model is
\begin{align*}
    &\prod_{i=1}^n\prod_{j=1}^{m_i} {\rm Poisson}\left(\tilde{N}_{ij}; \frac{s_i(\boldsymbol{\theta})}{m_i}\right)\propto  \prod_{i=1}^n\prod_{j=1}^{m_i} \left(\left[\frac{s_i(\boldsymbol{\theta})}{m_i}\right]^{\tilde{N}_{ij}} \exp\left[-\frac{s_i(\boldsymbol{\theta})}{m_i}\right]\right)\\
    &\propto \prod_{i=1}^n \left[{s_i(\boldsymbol{\theta})}\right]^{N_i} \exp\left[-{s_i(\boldsymbol{\theta})}\right]\propto \prod_{i=1}^n {\rm Poisson} \left( N_i; s_i(\boldsymbol{\theta})\right).
\end{align*}
\end{proof}

\subsection{Regularity conditions and properties of the MLE}\label{append:subsec_MLE}

\subsubsection{Regularity conditions in practice}\label{appendix:conditionA}
In this section, we introduce a set of more practical regularity conditions, denoted as { Regularity Conditions (A)}. These conditions are used to derive the asymptotic normality of $\hat{\boldsymbol{\theta}}$, as required by the { Regularity Condition (B1)} given in Section~\ref{sec:asymptotic}. Similar conditions can be found in \cite{hoadley1971asymptotic} and \citet{dale1986JRSSB}.

\noindent
{\bf Regularity Conditions (A)}
\begin{description}
    \item (A1) $N_i\sim f_i(\cdot\vert\boldsymbol{\theta})$ independently for $i=1,\ldots, n$.
    \item (A2) The model is identifiable, i.e.
    (i) $\lim\limits_{n\to\infty}\frac{1}{n}\sum^n_{i=1}\left(s_i(\boldsymbol{\theta}^\star)\log\frac{s_i(\boldsymbol{\theta})}{s_i(\boldsymbol{\theta}^\star)}-(s_i(\boldsymbol{\theta})-s_i(\boldsymbol{\theta^\star}))\right)<0$ for $\boldsymbol{\theta}\neq\boldsymbol{\theta}^\star$; (ii) there exists a $\rho>0$ such that 
    $$\lim_{n\to\infty}\frac{1}{n}\sup_{\Vert\boldsymbol{\theta}-\boldsymbol{\theta}^\star\Vert=\rho}\sum_{i=1}^n\left(s_i(\boldsymbol{\theta}^\star)\log\frac{s_i(\boldsymbol{\theta})}{s_i(\boldsymbol{\theta}^\star)}-(s_i(\boldsymbol{\theta})-s_i(\boldsymbol{\theta^\star}))\right)<0.$$
    \item (A3) For any $\boldsymbol{\theta}\in\boldsymbol{\Theta}$ and $i$, $f_i(x\vert\boldsymbol{\theta})$ has the same support and is differentiable about $\boldsymbol{\theta}.$
    \item (A4) There exists an open set $\mathcal{O}\subset\boldsymbol{\Theta}$ such that the true value of parameter $\boldsymbol{\theta^\star}\in\mathcal{O}$.
    \item (A5) For any $x\in\mathcal{X}$, where $\mathcal{X}$ is the state space, the density function $f_i(x\vert\boldsymbol{\theta})$ has third-order continuous derivative on $\boldsymbol{\theta}$, uniformly in $i$. 
    \item (A6) For any $\boldsymbol{\theta}\in\boldsymbol{\Theta}$, there exists a constant $c>0$ and a function $M(x)$, both of which may depend on $\boldsymbol{\theta}$, such that for any $x\in\mathcal{X}$, $i$ and $\left\Vert\boldsymbol{\theta}-\boldsymbol{\theta}^\star\right\Vert\le c$,
    $$\Bigg\Vert\frac{\partial^3}{\partial\boldsymbol{\theta}^3}\log f_i(x\vert\boldsymbol{\theta})\Bigg\Vert\le M(x),\quad\mathrm{and}\quad \mathbb{E}_{\boldsymbol{\theta}^\star}\vert M(x)\vert<\infty.$$ 
\end{description}

Among Regularity Conditions (A), only (A2) does not follow directly under our Poisson setup (\ref{eqn:Poiss_ind_intro}) with smooth link functions $s_i(\boldsymbol{\theta})$. To verify  (A2) in commonly used models,  we use a stronger but  simpler condition: 

\begin{lemma}[Regularity Condition (C)]
\label{lem:sc_for_a2}
    Assume that $N_i\indep {\rm Poisson}(s_i(\boldsymbol{\theta}))$ for all $i$ and each $s_i(\boldsymbol{\theta})$ belongs to a function family $\mathcal{G}(\boldsymbol{\theta};\alpha)$, i.e. $s_i(\boldsymbol{\theta})=\mathcal{G}(\boldsymbol{\theta};\alpha_i)$, where $\alpha$ lies in a compact set $\mathcal{F}$. If (i) $\mathcal{G}(\boldsymbol{\theta};\alpha)$ is continuous with respect to $\alpha$ when $\boldsymbol{\theta}$ is fixed or (ii) $\alpha$ takes values in a finite set, then condition (A2) is satisfied.  
\end{lemma}

\label{appendix:proof_sc_for_a2}
\begin{proof}
     For any density function $f$, $g$ with the same support, the Jensen's inequality gives
$$\mathbb{E}_f\left[\log\frac{g(X)}{f(X)}\right]\le \log \mathbb{E}_f\left[\frac{g(X)}{f(X)}\right]=0,$$
with equality if and only if $f=g$. 

First we consider the case that $\mathcal{G}(\boldsymbol{\theta};\alpha)$ is continuous with respect to $\alpha$ when $\boldsymbol{\theta}$ is fixed. Denote the density function of each $N_i$ as $f(x\vert\boldsymbol{\theta};\alpha_i)$. Since $\mathbb{E}_{\boldsymbol{\theta}^\star}\left[\log\frac{f(x\vert\boldsymbol{\theta};\alpha)}{f(x\vert\boldsymbol{\theta}^\star;\alpha)}\right]=\mathcal{G}(\boldsymbol{\theta}^\star;\alpha)\log\frac{\mathcal{G}(\boldsymbol{\theta};\alpha)}{\mathcal{G}(\boldsymbol{\theta}^\star;\alpha)}-(\mathcal{G}(\boldsymbol{\theta};\alpha)-\mathcal{G}(\boldsymbol{\theta^\star};\alpha))$ is continuous with respect to the parameter $\alpha$ and $\alpha$ lies in a compact set, we may assume it takes maximum value at $\alpha_0$; then for any fixed $\boldsymbol{\theta}\neq\boldsymbol{\theta}^\star$, we have
\begin{align*}
    &\lim\limits_{n\to\infty}\frac{1}{n}\sum^n_{i=1}\left(s_i(\boldsymbol{\theta}^\star)\log\frac{s_i(\boldsymbol{\theta})}{s_i(\boldsymbol{\theta}^\star)}-(s_i(\boldsymbol{\theta})-s_i(\boldsymbol{\theta^\star}))\right)\\
    =&\lim_{n\to\infty}\frac{1}{n}\sum^n_{i=1}\mathbb{E}_{\boldsymbol{\theta}^\star}\left[\log\frac{f(x\vert\boldsymbol{\theta};\alpha_i)}{f(x\vert\boldsymbol{\theta}^\star;\alpha_i)}\right]
   \le\mathbb{E}_{\boldsymbol{\theta}^\star}\left[\log\frac{f(x\vert\boldsymbol{\theta};\alpha_0)}{f(x\vert\boldsymbol{\theta}^\star;\alpha_0)}\right]<0,
\end{align*}
which establishes the condition (A2)(i). Because for any $\rho>0$, the sphere $B(\boldsymbol{\theta}^\star,\rho)$ is compact and the expectation of the log-likelihood is continuous with respect to $\boldsymbol{\theta}$ when $\alpha_0$ is fixed (condition A3), we have
\begin{align*}
    &\lim_{n\to\infty}\frac{1}{n}\sup_{\Vert\boldsymbol{\theta}-\boldsymbol{\theta}^\star\Vert=\rho}\sum_{i=1}^n\left(s_i(\boldsymbol{\theta}^\star)\log\frac{s_i(\boldsymbol{\theta})}{s_i(\boldsymbol{\theta}^\star)}-(s_i(\boldsymbol{\theta})-s_i(\boldsymbol{\theta^\star}))\right)\\
    =&\lim_{n\to\infty}\frac{1}{n}\sup_{\Vert\boldsymbol{\theta}-\boldsymbol{\theta}^\star\Vert=\rho}\sum^n_{i=1}\mathbb{E}_{\boldsymbol{\theta}^\star}\left[\log\frac{f(x\vert\boldsymbol{\theta};\alpha_i)}{f(x\vert\boldsymbol{\theta}^\star;\alpha_i)}\right]
    \le\sup_{\Vert\boldsymbol{\theta}-\boldsymbol{\theta}^\star\Vert=\rho}\mathbb{E}_{\boldsymbol{\theta}^\star}\left[\log\frac{f(x\vert\boldsymbol{\theta};\alpha_0)}{f(x\vert\boldsymbol{\theta}^\star;\alpha_0)}\right]<0,
\end{align*}
which establishes the condition (A2)(ii).


When $\alpha$ only takes value in a finite set, for any given $\boldsymbol{\theta}$, the expectation of the likelihood will also have maximum value and $s_i(\boldsymbol{\theta})$ will be uniformly bounded from below and above too, so similar steps from above can be adopted to complete the proof.
\end{proof}

For most commonly adopted spectral models, Regularity Condition (C) is 
satisfied. For example, the spectral model $f(E;\boldsymbol{\theta})$ in \eqref{eq:po} is characterized by the energy $E$, which ranges from ${E}_{\min}$ to ${E}_{\max}$ and it is easy to see that when $\boldsymbol{\theta}$ is fixed, the functions are continuous with respect to $E$.

\subsubsection{Consistency and Asymptotic Normality of $\hat{\boldsymbol{\theta}}$}
In this section, we establish the asymptotic normality of $\hat{\boldsymbol{\theta}}$ under Regularity Conditions (A) given in Section~\ref{appendix:conditionA} and $H_0$ in Equation~\eqref{eqn:null}.

\begin{theorem}[Consistency and Asymptotic Normality of $\hat{\boldsymbol{\theta}}$]\label{theorem:CAN of theta}
Under Regularity Conditions (A) and $H_0$ in~\eqref{eqn:null}, assuming $\sup\limits_{1\le i\le n}\mathbb{E}_{\boldsymbol{\theta}^\star}\vert\log f_i(N_i\vert\boldsymbol{\theta})\vert<\infty$, we have for any $\epsilon>0$,
$$P_{\boldsymbol{\theta}^\star}(\vert\hat{\boldsymbol{\theta}}-\boldsymbol{\theta}^\star\vert\le\epsilon)\to 1,\quad \mathrm{as }\quad n\to\infty.$$
    Furthermore, if the expected Fisher information matrix $I(\boldsymbol{\theta}^\star)$
    exists and is positive definite, 
    then  $\hat{\boldsymbol{\theta}}$ converges in distribution to Gaussian:
$$\sqrt{n}(\hat{\boldsymbol{\theta}}-\boldsymbol{\theta}^\star)\stackrel{D}{\to} N(\boldsymbol{0},I^{-1}(\boldsymbol{\theta}^\star)),\quad \mathrm{as}\quad n\to\infty.$$
\end{theorem}

\label{proof:CAN of theta}
\begin{proof}This proof is quite similar to the proofs in \citet{hoadley1971asymptotic} (Theorem 1 and 2){, who extends the standard properties of
consistency and asymptotic normality for \textit{i.i.d.} variables \citep[e.g., Sec.~5f of][]{rao1973} \citep[Sec.~33.3 of][]{cramer1946} to non-\textit{i.i.d.} measurements}.

First we prove the consistency of $\hat{\boldsymbol{\theta}}$.

Since $\sup\limits_{1\le i \le n}\mathbb{E}_{\boldsymbol{\theta}^\star}\vert\log f_i(N_i\vert\boldsymbol{\theta})\vert<\infty$ and our model is identifiable, the Kolmogorov's strong law of large numbers implies the convergence in probability
$$\frac{\log L_n(\boldsymbol{\theta})-\log L_n(\boldsymbol{\theta}^\star)}{n}\to \lim_{n\to\infty}\frac{1}{n}\mathbb{E}_{\boldsymbol{\theta}^\star}\left[\log\frac{  L_n(\boldsymbol{\theta})}{L_n(\boldsymbol{\theta}^\star)}\right]<0,$$
which establishes consistency of $\hat{\boldsymbol{\theta}}$.

Next we prove the asymptotic normality of $\hat{\boldsymbol{\theta}}$.
Denote $\ell_i(\boldsymbol{\theta}\vert X)=\log f_i(X\vert\boldsymbol{\theta})$ and $S^{(i)}_{\boldsymbol{\theta}}(X)=\boldsymbol{D}\ell_i(\boldsymbol{\theta}\vert X)=\left(\frac{\partial\ell_i(\boldsymbol{\theta}\vert X)}{\partial\theta_1},\cdots,\frac{\partial\ell_i(\boldsymbol{\theta}\vert X)}{\partial\theta_d} \right)^\top$. The Fisher information matrix $I_i(\boldsymbol{\theta})={\rm Var}_{\boldsymbol{\theta}}(S^{(i)}_{\boldsymbol{\theta}}(X))=\mathbb{E}_{\boldsymbol{\theta}}[S^{(i)}_{\boldsymbol{\theta}}(X)S^{(i)}_{\boldsymbol{\theta}}(X)^\top].$ Since $\boldsymbol{D}\ell(\hat{\boldsymbol{\theta}})=\sum^n_{i=1}S^{(i)}_{\hat{\boldsymbol{\theta}}}(N_i)=\boldsymbol{0}$, by Taylor's expansion, we have
$$\boldsymbol{0}=\frac{1}{n}\sum^n_{i=1}S^{(i)}_{\hat{\boldsymbol{\theta}}_n}(N_i)=\frac{1}{n}\sum^n_{i=1}S^{(i)}_{\boldsymbol{\theta}^\star}(N_i)+\frac{1}{n}\sum^n_{i=1}H_i(\boldsymbol{\theta}^\star)(\hat{\boldsymbol{\theta}}_n-\boldsymbol{\theta}^\star)+R_n,$$
where $H_i(\boldsymbol{\theta})$ is the Hessian matrix of $\ell_i(\boldsymbol{\theta},X)$.
Since $\hat{\boldsymbol{\theta}}_n\to\boldsymbol{\theta}^\star$ in probability and $R_n$ is quadratic in $\hat{\boldsymbol{\theta}}_n-\boldsymbol{\theta}^\star$, we have $\frac{\Vert R_n\Vert}{\Vert\hat{\boldsymbol{\theta}}_n-\boldsymbol{\theta}^\star \Vert}=o_p(1)\to 0$. Because for any $j,k$, $\sup\limits_i\mathbb{E}_{\boldsymbol{\theta}^\star}\vert\frac{\partial^2}{\partial\theta_j\partial\theta_k}\log f_i(x\vert\boldsymbol{\theta})\vert<\infty$, applying the Kolmogorov's law of large numbers gives
$$\lim_{n\to\infty}\frac{1}{n}\sum^n_{i=1}H_i(\boldsymbol{\theta}^\star)=\lim_{n\to\infty}\frac{1}{n}\sum^n_{i=1}\mathbb{E}_{\boldsymbol{\theta}^\star}[H_i(\boldsymbol{\theta}^\star)]=-\lim_{n\to\infty}\frac{1}{n}\sum^n_{i=1}\mathbb{E}_{\boldsymbol{\theta}^\star}[S^{(i)}_{\boldsymbol{\theta}^\star}(N_i)S^{(i)}_{\boldsymbol{\theta}^\star}(N_i)^\top]=-I(\boldsymbol{\theta}^\star). $$
Since $\mathbb{E}_{\boldsymbol{\theta}^\star}[S^{(i)}_{\boldsymbol{\theta}^\star}(N_i)]=\boldsymbol{0}$, the central limit theorem yields:
$$\frac{1}{\sqrt{n}}\sum^n_{i=1}S^{(i)}_{\boldsymbol{\theta}^\star}(N_i)\to N_d(\boldsymbol{0},I(\boldsymbol{\theta}^\star)). $$
Combining all these results and applying Slutsky's theorem\footnote{ This convergence theorem is due \cite{slutsky1925} in German, 
and an
equivalent theorem is given in Sec.~20.6 of \cite{cramer1946} in English.}, we have
$$\sqrt{n}(\hat{\boldsymbol{\theta}}_n-\boldsymbol{\theta}^\star)\to N_d(0,I^{-1}(\boldsymbol{\theta}^\star))\quad \mathrm{as}\quad  n\to\infty.$$\end{proof}

\subsection{Proofs about the $C$ function and $C$ statistics}
Proofs of the properties about $C_n(\boldsymbol{\theta}^*)$ and $C_n(\hat{\boldsymbol{\theta}})$ are given in this section.

\subsubsection{A Careful Application of Wilk's Theorem}
We begin with a review of the classic Wilks' Theorem. 

\begin{theorem}[Wilks' theorem]\label{Wilk's theorem}
    Consider the following hypothesis test:
$$H_0:\boldsymbol{\theta}\in\boldsymbol{\Theta}_0 
\ \hbox{ versus } \ 
H_1:\boldsymbol{\theta}\notin\boldsymbol{\Theta}_0, 
\ \hbox{ where }  \
\boldsymbol{\Theta}_0\subset\boldsymbol{\Theta},$$
where the parameter spaces, $\boldsymbol{\Theta_0}$ and $\boldsymbol{\Theta}$, are $d$ and
$k$-dimensional subsets of $\mathbb{R}^k$, respectively. 
Under Regularity Conditions (A) and assuming 
$H_0$,
    \begin{equation}
        -2\log\Lambda_n=-2\log \frac{\sup_{\boldsymbol{\Theta}_0} L(s_1,\ldots, s_n\mid N_1,\ldots, N_n)}{\sup_{\boldsymbol{\Theta}} L(s_1,\ldots, s_n\mid N_1,\ldots, N_n)}\to\chi^2_{k-d},
        \ \hbox{ as } \ n\to\infty, 
    \end{equation}
where $\chi_{k-d}^2$ denotes a chi-squared random variable with $k$--\,$d$ degrees of freedom.
\end{theorem}

Then we introduce a key lemma which characterizes the relationship between $C_n(\boldsymbol{\theta}^*)$ and $C_n(\hat{\boldsymbol{\theta}})$.

\begin{lemma}  
[Wilk's Theorem for C-statistics]
    \label{prop:wilks}
        Under Regularity Condition (B1) and under the null hypothesis $H_0$ given in~\eqref{eqn:null}
    \begin{equation}
        \label{eqn:wilks}
       \Gamma_n := C_n(\boldsymbol{\theta}^\star) -C_n(\hat{\boldsymbol{\theta}}) \rightarrow \chi_d^2 \quad {\rm as}\quad n\rightarrow\infty,
    \end{equation}
where $d$ is the number of parameters under the null.
\end{lemma}

\label{proof:wilks}
\begin{proof}
We have 
$$\Gamma_n=C_n(\boldsymbol{\theta}^\star)-C_n(\hat{\boldsymbol{\theta}})=-2\log\frac{L_n(\boldsymbol{\theta}^\star)}{\sup_{\boldsymbol{\theta}\in\boldsymbol{\Theta}}L_n(\boldsymbol{\theta})}.$$
Since the dimension of $\boldsymbol{\Theta}$ is fixed, we can directly apply the Wilks' Theorem to obtain 
$$\Gamma_n=C_n(\boldsymbol{\theta}^\star)-C_n(\hat{\boldsymbol{\theta}})=-2\log\Lambda_n\to\chi^2_d,\quad\mathrm{as}\quad n\to\infty.$$\end{proof}

\subsubsection{Proof of Theorem \ref{theorem:CAN of C_true}}\label{proof:CAN of C_true}

\begin{proof} (i) 
In this proof, because the measures we use are all unconditional measure, for simplicity, we omit the subscripts. Condition (B2) and Lemma~\ref{lemma:divide_into_infinite_bins} indicate that $s_i$ are uniformly bounded away from $0$ and above. Since ${\rm Var}(C^{(i)}(\boldsymbol{\theta}^{\star}))$ and $\mathbb{E}\vert C^{(i)}(\boldsymbol{\theta})-\mathbb{E}[C^{(i)}(\boldsymbol{\theta})]\vert^3$ are continuous functions with respect to $s_i$, there exists $M_1>0$ and $M_2>0$ such that ${\rm Var}(C^{(i)}(\boldsymbol{\theta}^{\star}))>M_1$ and $\mathbb{E}\vert C^{(i)}(\boldsymbol{\theta})-\mathbb{E}[C^{(i)}(\boldsymbol{\theta})]\vert^3<M_2$. Therefore,
\begin{align*}\nonumber
    0\le\frac{\sum_{i=1}^n\mathbb{E}\vert C^{(i)}(\boldsymbol{\theta}^\star)-\mathbb{E}[C^{(i)}(\boldsymbol{\theta}^\star)]\vert^3}{\big(\sum_{i=1}^n{\rm Var}(C^{(i)}(\boldsymbol{\theta}^\star))\big)^{3/2}}\le \frac{nM_2}{\left(nM_1\right)^{3/2}}\to0,
\end{align*}
which is the Lyapunov condition when $\delta=3$.

By applying the Lindeberg-Feller Central Limit Theorem, we get
$$\frac{C_n(\boldsymbol{\theta}^{\star})-\mathbb{E}[C_n(\boldsymbol{\theta}^{\star})]}{\sqrt{{\rm Var}(C_n(\boldsymbol{\theta}^{\star}))}}\to N(0,1).$$

(ii) 
From Lemma~\ref{prop:wilks}, $C_n(\hat{\boldsymbol{\theta}})-C_n(\boldsymbol{\theta}^{\star})\to-\chi^2_d$ as $n\to\infty$, where $d$ is the dimensionality of $\boldsymbol{\theta}$. And condition (B2) implies that $\sqrt{{\rm Var}(C_n(\boldsymbol{\theta}^{\star}))}=\sqrt{\sum^n_{i=1}{\rm Var}(C^{(i)}(\boldsymbol{\theta}^{\star}))}\to\infty$ as $n\to\infty$. Hence we have 
$$\frac{C_n(\hat{\boldsymbol{\theta}})-C_n(\boldsymbol{\theta}^{\star})}{\sqrt{{\rm Var}(C_n(\boldsymbol{\theta}^{\star}))}}\to0\quad \text{in probability.}$$
So by the Slutsky's theorem, 
$$\frac{C_n(\hat{\boldsymbol{\theta}})-\mathbb{E}[C_n(\boldsymbol{\theta}^{\star})]}{\sqrt{{\rm Var}(C_n(\boldsymbol{\theta}^{\star}))}}\to N(0,1).$$
Similarly, denote $C_n(\hat{\boldsymbol{\theta}})-C_n(\boldsymbol{\theta}^{\star})=\Gamma_n$, we have the following results:
$$\frac{\mathbb{E}[C_n(\hat{\boldsymbol{\theta}})]-\mathbb{E}[C_n(\boldsymbol{\theta}^{\star})]}{\sqrt{{\rm Var}(C_n(\boldsymbol{\theta}^{\star}))}}=\frac{\mathbb{E}[\Gamma_n]}{\sqrt{{\rm Var}(C_n(\boldsymbol{\theta}^{\star}))}}\to0\quad \text{in probability,}$$
and
\begin{equation}
\begin{aligned}\nonumber
\left\vert\frac{{\rm Var}(C_n(\hat{\boldsymbol{\theta}}))}{{\rm Var}(C_n(\boldsymbol{\theta}^{\star}))}-1\right\vert&=\left\vert\frac{{\rm Var}(\Gamma_n+C_n(\boldsymbol{\theta}^{\star}))}{{\rm Var}(C_n(\boldsymbol{\theta}^{\star}))}-1\right\vert=\left\vert\frac{{\rm Var}(\Gamma_n)+2{\rm Cov}(C_n(\boldsymbol{\theta}^{\star}),\Gamma_n)}{{\rm Var}(C_n(\boldsymbol{\theta}^{\star}))}\right\vert\\
&\le\left\vert\frac{{\rm Var}(\Gamma_n)}{{\rm Var}(C_n(\boldsymbol{\theta}^{\star}))}\right\vert+\left\vert\frac{2\sqrt{{\rm Var}(C_n(\boldsymbol{\theta}^{\star})){\rm Var}(\Gamma_n)}}{{\rm Var}(C_n(\boldsymbol{\theta}^{\star}))}\right\vert\quad(\text{Cauchy's Inequality})\\
&\to0\quad \text{in probability,}
\end{aligned}
\end{equation}

So by the continuous mapping theorem, we have
$$\sqrt{\frac{{\rm Var}(C_n(\boldsymbol{\theta}^{\star}))}{{\rm Var}(C_n(\hat{\boldsymbol{\theta}}))}}\to1\quad \text{in probability.}$$
Again, by the Slutsky's theorem, we get,
$$\frac{C_n(\hat{\boldsymbol{\theta}})-\mathbb{E}[C_n(\hat{\boldsymbol{\theta}})]}{\sqrt{{\rm Var}(C_n(\hat{\boldsymbol{\theta}}))}}\to N(0,1).$$\end{proof}

\subsubsection{Proof of Theorem \ref{thm:cmin}}
\label{proof:cmin}
\begin{proof}

(i) Using a first-order expansion of $\boldsymbol{D}\ell$ around $\boldsymbol{\theta}^\star$, we have
\begin{equation}
\label{eq:expansion_Dl}
\hat{\boldsymbol{\theta}}-\boldsymbol{\theta}^\star={I}_n^{-1}(\boldsymbol{\theta}^\star)\boldsymbol{D}\ell(\boldsymbol{\theta}^\star)+O_p(n^{-1}).
\end{equation}
Denote $Z(\boldsymbol{\theta})=C_n(\boldsymbol{\theta})-\mu(\boldsymbol{\theta})$. By expansion, we have
\begin{equation}\label{eq:expansion_Z}
    Z(\hat{\boldsymbol{\theta}})-Z(\boldsymbol{\theta}^\star)=\left\{\boldsymbol{D}Z(\boldsymbol{\theta}^\star)\right\}^\top[\hat{\boldsymbol{\theta}}-\boldsymbol{\theta}^\star]+O_p(1),
\end{equation}
where $\boldsymbol{D}Z(\boldsymbol{\theta})=\boldsymbol{D}C_n(\boldsymbol{\theta})-{\rm Cov}_{\boldsymbol{\theta}}(C_n(\boldsymbol{\theta}),\boldsymbol{D}\ell(\boldsymbol{\theta}))=-{\rm Cov}_{\boldsymbol{\theta}}(C_n(\boldsymbol{\theta}),\boldsymbol{D}\ell(\boldsymbol{\theta}))+O_p(n^{1/2})$, because $\mathbb{E}[\boldsymbol{D}C_n(\boldsymbol{\theta})]=0$.
In combination with \eqref{eq:expansion_Dl}, we get the following representation
\begin{equation}
\label{eq:z_theta}
Z(\hat{\boldsymbol{\theta}})=\Phi_{+}+O_p(1),
\end{equation}
where $\Phi_{+}=Z(\boldsymbol{\theta}^\star)-\boldsymbol{c}^\top(\boldsymbol{\theta}^\star){I}_n^{-1}(\boldsymbol{\theta}^\star)\boldsymbol{D}\ell(\boldsymbol{\theta}^\star)$, which is the sum of independent random variables. Since $\mathbb{E}[\Phi_{+}]=0$ and ${\rm Var}(\Phi_{+})=\sigma^2(\boldsymbol{\theta}^\star)={\rm Var}(C_n(\boldsymbol{\theta}^\star))-Q(\boldsymbol{\theta}^\star)$, with the regularity condition (B2), the Lyapunov condition is satisfied, so the central limit theorem yields
\begin{equation}
\label{eq:phi_norm}
\Phi_{+}/\sigma(\boldsymbol{\theta}^\star)\to N(0,1)\quad{\rm as}\quad n\to\infty.
\end{equation}
Using $\sigma(\boldsymbol{\theta})\to\infty$ as a consequence of condition (B2), we get from \eqref{eq:z_theta} and \eqref{eq:phi_norm}
$$Z(\hat{\boldsymbol{\theta}})/\sigma(\boldsymbol{\theta}^\star)=T\cdot\sigma(\hat{{\boldsymbol{\theta}}})/\sigma(\boldsymbol{\theta}^\star)\to N(0,1)\quad{\rm as}\quad n\to\infty.$$ 
The condition (B1) indicates $\sigma(\hat{{\boldsymbol{\theta}}})/\sigma(\boldsymbol{\theta}^\star)
\to1$ in probability, which establishes the results.

(ii) For the conditional asymptotic distribution of $C_n(\hat{\boldsymbol{\theta}})$ given $\hat{\boldsymbol{\theta}}$, since  $C_n(\hat{\boldsymbol{\theta}})-C_n(\boldsymbol{\theta}^\star)$ is conditionally constant based on Lemma~\ref{prop:wilks}, we only need to consider the conditional distribution of $C_n(\boldsymbol{\theta}^\star)$ given $\hat{\boldsymbol{\theta}}$. First we consider joint distribution of $\left(C_n(\boldsymbol{\theta}^\star),{I}_n^{-1}(\boldsymbol{\theta}^\star)\boldsymbol{D}\ell(\boldsymbol{\theta}^\star)\right)$, which is the sum of independent random vectors.  By the central limit theorem, the joint distribution is $(d+1)$-variate normal. Therefore, with \eqref{eq:expansion_Dl}, the joint distribution of $(C_n(\boldsymbol{\theta}^\star),\hat{\boldsymbol{\theta}})$ is also $(d+1)$-variate normal.

\end{proof}

\subsubsection{Proof of Proposition~\ref{rmk:1_order_same} }
\label{proof:con_uncon_same}
\begin{proof}
 Based on the joint asymptotic normality of $(C_n(\boldsymbol{\theta}^\star),\hat{\boldsymbol{\theta}})$, to the first order, it follows that, conditionally on $\hat{\boldsymbol{\theta}}$, $C_n(\boldsymbol{\theta}^\star)$ is asymptotically normal. Therefore, we have
\begin{equation}
\begin{aligned}
    \mathbb{E}[C_n(\hat{\boldsymbol{\theta}})\vert\hat{\boldsymbol{\theta}}]&=\mathbb{E}_{\boldsymbol{\theta}^*}[C_n(\boldsymbol{\theta}^*)]+\left\{{\rm Cov}_{\boldsymbol{\theta}^*}(C_n(\boldsymbol{\theta}^*),\hat{\boldsymbol{\theta}})\right\}^\top{\rm Cov^{-1}_{\boldsymbol{\theta}^*}(\hat{\boldsymbol{\theta}})}\left(\hat{\boldsymbol{\theta}}-\mathbb{E}_{\boldsymbol{\theta}^*}[\hat{\boldsymbol{\theta}}]\right)+O_p(1)\\
    &=\mu(\hat{\boldsymbol{\theta}})+O_p(1),\\
    {\rm Var}(C_n(\hat{\boldsymbol{\theta}})\vert\hat{\boldsymbol{\theta}})&={\rm Var}_{\boldsymbol{\theta}^*}(C_n(\boldsymbol{\theta}^*))-\left\{{\rm Cov}_{\boldsymbol{\theta}^*}(C_n(\boldsymbol{\theta}^*),\hat{\boldsymbol{\theta}})\right\}^\top{\rm Cov^{-1}_{\boldsymbol{\theta}^*}(\hat{\boldsymbol{\theta}})}\left\{{\rm Cov}_{\boldsymbol{\theta}^*}(C_n(\boldsymbol{\theta}^*),\hat{\boldsymbol{\theta}})\right\}+O_p(1)\\
    &=\sigma^2(\hat{\boldsymbol{\theta}})+O_p(n^{1/2}),
\end{aligned}    
\end{equation}
where for each equality, the first equality holds due to Wilks' Theorem and properties of multivariate normal distribution; the second equality holds because of \eqref{eq:expansion_Dl}, \eqref{eq:expansion_Z} and the asymptotic normality of $\hat{\boldsymbol{\theta}}$.     Hence, to first order the conditional moments are the same as these unconditional moments evaluated at $\hat{\boldsymbol{\theta}}$.

\end{proof}
\begin{remark}
    Moreover, based on the results in Theorem~\ref{thm:highorder}, and observing that 
\begin{equation}\label{eq:covariance}
\begin{aligned}
    \boldsymbol{c}(\boldsymbol{\theta})&={\rm Cov}_{\boldsymbol{\theta}}(C_n(\boldsymbol{\theta}),\boldsymbol{D}\ell(\boldsymbol{\theta}))=\sum_{i=1}^n{\rm Cov}_{\boldsymbol{\theta}}(C^{(i)}(\boldsymbol{\theta}),\frac{N_i}{s_i}\boldsymbol{D}s_i)=X^\top\kappa_{11},\\
{I}_n(\boldsymbol{\theta})&=\mathbb{E}_{\boldsymbol{\theta}}\left[\frac{\partial\ell(\boldsymbol{\theta})}{\partial\boldsymbol{\theta}}\frac{\partial\ell(\boldsymbol{\theta})}{\partial\boldsymbol{\theta}}^\top\right]={\rm Cov}_{\boldsymbol{\theta}}(\boldsymbol{D}\ell(\boldsymbol{\theta}))=X^\top{\rm Cov}(\boldsymbol{N}/\boldsymbol{s}^2)X=X^\top V^{-1}X,\\
    \mu({\boldsymbol{\theta}})&=\mathbb{E}_{{\boldsymbol{\theta}}}[C_n(\boldsymbol{\theta})]=\kappa_1^{(\cdot)},\\
    \sigma^2({\boldsymbol{\theta}})&={\rm Var}_{{\boldsymbol{\theta}}}(C_n(\boldsymbol{\theta}))-Q({\boldsymbol{\theta}})=\kappa_2^{(\cdot)}-\kappa_{11}^\top X(X^\top V^{-1}X)^{-1}X^\top\kappa_{11} ,       
\end{aligned}
\end{equation}
    we can further reduce the error of variance to $O_p(1)$.
\end{remark}

\subsubsection{Proof of Proposition~\ref{prop:bootstrap_invalid}}
\label{proof:bootstrap_invalid}
\begin{proof}
    Combining the results regarding the unconditional moments of $C_n(\hat{\boldsymbol{\theta}})$ in \eqref{eq:moments_uncon_Cmin}, the expectation and variance given by methods based on Theorem~\ref{theorem:CAN of C_true}, but with the mean and variance in \eqref{eq:marginal_normal_cmin}
computed assuming $\boldsymbol{\theta}^\star =\hat{\boldsymbol{\theta}}$ can be expressed as $\hat\mu=\hat{\kappa}_1^{(\cdot)}+O_p(1)$ and $\hat\sigma^2=\hat{\kappa}_2^{(\cdot)}+O_p(1)$. The expectation agrees with the results in \eqref{eq:T_normal} but the variance fails. The difference is $Q(\hat{\boldsymbol{\theta}})={\boldsymbol{c}}^\top(\hat{\boldsymbol{\theta}})n^{-1}{\boldsymbol{I}}^{-1}(\hat{\boldsymbol{\theta}}){\boldsymbol{c}}(\hat{\boldsymbol{\theta}})=\hat\kappa_{11}^\top \hat{X}(\hat{X}^\top\hat{V}^{-1}\hat{X})^{-1}\hat{X}^\top\hat\kappa_{11}$, which is $O_p(n)$ according to our assumption. Therefore, using the Slutsky theorem, \eqref{eq:T_normal} implies using expectation and variance given by these na\"ive methods will lead to bias with order $O_p(1)$ as a result of the difference ${Q}(\hat{\boldsymbol{\theta}})$. Because $\kappa_{11}=(\kappa_{11}^{(1)}/V_1,\cdots,\kappa_{11}^{(n)}/V_n)^\top$ and $\kappa_{11}^{(i)}\to0$ as $s_i\to\infty$, we have $Q(\boldsymbol{\theta})\to 0$ as $s_i \to\infty$ uniformly. Hence again, the Slutsky theorem implies $Q(\hat{\boldsymbol{\theta}})\to 0$ as $s_i \to\infty$ uniformly; so the bias will diminish.
\end{proof}

\subsubsection{Proof of Theorem \ref{thm:uncon_highorder}}
\begin{proof}
    
In this section, we use the notation $h_i^{-1}(\mu_i)=\boldsymbol{\beta}\in\mathbb{R}^p$ and unless otherwise specified with subscripts, the measure is unconditional measure given by the true value of parameter.
Here we introduce the modified Cash statistics
$$C_n(\hat{\boldsymbol{\theta}})^\star=\sum^n_{i=1}\frac{\hat{C}^{(i)}}{\hat{\kappa}_1^{(i)}},$$
and a supplementary estimating equation $g_q(\hat{\boldsymbol{\beta}},\hat{\phi})=0$, where
\begin{equation}
g_q(\boldsymbol{\beta},\phi)=\sum^n_{i=1}\left\{ \frac{C^{(i)}}{\kappa_1^{(i)}}-\phi\right\}
    \label{eq:estimating_eq}
\end{equation}
and $q=p+1$. In \eqref{eq:estimating_eq}, $\phi$ represents a notional dispersion parameter $\mathbb{E}(C^{(i)})/\kappa_1^{(i)}$ whose value is 1 under the model, and $C_n(\hat{\boldsymbol{\theta}})^\star=n\hat{\phi}$. The replacement of $\kappa_1^{(i)}$ by 1 in the definition of $C_n(\hat{\boldsymbol{\theta}})^\star$ would yield the usual statistic $C_n(\hat{\boldsymbol{\theta}})$. Let $\boldsymbol{\theta}$ denote the true parameter value $(\beta_1,\cdots,\beta_p,1)^\top$ and $g=(g_1,\cdots,g_p,g_q)^\top$.

Then, the Taylor expansion of $g_r,r=1,...,q$, about $\boldsymbol{\theta}$ leads to
\begin{equation}
0=n^{-1/2}g_r(\boldsymbol{\theta})+n^{-1/2}(e_r+n\delta_r)(\hat{\boldsymbol{\theta}}-\boldsymbol{\theta})+\frac{1}{2}n^{-1/2}(\hat{\boldsymbol{\theta}}-\boldsymbol{\theta})^\top(u_r+n\nu_r)(\hat{\boldsymbol{\theta}}-\boldsymbol{\theta})+O_p(n^{-1})
\label{eq:taylor}
\end{equation}
where 
$$\delta_{rs}=\frac1n\mathbb{E}(\frac{\partial g_r}{\partial \theta_s}\vert_{\boldsymbol{\theta}}),\quad \delta_r=(\delta_{r1},...,\delta_{rq}),$$
$$e_{rs}=\frac{\partial g_r}{\partial \theta_s}\vert_{\boldsymbol{\theta}}-n\delta_{rs},\quad e_r=(e_{r1},...,e_{rq}),$$
$$\nu_{r,st}=\frac1n\mathbb{E}(\frac{\partial^2g_r}{\partial\theta_s\partial\theta_t}\vert_{\boldsymbol{\theta}}),\quad \nu_r=[\nu_{r,st}],$$
$$u_{r,st}=\frac{\partial^2g_r}{\partial\theta_s\partial\theta_t}\vert_{\boldsymbol{\theta}}-n\nu_{r,st},\quad u_r=[u_{r,st}].$$
Since $\hat{\boldsymbol{\theta}}$ has asymptotic normality, writing 
$$\hat{\boldsymbol{\theta}}-\boldsymbol{\theta}=n^{-1/2}Z_1+n^{-1}Z_2+O_p(n^{-3/2}),$$
where $Z_1$ and $Z_2$ are $q$-dimensional vectors of $O_p(1)$, and inverting \eqref{eq:taylor} gives
$$Z_1=-n^{-1/2}\Delta^{-1}g,$$
$$Z_2=-\frac{1}{2}\Delta^{-1}c-n^{-1/2}EZ_1,$$
where $\Delta=[\delta_{rs}]$, $c_r=Z_1^\top\nu_rZ_i$ and $E=[e_{rs}]$. Hence
\begin{equation}
   \begin{aligned} 
    \mathbb{E}(\hat{\boldsymbol{\theta}}-\boldsymbol{\theta})=n^{-1}\mathbb{E}(Z_2)+O(n^{-2})    \\
    \mathrm{Var}(\hat{\boldsymbol{\theta}})=n^{-1}\mathbb{E}(Z_1Z_1^\top)+O(n^{-2}).
\end{aligned}\label{eq:aysp_theta}
\end{equation}
Partition $\Delta$ as $(p+1)\times(p+1)$ matrix and write
$$\Delta=\left(
\begin{array}{cc}
    \Delta_{11} & \Delta_{12}  \\
     \Delta_{21}& \Delta_{22}
\end{array}\right),
$$
where 
$$\Delta_{11,rs}=-\frac{1}{n}\sum^n_{i=1}\frac{1}{V_i}\frac{\partial \mu_i}{\partial \beta_r}\frac{\partial \mu_i}{\partial \beta_s},$$
$$\Delta_{12,s}=\Delta_{21,r}=0,\quad\Delta_{22}=-1.$$
Similarly we have
$$e_{qs}=2\sum^n_{i=1}\frac{y_i-\mu_i}{V_i\kappa_1^{(i)}}\frac{\partial\mu_i}{\partial\beta_s},$$
$$e_{rq}=e_{qq}=0.$$
Partitioning $E=[e_{rs}]$ as 
$$\left(
\begin{array}{cc}
    E_{11} & 0 \\
   E_{21}  & 0
\end{array}\right)$$
and $c$ as $(c^1,c^2)^\top$, it follows from \eqref{eq:aysp_theta} that
\begin{equation}
    \begin{aligned}
\mathbb{E}(\hat{\phi})&=1-\frac{1}{n}\mathbb{E}\left(-\frac{1}{2}c^2+\frac{1}{n}E_{21}\Delta_{11}^{-1}g^1\right)+O(n^{-2}),\\
\mathrm{Var}(\hat{\phi})&=n^{-2}\mathbb{E}[(g^2)^2]+O(n^{-2}).
\end{aligned}\label{eq:unconditional}
\end{equation}
Now, using tensor notation and omitting summation, we have
\begin{equation}\nonumber
\begin{aligned}
    \mathbb{E}(c^2)&=\mathbb{E}(Z_1^\top\nu_qZ_1)=\frac{1}{n}\mathbb{E}(g_1^\top\Delta_{11}^{-1}\nu_q^1\Delta_{11}^{-1}g_1)\\
    &=n^{-1}\mathbb{E}\left\{g_r(-nb_{rr'})\nu_{qr's'}(-nb_{s's})g_s\right\}\\
    &=\frac{2}{V_i\kappa_1^{(i)}}Q_{ii}
\end{aligned}
\end{equation}
and
\begin{equation}\nonumber
\begin{aligned}
    \frac{1}{n}\mathbb{E}(E_{21}\Delta_{11}^{-1}g^1)&=\frac{1}{n}\mathbb{E}\left\{e_{21,r}(-nb_{rs})g_s\right\}\\
    &=b_{rs}\mathbb{E}\left\{\frac{2(y_i-\mu_i)}{V_i\kappa_1^{(i)}}\frac{\partial\mu_i}{\partial\beta_r}g_s \right\}=\frac{2}{V_i\kappa_1^{(i)}}Q_{ii}.
\end{aligned}
\end{equation}
Then, using the above equations in \eqref{eq:unconditional}, after simplification, we obtain
\begin{equation}\label{eq:Cmin_expectation_unconditional}
    \mathbb{E}(\hat{\phi})=1-\frac{1}{nV_i\kappa_1^{(i)}}Q_{ii}+O(\frac{1}{n^2}).
\end{equation}
Similarly it can be shown that
\begin{equation}\label{eq:Cmin_variance_unconditional}
    \mathrm{Var}(\hat{\phi})=\frac{1}{n^2}\frac{\kappa_2^{(i)}}{{\kappa_1^{(i)}}^2}+O(\frac{1}{n^2}).
\end{equation}
Replace $\kappa_1^{(i)}$ with $1$ and multiply $n$. Observing that $\sum^n_{i}\frac{Q_{ii}}{V_i}={\rm trace}(VQ)=p$, we get
\begin{equation}
\begin{aligned}\label{eq:moments_uncon_Cmin_proof}
    &\mathbb{E}[C_n(\hat{\boldsymbol{\theta}})]=\kappa_1^{(i)}-\frac{Q_{ii}}{V_i}+O(\frac{1}{n})=\mathbb{E}[C_n({\boldsymbol{\theta}}^*)]-p+O(\frac{1}{n}),\\
    &{\rm Var}(C_n(\hat{\boldsymbol{\theta}}))={\rm Var}(C_n({\boldsymbol{\theta}}^*))+O(1),
\end{aligned}
\end{equation}
 which match the results implied by Wilks' Theorem, i.e. $C_n(\boldsymbol{\theta}^*)-C_n(\hat{\boldsymbol{\theta}})\to\chi^2_p$ as $n\to\infty$.
\end{proof}

\subsubsection{Proof of Theorem~\ref{thm:highorder}}
\begin{proof}

To obtain the asymptotic expression of conditional expectation and conditional variance of $\hat{\phi}$, we need the conditional cumulant formula in \cite{mccullagh1984}, section 6.2 and \cite{mccullagh1987tensor}, section 5.6. Here we restate it with our notations in the following Lemma.

\begin{lemma}\label{lemma:PeterM's_con_formula}
    Under regularity condition (B), the conditional log density of $\hat{\phi}$ given $\hat{\boldsymbol{\beta}}$ has a formal Edgeworth expansion in which the first two cumulants are
    \begin{align}
        \mathbb{E}(\hat{\phi}\vert\hat{\boldsymbol{\beta}})&=\hat\kappa^{d}+\hat\kappa^{d,r}\hat h_r+\hat\kappa_1^{d,r,s}\frac{\hat h_{rs}}{2\sqrt{n}}+O(n^{-3/2}),\\
        \mathrm{Var}({\hat\phi}\vert\hat{\boldsymbol{\beta}})&=\hat\kappa_1^{d,d}+\hat\kappa_1^{d,d,r}\hat h_r+O(n^{-2}).
    \end{align}
Here $\kappa$'s are the cumulants of $(\hat\phi,\hat{\boldsymbol{\beta}})$ and $\kappa_1$'s are the cumulants of uncorrelated variables $Y$, where $Y^q=\hat\phi-\sum^p_{j=1}\hat\eta_j\hat\beta_j$ and $Y^j=\hat\beta_j$, $j=1,\dots,p$. We have $\kappa_1^{d,d}=\kappa^{d,d}-\eta_r\eta_s\kappa^{r,s}$, $\kappa_1^{d,r,s}=\kappa^{d,r,s}-\eta_t\kappa^{r,s,t}$ and $\kappa_1^{d,d,r}=\kappa^{d,d,r}-\eta_s\kappa^{d,r,s}[2]+\eta_s\eta_t\kappa^{r,s,t}$.
\end{lemma}
So we have
\begin{equation}
    \begin{aligned}
        \mathbb{E}(\hat{\phi}\vert\hat{\boldsymbol{\beta}})&=\hat{\kappa}^d+\hat{\kappa}^{d,r}\hat{h}_r+(\hat{\kappa}^{d,r,s}-\hat{\eta}_t\hat{\kappa}^{r,s,t})\frac{\hat{h}_{rs}}{2\sqrt{n}}+O(n^{-3/2})\\
        \mathrm{Var}(\hat{\phi}\vert\hat{\boldsymbol{\beta}})&=\hat{\kappa}^{d,d}-\hat{\eta}_r\hat{\eta}_s\hat{\kappa}^{r,s}+(\hat{\kappa}^{d,d,r}-\hat{\eta}_s\hat{\kappa}^{d,r,s}[2]+\hat{\eta}_s\hat{\eta}_t\hat{\kappa}^{r,s,t})\frac{\hat{h}_r}{\sqrt{n}}+O(n^{-2}),
    \end{aligned}\label{eq:conditional}
\end{equation}
where $h_r=\lambda_{r,s}(x^s-\lambda^s)=\lambda_{r,s}(\hat{\boldsymbol{\beta}}\sqrt{n}-\mathbb{E}(\hat{\boldsymbol{\beta}}\sqrt{n}))$, $h_{rs}(x,\lambda)=h_rh_s-\lambda_{r,s}$, $\eta_t=\kappa^{d,s}(\kappa^{s,t})^{-1}$, $\kappa^d=\mathbb{E}[\phi]$,  $\kappa^{d,d}=\mathrm{cov}(\hat{\phi},\hat{\phi})$,  $\kappa^{d,r}=\mathrm{cov}(\hat{\phi},{\hat{\beta}_r})$, $\kappa^{r,s}=\mathrm{cov}({\hat\beta_r},{\hat\beta_s})$, $\kappa^{r,s,t}=\mathrm{cum}({\hat\beta_r},{\hat\beta_s},{\hat\beta_t})\sqrt{n}$, $\kappa^{d,r,s}=\mathrm{cum}({\hat\phi},{\hat\beta_r},{\hat\beta_s})\sqrt{n}$, $\kappa^{d,d,r}=\mathrm{cum}({\hat\phi},{\hat\phi}, {\hat\beta_r})\sqrt{n}$ and $\lambda_{r,s}=-\delta_{rs}$.
Using \eqref{eq:aysp_theta}, after detailed calculations, we have
$$h_r(x,\lambda)=\frac{1}{\sqrt{n}}g_r+O_p(n^{-1/2}),$$
$$h_{rs}(x,\lambda)=\frac{1}{n}g_rg_s+\delta_{rs}+O_p(n^{-1/2}),$$
$$\kappa^r=\frac{1}{n}\frac{\kappa_{11}^{(i)}}{\kappa_1^{(i)}V_i}\frac{\partial\mu_i}{\partial\beta_t}b_{tr}+O(\frac{1}{n^2}),$$
$$\kappa^{r,s,t}=n^{1/2}b_{rr'}b_{ss'}b_{tt'}\frac{\partial\mu_i}{\partial\beta_{r'}}\frac{\partial\mu_i}{\partial\beta_{s'}}\frac{\partial\mu_i}{\partial\beta_{t'}}\frac{\kappa_{03}^{(i)}}{V_i^3}+O(n^{-3/2}),$$
$$\kappa^{d,r,s}=n^{-1/2}\frac{\kappa_{12}^{(i)}}{\kappa_1^{(i)}V_i^2}\frac{\partial\mu_i}{\partial\beta_{r'}}\frac{\partial\mu_i}{\partial\beta_{s'}}b_{rr'}b_{ss'}+O(n^{-3/2}).$$
Then, using the above equations in \eqref{eq:conditional} and recall that $X=(\nabla h_1^\top,\cdots,\nabla h_n^\top)^\top$, we obtain
$$\mathbb{E}(\hat{\phi}\vert\hat{\boldsymbol{\beta}})=1-\frac{1}{2n}\bm{1}^\top\hat{X}^\top\{\hat{\kappa}_{12}-\hat{\kappa}_{11}^1\hat{\kappa}_{03}^3\}\hat{X}(\hat{X}^\top\hat{V}^{-1}\hat{X})^{-1}\bm{1}+O(n^{-3/2}),$$
where $\kappa_{12}=\mathrm{diag}(\kappa_{12}^{(i)}/\kappa_1^{(i)}V_i^2)$, $\kappa_{11}^1=\mathrm{diag}(\sum^n_{j=1}\kappa_{11}^{(j)}Q_{ji}/\kappa_1^{(j)}V_j)$, $\kappa_{03}^{3}=\mathrm{diag}(\kappa_{03}^{(i)}/V_i^3).$\\
Following similar calculations it can be shown that 
$$\mathrm{Var}(\hat{\phi}\vert\hat{\boldsymbol{\beta}})=\frac{1}{n^2}\hat{\kappa}_2-\frac{1}{n^2}\hat{\kappa}_{11}^\top\hat{X}(\hat{X}^\top\hat{V}^{-1}\hat{X})^{-1}\hat{X}^\top\hat{\kappa}_{11}^\top+O(n^{-2}), $$
where $\kappa_2=\sum^n_{j=1}\kappa_2^{(i)}/{\kappa_1^{(i)}}^2$. Replacing $\kappa_1^{(i)}$ with $1$ in the above equations, we obtain the results in Theorem \ref{thm:highorder}.\end{proof}

\section{Additional Numerical Results}
\label{results_appendix}

In this section, we present additional numerical results. Section~\ref{sec:Label_time_appendix} compares the computation time of our proposed algorithms and the existing algorithms discussed in Section~\ref{sec:algorithm} and Appendix~\ref{sec:algorithms_description}. Section~\ref{sec:sim_RMF} compares the true and nominal null distributions and the computational efficiency of the various algorithms in simulation studies where the RMF is varied.

\subsection{Computational Time Comparison}
\label{sec:Label_time_appendix}

Figure~\ref{fig:time} compares the computational time of the algorithms described in Appendix~\ref{sec:algorithms_description} and shows that our corrected Z-test is orders of magnitude faster than bootstrap-based algorithms (time is plotted on the log scale). 
\begin{figure}[tph]
    \centering
    \includegraphics[width=1\linewidth]{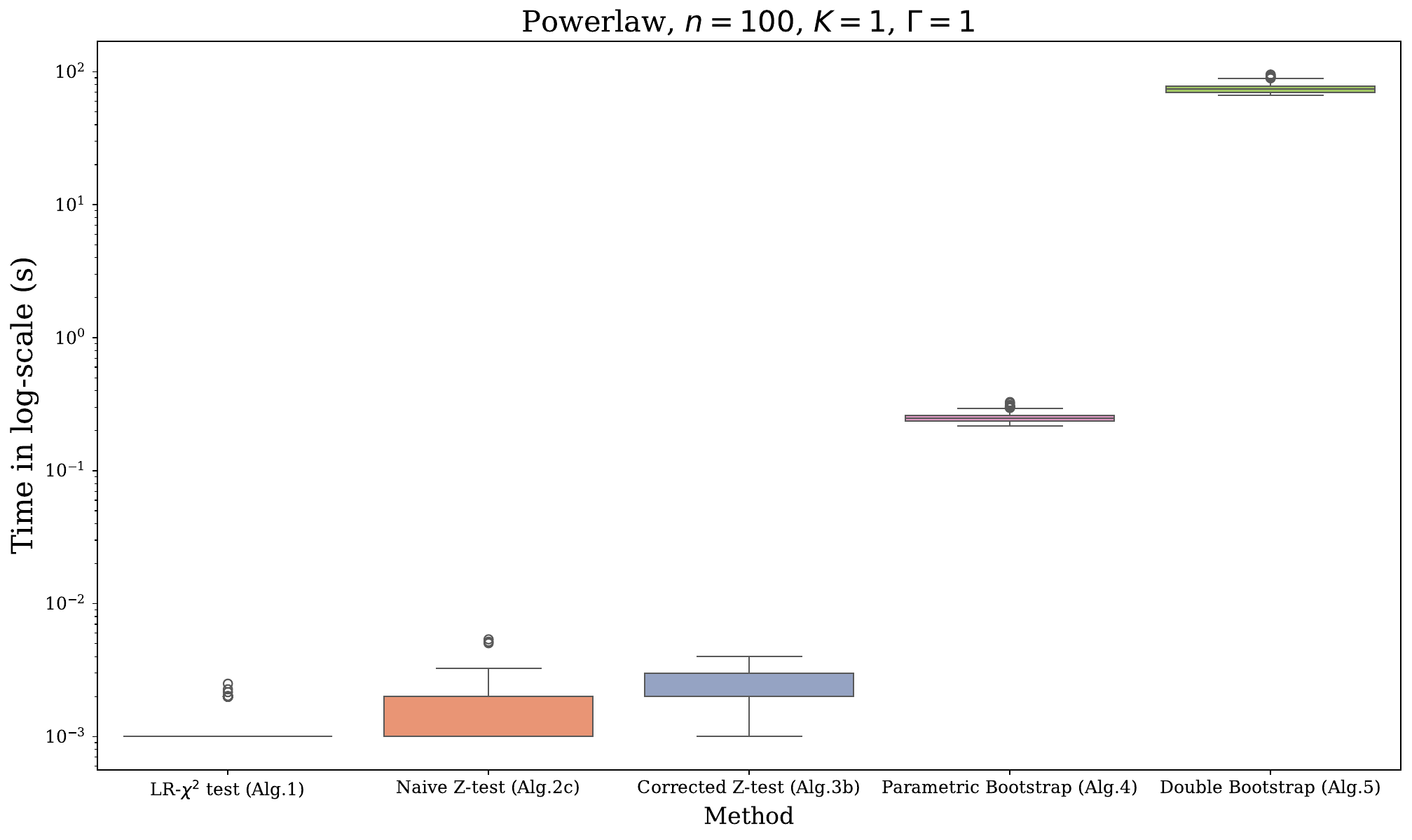}
    \caption{Comparison on the computational time of the algorithms. We simulate 100 replicate data sets under a Powerlaw with $\Gamma=1$, $K=1$ and $n=100$.  We used $B=300$ bootstrap replicates. For visualization purposes, a constant offset $(10^{-3})$ was added to all data points to accommodate the logarithmic scale; this transformation does not affect the relative comparisons.}
    \label{fig:time}
\end{figure}

\subsection{Tests With RMF}
\label{sec:sim_RMF}

{As shown in Equation~\eqref{eq:si2}, the evaluation of the Poisson rates $\{s_i(\boldsymbol{\theta})\}$ involves a summation resulting from the redistribution matrix file (RMF). The majority of the RMF's probability mass (around 80-90\%) typically lies along its diagonal and thus a simplified analysis might ignore the RMF altogether (as we do in Section~\ref{sec:sim}). Here we investigate the impact of this simplifying assumption by conducting simulation studies that full account for a realistic non-diagonal RMF.


We consider several RMFs $R(E,i)$ and analyze they effect the nominal null distributions of $C_n(\hat{\boldsymbol{\theta}})$ obtained from \texttt{Algorithms} 1,2 and 4. We consider the power-law spectral model, 
\begin{equation}
    \label{eq:spectral_model_rmf}
    g(\tilde{E},\boldsymbol{\theta})=K\cdot \tilde{E}^{-\Gamma},
\end{equation}
where $\boldsymbol{\theta}=\{K,\Gamma\}$, $n=50$, $A(\tilde{E_i})= 1$, $B_i=0.1$  and $\tilde{E}_i=(E_{j-1}+E_j)/2$, for $i=1,\dots,n$. We choose $E_j=1+\frac{i}{n}$ for $i=0,\dots,n$, $K/n\in\{1,10\}$ and $\Gamma=3$, and consider four choices of the RMF. Case 1 represents a well-refined observation in which the RMF is almost diagonal with diagonal entries close to $1$, while Case 2 represents a dispersed case where the diagonal entries of the RMF are close to $0.5$. We also consider two other extreme cases, $R=I$ and $R=\boldsymbol{1}\boldsymbol{1}^\top$, respectively. In this setup, the number of bins is large and the average expected count per channel is around $\frac{3}{8}K/n+B_i$.

\vspace{12pt}
\noindent
\textbf{Case 1}: Redistribution Matrix $R=\left(
\begin{array}{ccccc}
0.9&0.1& &\\
0.1&0.8&0.1&\\
&\ddots&\ddots&\ddots\\
& & 0.1&0.8&0.1\\
& & &0.1&0.9
\end{array}\right)\equiv D$,\\ which is almost diagonal.

\noindent
\textbf{Case 2}: Redistribution Matrix $R=(\bm{1}\bm{1}^\top+nI)/2n=\left(
\begin{array}{ccccc}
\frac{n+1}{2n}&\frac{1}{2n}&\cdots& \cdots &\frac{1}{2n}\\
\frac{1}{2n}&\frac{n+1}{2n}&\frac{1}{2n}& & \vdots\\
\vdots&\ddots&\ddots&\ddots&\vdots\\
\vdots & & \frac{1}{2n}&\frac{n+1}{2n}&\frac{1}{2n}\\
\frac{1}{2n}& \cdots &\cdots&\frac{1}{2n}&\frac{n+1}{2n}
\end{array}\right)$, which is dispersed. 

\begin{figure}[tbph]
    \centering
    \includegraphics[width=1\textwidth]{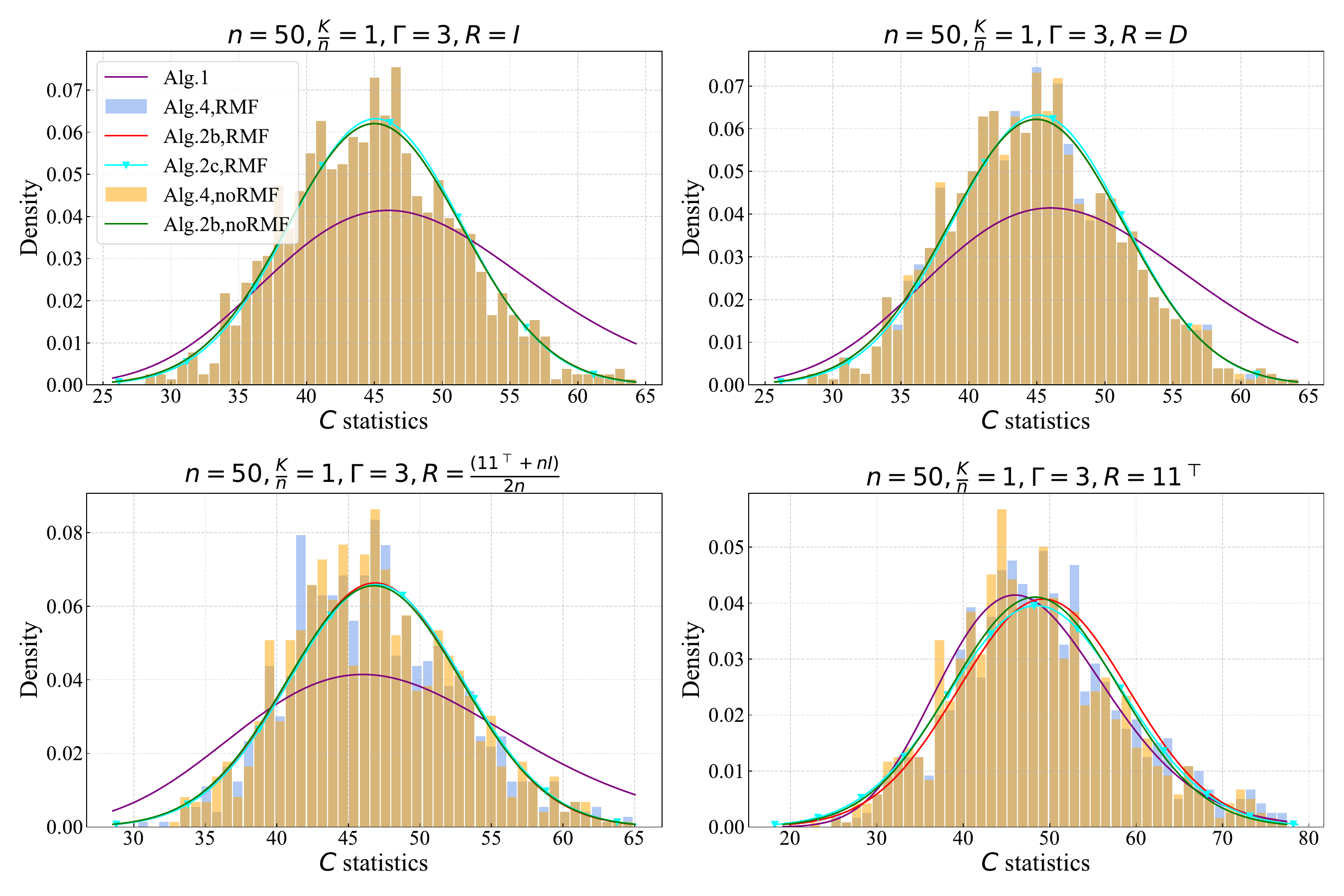}
    \caption{Histograms of the null distributions of $C_n(\hat{\boldsymbol{\theta}})$ from \texttt{Algorithms} 1,2 and 4 under four cases when $K/n=1$. {\texttt{Algorithm~4} was run with $B=1000$ bootstrap replicates.}} 
    \label{fig:hist_rmf_case1}
\end{figure}
\begin{figure}[tbph]
    \centering
    \includegraphics[width=1\textwidth]{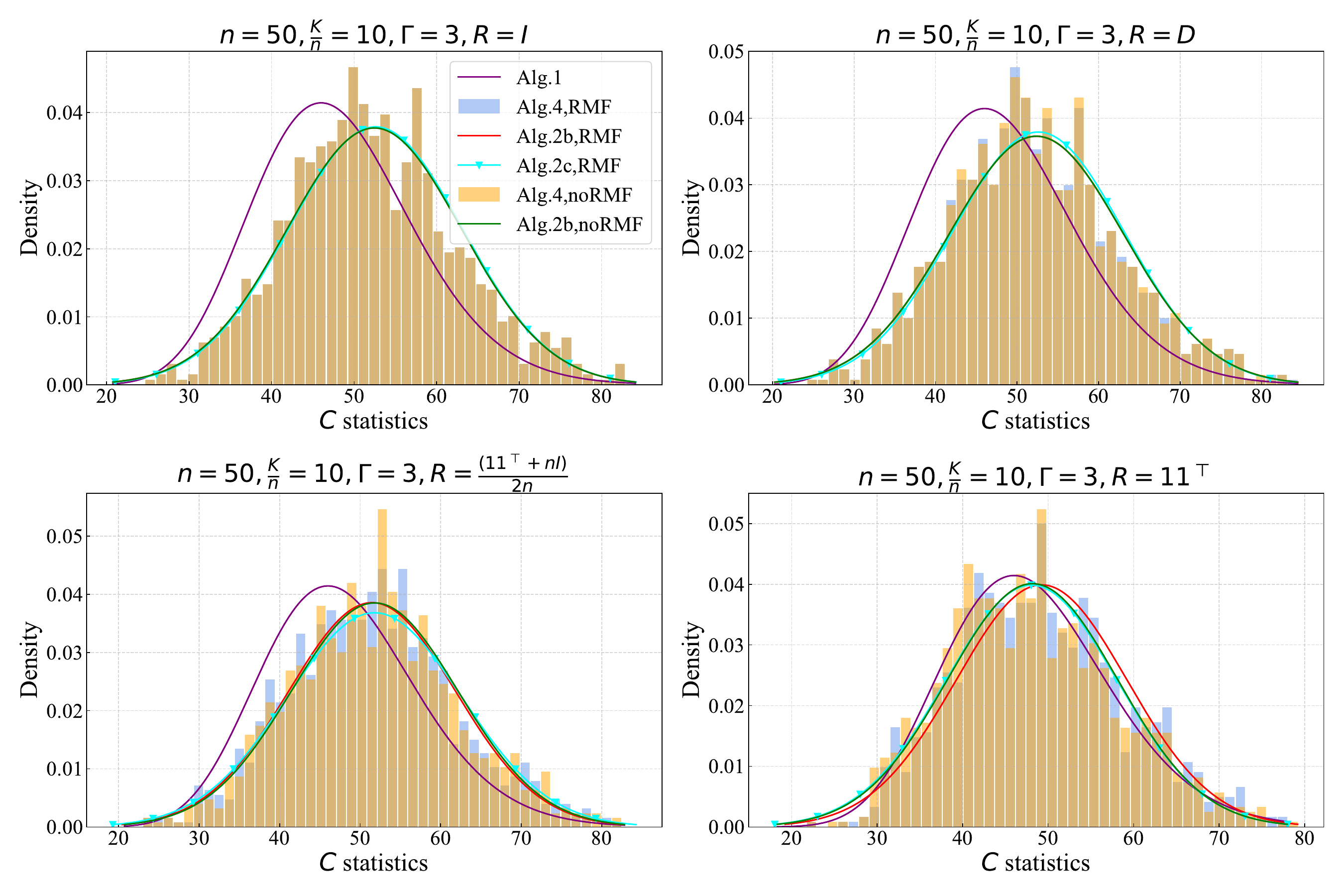}
    \caption{Histograms of the null distributions of $C_n(\hat{\boldsymbol{\theta}})$ from \texttt{Algorithms} 1,2 and 4 under four cases when $K/n=10$. {\texttt{Algorithm~4} was run with $B=1000$ bootstrap replicates.}} 
    \label{fig:hist_rmf_case2}
\end{figure}

The histograms of the null distributions of $C_n(\hat{\boldsymbol{\theta}})$ obtained from \texttt{Algorithms} 1,2 and 4 are illustrated in Figure \ref{fig:hist_rmf_case1} and \ref{fig:hist_rmf_case2}. Obviously, two histograms in each panel of Figures~\ref{fig:hist_rmf_case1} and \ref{fig:hist_rmf_case2} follow a similar pattern, demonstrating that while the RMF does impact the MLE, its effect on the distribution of $C_n(\hat{\boldsymbol{\theta}})$ is minimal, as long as the RMF is not extremely dispersed.  

The average computational costs of the Parametric Bootstrap (\texttt{Algorithm 4}) and the Corrected Z-test (\texttt{Algorithm} 3b) are recorded in Table \ref{tab:computation_time}. The algorithm using the high-order asymptotic is significantly more effective than the bootstrap algorithm. The computation time of both algorithms increases quadratically with the number of channels $n$. Note that the double Bootstrap algorithm will be much more time consuming than the single Parametric Bootstrap, making it too costly and unrealistic in practice.

\begin{table}[h!]
    \centering

    \caption{Average run time (in CPU seconds) of \texttt{Algorithms} 4 and 3b under Case 1.} 
    
    \smallskip
    \label{tab:computation_time}
        \begin{tabular}{lccccc}
    \hline\hline
    $n$ &10&30&50&75&100 \\
    \hline
   \texttt{Algorithms} 4: Parametric Bootstrap$^a$ &0.90 & 7.07& 20.24&47.36&84.99\\
    \texttt{Algorithms} 3b: Corrected Z-test &0.02 &0.13 &0.36 &0.77 &1.38\\
    \hline
    \multicolumn{3}{l}{\footnotesize $^a$ \texttt{Algorithms} 4 was run with $B=100$ bootstrap replicates.}
    \end{tabular}
\end{table}

\subsection{Additional Numerical Results ($\Gamma=1$)}

We give several additional numerical results in Figures~\ref{fig:compare_K&M_Gamma=1_K=10},~\ref{fig:cover_width_mu_Gamma1},~\ref{fig:compare_n_vary}, and~\ref{fig:compare_type1_gamma1}. 

\begin{figure}[t!]
    \centering
    \includegraphics[width=.97\textwidth]{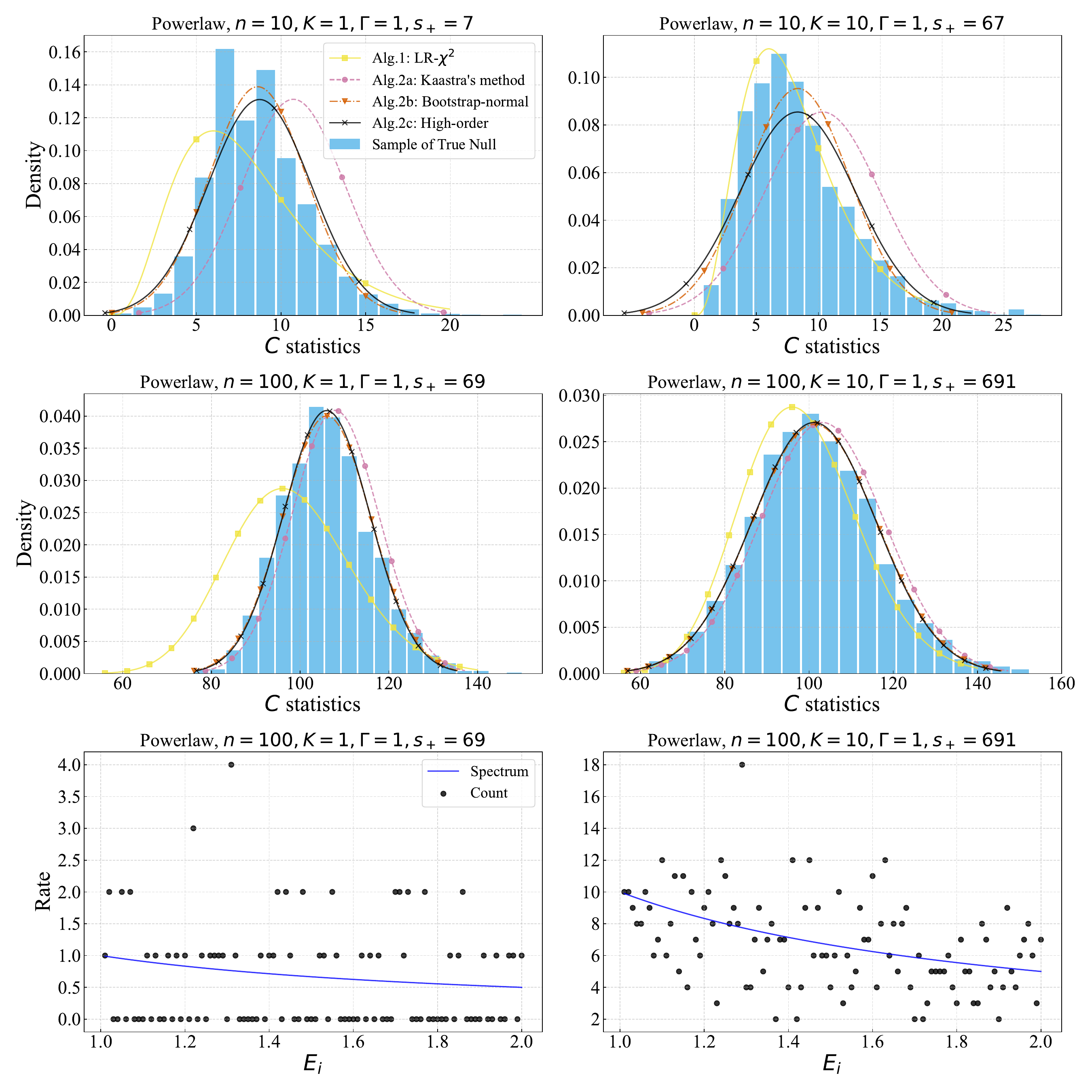}
    \spacingset{1.1}
    \caption{
    Comparison of a histogram of true null distribution of $C_n(\hat{\boldsymbol{\theta}})$ with the approximations of \texttt{Algorithm}~1 (LR-$\chi^2$) and three variants of \texttt{Algorithm}~2. Date are simulated from a Powerlaw model with $\Gamma=1$; the remaing parameters are listed in each panel; and the bootstrap size is $B=1000$. Row 3 shows the spectra and corresponding counts under these settings of the Powerlaw model.}
    \label{fig:compare_K&M_Gamma=1_K=10}
\end{figure}

\begin{figure}[t!]
    \centering
    \includegraphics[width=1\linewidth]{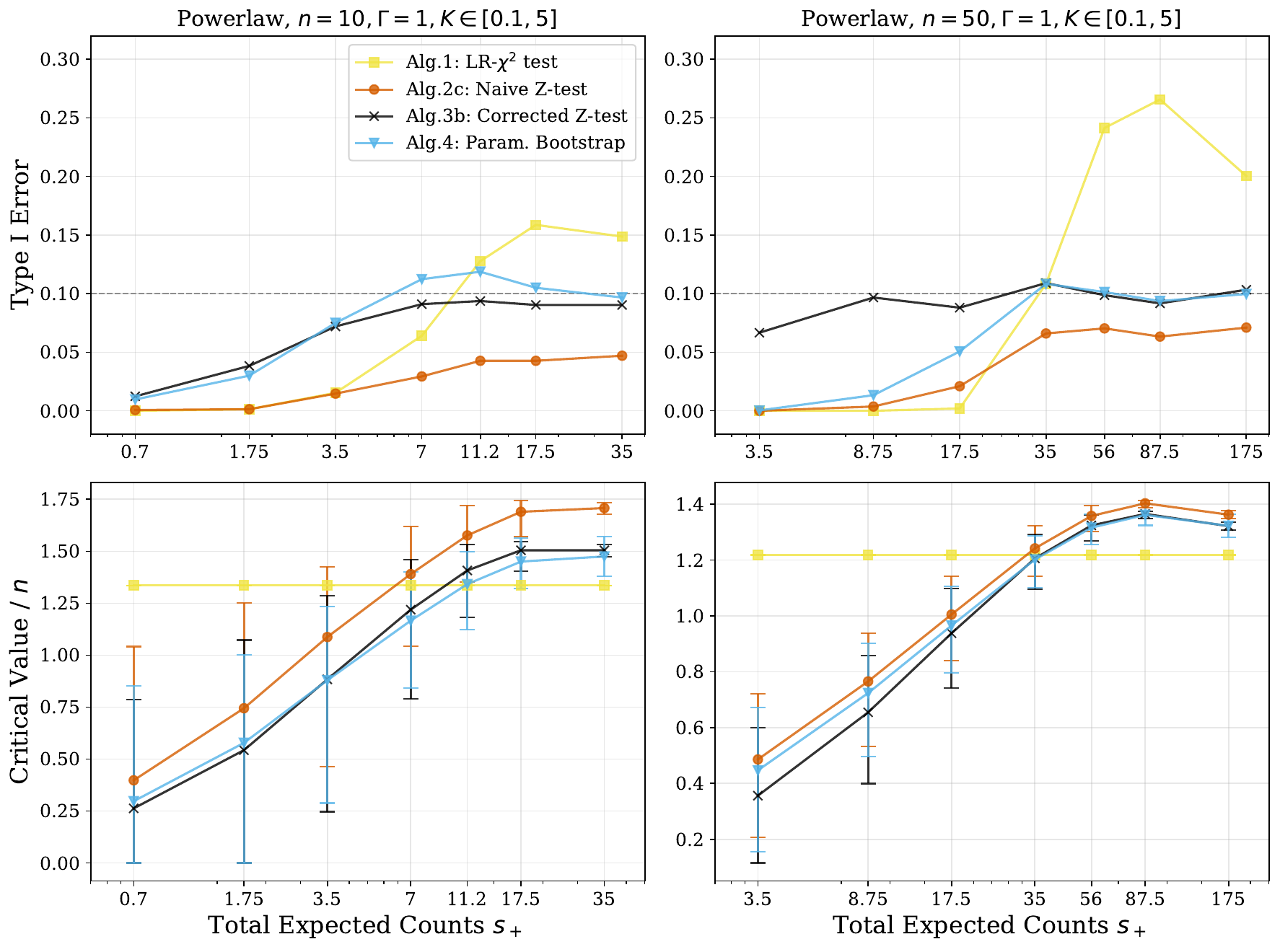}
    \spacingset{1.1}
    \caption{
    Performances of the four algorithms when $K\in\{0.1, 0.25, 0.5, 1, 1.6, 2.5, 5\}$ varies. The true models and null models are a Powerlaw with $\Gamma=1~(s_+\approx nK\log2 )$. We simulated 3000 replicate data sets and used $B=300$ bootstrap replicates.  The dashed line in the first row is the nominal Type~I error rate. Tests with Type~I error rates close to the nominal rate and small critical values are preferred. 
    The simulation shows the overall strong performance of our recommended Corrected $Z$-test.
    }
    \label{fig:cover_width_mu_Gamma1}
\end{figure}

\begin{figure}[t!]
    \centering
    \includegraphics[width=1\linewidth]{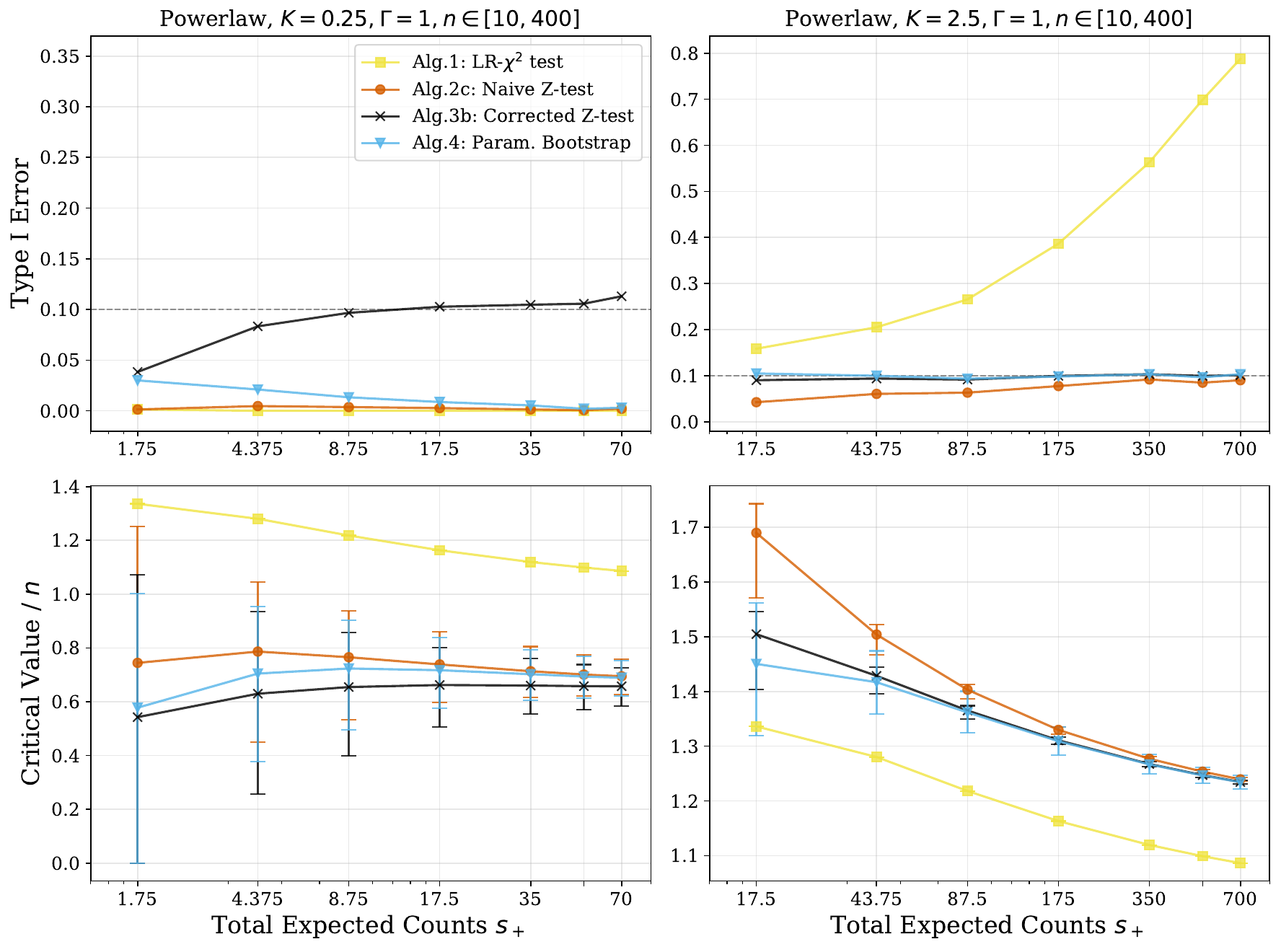}
    \spacingset{1.1}
    \caption{
    Performances of the four algorithms when $n\in\{10, 25, 50, 100, 200, 300, 400\}$ varies.  The simulations set up is as in Figure~\ref{fig:cover_width_mu_Gamma1}, but results are plotted as a function of $n$. The overall strong performance of our recommended Corrected $Z$-test is again evident. }
    \label{fig:compare_n_vary}
\end{figure}

\begin{figure}[t!]
    \centering
    \includegraphics[width=.95\linewidth]{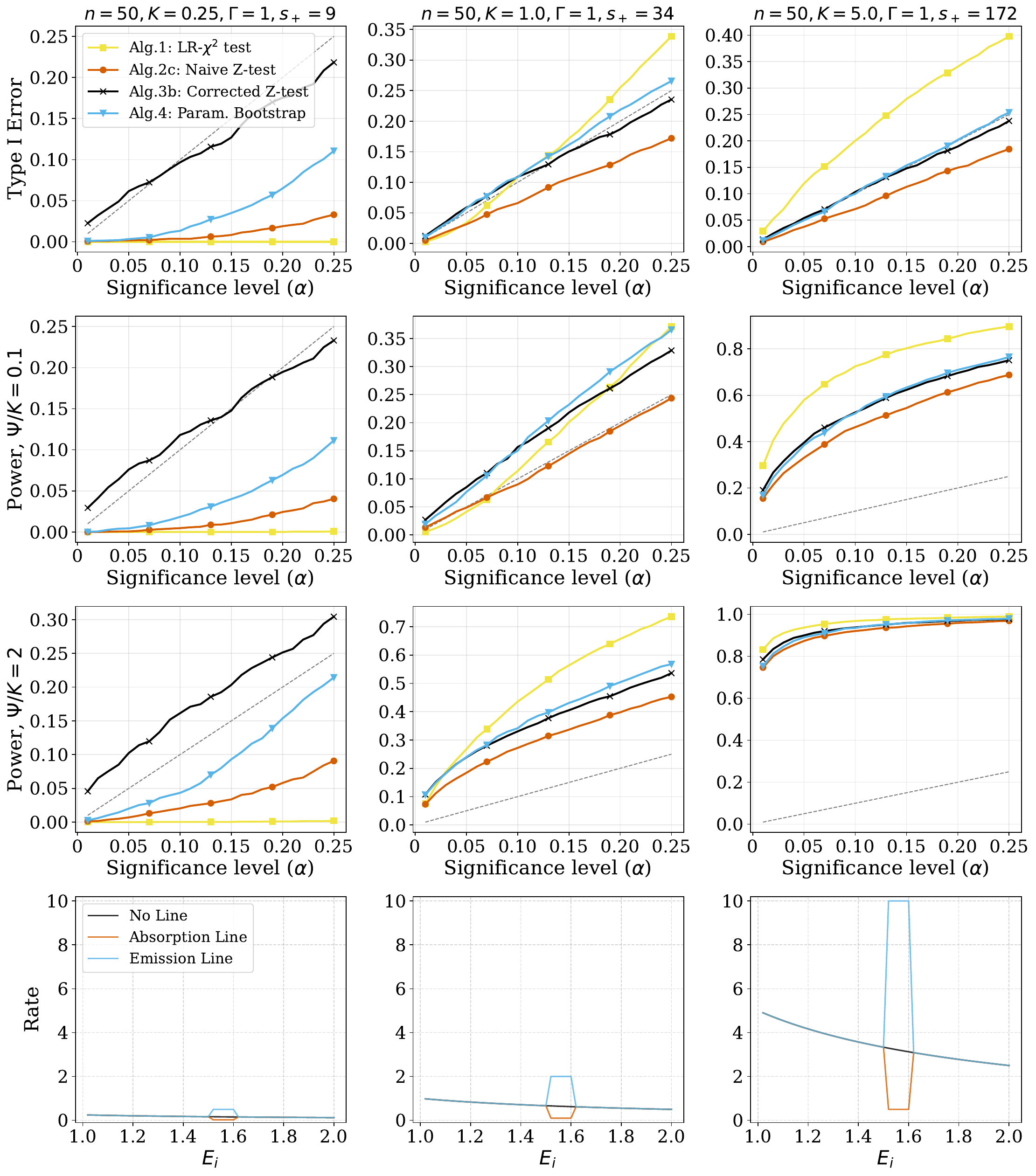}
    \spacingset{1.1}
    \caption{
    Comparison on the Type~I error rate and power of the four algorithms. We simulate 3000 replicate data sets under a Powerlaw (row 1), Powerlaw + Absorption Line  (row 2), and  Powerlaw + Emission Line (row 3).  In all cases $\Gamma=1$, $n=50$, we used $B=300$ bootstrap replicates, and $K$ varies among the columns. For models with an absorption/emission line, we choose $b=n/10$ and $m=n/2$. Each panels plots the probability of rejecting the null Powerlaw model (no emission/absorption line). This corresponds to Type~1 error in row 1 and power in rows 2 and 3.  
Ideally, the power should be as large as possible while maintaining the Type~I Error below $\alpha$. Overall, our recommended Corrected $Z$-test is best calibrated in terms of Type~I error rate and power.
}
\label{fig:compare_type1_gamma1}
\end{figure}

\end{document}